\documentclass[onecolumn,floatfix,aps,superscriptaddress]{revtex4}
\usepackage{graphicx}
\usepackage{amsmath}
\usepackage[colorlinks=true, citecolor=blue, urlcolor=blue, linkcolor = blue]{hyperref}
\usepackage{amssymb}
\usepackage{bm}
\usepackage{color}
\usepackage{bookmark}
\usepackage{tabularx}
\usepackage{mathtools}
\usepackage{microtype}
\usepackage{cancel}
\usepackage{relsize}

\usepackage{xcolor}

\begin{document}
\title{Nonlinear dynamical modelling of high frequency electrostatic drift waves using fluid theoretical approach in magnetized plasma}
\author{S. P. Acharya
\footnote{Electronic mail: siba.acharya@saha.ac.in and siba.acharya39@gmail.com}}
\affiliation{Saha Institute of Nuclear Physics, a Constituent Institute of Homi Bhabha National Institute (HBNI), 1/AF Bidhannagar, Kolkata-700064 (India)}
\author{M. S. Janaki
\footnote{Electronic mail: ms.janaki@saha.ac.in}}
\affiliation{Saha Institute of Nuclear Physics, a Constituent Institute of Homi Bhabha National Institute (HBNI), 1/AF Bidhannagar, Kolkata-700064 (India)}
\begin{abstract}
A novel third order nonlinear evolution equation governing the dynamics of the recently observed high frequency electrostatic drift waves has been derived in the framework of a plasma fluid model in an inhomogeneous magnetized plasma. The corresponding linear dispersion relation arising out of the fluid equations has been studied in detail for the conventional low frequency as well as the recently observed high frequency electrostatic drift waves along with the electrostatic ion cyclotron waves. The derived nonlinear evolution equation of third order is then decomposed into two second order equations as the order of the equation becomes reduced after this kind of decomposition under certain conditions, and one equation becomes linear whereas the other equation becomes nonlinear. The detailed analysis of fixed points leading to bifurcations has been performed using the qualitative theory of differential equations and the bifurcation theory of planar dynamical systems for the reduced second order nonlinear equation. The bifurcation curves for the individual fixed points are also shown for the variations of different parameters in the parameter space of this reduced nonlinear equation. Then some exact as well as approximate travelling wave solutions of this reduced nonlinear equation have been derived. As the other second order reduced equation is linear in nature, its exact oscillatory and exponential solutions are obtained. The intersection of the solutions of these two reduced second order equations provides the solutions of the original third order nonlinear evolution equation; it is verified that the solutions of the reduced nonlinear second order equation are subsets of the oscillatory solution of the reduced linear second order equation in most cases whereas the intersection is different if the exponential solution of the reduced linear second order equation is considered. This further implies that the solutions of the reduced second order nonlinear equation can directly represent the solutions of the original third order nonlinear evolution equation in most cases representing the dynamics of the nonlinear high frequency electrostatic drift waves. These solutions of the novel third order nonlinear evolution equation represent certain vortex-like structures as our work is restricted to $(1+1)$ dimensions only. The possible applications of our novel results on the dynamics of the high frequency electrostatic drift waves are discussed along with future directions for research in this field.
\end{abstract}
\maketitle

\section{Introduction}
The research on drift waves in magnetically confined inhomogeneous plasmas, in devices such as tokamaks, Q-machines, helicon discharges and various other fusion devices etc. as well as other laboratory, space and astrophysical plasma systems, is of profound importance due to its significant contributions in the understanding of transport phenomena in these systems. The drift waves are universal modes predominantly observed in the regions where the density gradients are maximum in magnetized plasmas. Especially the conventional low frequency unstable drift waves have a significant role in the anomalous transport phenomenon in plasmas. Recently, a detailed review dealing with the theory of drift waves for transport in fusion devices like tokamaks has been reported by Weiland and Zagorodny \cite{Weiland}. The frequencies of these conventional drift waves are low in the sense that they are very much less than the ion cyclotron frequency, i.e. $\omega<<\omega_{ci}$ where $\omega$ and $\omega_{ci}$ denote the drift wave frequency and ion cyclotron wave frequency respectively. These drift waves are usually observed in the regions corresponding to non-vanishing values of the plasma density gradients. Various studies on these drift waves have been carried out extensively since their existence was first demonstrated in laboratory plasmas \cite{Angelo}. Most of the previous experimental works on drift waves have been performed in the low frequency domain $\omega<<\omega_{ci}$ due to the fact that the experimental devices generally have magnetic fields of the order of several kilogauss or higher; however, one cannot avoid the possibility of having $\omega \geq \omega_{ci}$, particularly, if the magnetic field is low, i.e. of the order of hundreds of gauss \cite{Ghosh2015, Ghosh2017}.

Mikhailovskii and Timofeev \cite{Mikhailovskii} have investigated the stability of a non-uniform plasma at frequencies that are multiples of the ion cyclotron frequency long ago and reported a dispersion relation which represents the intersection of a drift branch and a cyclotron branch. Hendel and Yamada \cite{Hendel} have experimentally identified ion cyclotron drift waves (ICDW) with discrete and continuous spectra having frequencies in the range $\omega \geq \omega_{ci}$, the effects of which on the plasma are similar to that for the ion sound waves. These waves are shown to exist in plasmas which are ion sound stable \cite{Hendel}. Drift waves in the presence of finite density gradients and magnetic shears have been explored by Vranjes and Poedts \cite{Vranjes2004} while dealing with the global drift and oppositely propagating ion cyclotron waves in a plasma with Gaussian density distribution. The resistive drift Alfven wave, localized in the region corresponding to the largest plasma density gradient in a magnetized helicon plasma, has been explored in \cite{Sun}. The coronal heating mechanisms caused by the drift waves in the solar atmosphere has been explored by Vranjes and Poedts \cite{Vranjes2009} in the context of coupled unstable drift cyclotron modes using both fluid and kinetic theories. Different studies on the drift wave turbulence in linear plasmas have been performed by Yamada et al. \cite{Yamada2007,Yamada2008} using multi-probe systems. A pair of model equations dealing with the nonlinear coupling between electrostatic ion cyclotron and drift waves has been derived by Pokhotelov et al. \cite{Pokhotelov} in a non-uniform magnetoplasma. The high frequency magnetic drift wave is investigated by Huba \cite{Huba1991} in the frameworks of a fluid theory based on the resistive magnetohydrodynamic (MHD) equations as well as a kinetic theory based on the Vlasov equation; the fluid theory is valid in the frequency range $\omega<\omega_{lh}$ whereas the kinetic theory is valid in the frequency range $\omega>\omega_{lh}$ where $\omega_{lh}$ represents the lower hybrid frequency. It is shown in \cite{Huba1991} that the drift wave mode transitions into the lower hybrid wave mode in the limit $\omega \geq \omega_{lh}$. Kaneko et al. \cite{Kaneko} have explored the simultaneous excitations of several azimuthal drift wave modes driven by the field-aligned flow velocity shear in a Q-machine plasma.

Besides the above explorations, the research endeavours comprising various aspects of the excitations of drift waves have progressed a lot in the past. Rosenberg and Krall \cite{Rosenberg} have investigated high frequency drift wave instabilities in the frequency range  $\omega_{ci}<<\omega<<\omega_{ce}$, where $\omega_{ce}$ denotes the clectron cyclotron frequency, in a dusty magnetized plasma and discussed their applications to the Saturn's F-ring. A general dispersion relation governing the dynamics of the high frequency, i.e. $\omega_{ci}<<\omega<<\omega_{ce}$, electromagnetic drift waves has been derived long ago by Detyna and Wooding \cite{Detyna} along with the stability analysis. Detyna and Wooding \cite{Detyna1972} have also derived a dispersion relation for the ultra high frequency drift waves due to an electron stream drifting through a density gradient in a plasma. Rosenberg and Merlino \cite{Rosenberg2013} have discussed the drift wave instability in a plasma consisting of both positive and negative ions using the linear kinetic theory. Shukla and Rosenberg \cite{ShuklaRos2012} have explored dissipative drift wave instability in a radially bounded dusty magnetoplasma. In their subsequent work \cite{ShuklaRos}, coupled dust ion-acoustic drift wave instability in a radially bounded dusty magnetoplasma with ion velocity shear has been investigated using fluid models. The very low frequency electrostatic drift waves in a magnetized dusty plasma have been investigated by Salimullah et al. \cite{Salimullah} in the framework of the Vlasov-kinetic model. Gravier et al. \cite{Gravier} have studied experimentally the chaotic regimes of the nonlinear drift waves in a magnetized plasma using time-delay autosynchronization method and power spectrum. Garcia et al. \cite{Garcia} have explored different models for dealing with the electrostatic drift waves with the density gradients along the magnetic field directions in gravitational ionospheres. Various dynamical regimes of collisional drift waves in a magnetized plasma are experimentally studied by Gravier et al. \cite{Gravier2000}. In particular, regular, chaotic and turbulent regimes corresponding to these unstable low-frequency electrostatic drift waves have been analyzed, and the time-delay auto-synchronization method of controlling chaos has been applied in dealing with the dynamics. They \cite{Gravier2000} have also performed a numerical study based on the coupled nonlinear oscillators exhibiting chaos and compared the results of this model to the experimental results. Recently, Hassan et al. \cite{Hassan} have investigated the toroidal ion temperature gradient (TITG) driven pure drift modes leading to the formation of solitons and shocks in the presence of non-Maxwellian electrons in an electron-ion inhomogeneous magnetized plasma. Wu et al. \cite{Wu} have recently investigated the spectral modification of the lower hybrid wave caused by low frequency drift-wave-type density fluctuations in the framework of the wave scattering model via a Monte Carlo method. The resistive drift wave instabilities in a non-uniform magnetized plasma have been explored by Rehman \cite{Rehman2018} along with the considerations of both thermal and non-thermal electron distributions. The electromagnetic resistive drift wave instabilities have been analysed in the framework of a two fluid model by Rehman \cite{Rehman} in fusion plasmas by taking the effects of the ion viscosity into account. 


Most of the investigations of drift waves mainly concentrate on the conventional low frequency domain $\omega<<\omega_{ci}$; although, some works on ion cyclotron drift waves have been done in the frequency range $\omega \geq \omega_{ci}$. As far as our knowledge goes, there have not been reported any work on electrostatic drift waves in the frequency regime $\omega \geq \omega_{ci}$ except the recent experimental observations of Ghosh et al. \cite{Ghosh2015,Ghosh2017}. In the last decade, Ghosh et al. \cite{Ghosh2015} have experimentally investigated the excitations of electrostatic drift waves having frequencies greater than the ion cyclotron frequency in an Argon plasma. In particular, they have investigated the excitations of different electrostatic drift modes having azimuthal propagations in the direction of electron diamagnetic drift in a radio frequency (RF) produced magnetized Argon plasma expanding into a large expansion chamber; which is equipped in the double layer experimental (DLX) device \cite{SahaSK}. The maximum amplitudes of the drift wave modes are observed at the locations in the radial direction corresponding to the maximum radial density gradient. Experimentally it is also concluded by Ghosh et al. \cite{Ghosh2015} that the mode frequency is independent of the location of measurement provided the neutral pressure and magnetic field are fixed. They \cite{Ghosh2015} have also derived a local dispersion relation in the framework of a plasma fluid model predicting destabilization of the drift wave modes with frequencies greater than the ion cyclotron frequency caused by the electron-neutral collisions. In their subsequent work \cite{Ghosh2017}, they have reported the excitations of electrostatic drift waves having frequencies less than the ion cyclotron frequency but greater than that of the conventional low frequency drift waves in a magnetized Helium plasma. The observed drift wave modes mainly propagate in the azimuthal direction, which coincides with electron diamagnetic drift direction as in \cite{Ghosh2015}, along with weak axial propagations. It should be noted that these drift waves, having frequencies less than or greater than or of the order of the ion cyclotron frequency but always greater than the conventional low frequency drift waves, have been referred to as 'upper drift waves' in \cite{Ghosh2015,Ghosh2017} to distinguish them from the conventional low frequency drift waves. In this work \cite{Ghosh2017} on Helium plasma, the simultaneous excitations of two high frequency drift waves also termed as 'dual upper drift waves' have been found whereas this was not possible in case of Argon plasma \cite{Ghosh2015}. This is due to the larger value of the ion Larmor radius in Argon plasma than that in Helium plasma. Also a local dispersion relation in the framework of the plasma fluid model has been derived in \cite{Ghosh2017} resulting in an instability for the dual upper drift waves. 

 As the high frequency electrostatic drift waves have already been observed in experiments as discussed earlier \cite{Ghosh2015, Ghosh2017}, a detailed theoretical investigation of this spectacular phenomenon is further warranted in the frameworks of the appropriate fluid and kinetic models. As far as our knowledge goes, there have not been reported any detailed theoretical work which can explain the dynamics of the novel high frequency electrostatic drift waves. In this context, it is noteworthy that the pseudo-three-dimensional dynamics of the conventional low frequency, i.e. $\omega<<\omega_{ci}$, electrostatic drift waves can be described by the Hasegawa-Mima (HM) equation, which is a simple nonlinear equation derived by Hasegawa and Mima \cite{Hasegawa} for the non-uniform magnetized plasma by considering the three-dimensional electron but two-dimensional ion dynamics. Horton and Hasegawa \cite{Horton} have also performed a detailed review concerning the linear and nonlinear quasi-two-dimensional dynamics of plasmas and fluids. In particular, they \cite{Horton} have discussed the Charney and Hasegawa-Mima (CHM) equation, which is a common convective vortex equation describing the Rossby waves in planetary atmospheres and low frequency conventional drift waves in plasmas. Taking these interesting facts into account, we have proposed a novel nonlinear evolution equation (NLEE) governing the dynamics of the electrostatic drift waves in the high frequency regime in this article. Here the high frequency regime implies that the frequencies are greater than that of the conventional low frequency drift waves. In this work, fluid theory has been employed instead of kinetic theory even for the high frequency regime because the conditions for the validity of the kinetic theory, i.e., ${k^2}_{\perp} \rho^2_i>>1$, is violated for the experimental values of $k_{\perp}$ and $\rho_i$ corresponding to the excitations of the high frequency electrostatic drift waves \cite{Ghosh2015, Ghosh2017}. Here $k_{\perp}$ and $\rho_i$ denote the wave number perpendicular to the magnetic field direction and ion Larmor radius respectively.

 This article is organized in the following manner. The detailed derivation of a novel third order nonlinear evolution equation governing the dynamics of the high frequency electrostatic drift waves is provided in Sec. \ref{Der}. The linear dispersion relations corresponding to the conventional low frequency as well as the recently observed high frequency electrostatic drift waves and the electrostatic ion cyclotron waves have been analyzed in Sec. \ref{DR} in detail. The fixed points leading to bifurcations of the phase portraits of the reduced second order nonlinear equation have been investigated in Sec. \ref{PPBifur} employing the theory of planar dynamical systems. Some exact and approximate travelling wave solutions of the reduced nonlinear second order equation are derived in Sec. \ref{Solution} using the first integral whereas the exact solutions of the reduced linear second order equation are provided in Sec. \ref{ExSec}. The final solutions of the original third order nonlinear evolution equation are explored in Sec. \ref{Intersection}. The novel findings of our theoretical study on the dynamics of the high frequency electrostatic drift waves are discussed along with possible applications and future research plans in Sec. \ref{DisApp}. The concluding remarks are provided in Sec. \ref{Con} followed by acknowledgements and bibliography.
\section{Derivation of novel nonlinear evolution equation for high frequency electrostatic drift waves} \label{Der}
We consider a non-uniform collisionless electron-ion plasma to be immersed in a homogeneous magnetic field $B_0 \hat{z}$ where $B_0$ is the strength of the external homogeneous magnetic field. The ion dynamics in the system is governed by the following model equations \cite{Horton}:
\begin{equation}
\vec{\nabla}.\vec{v}=-\frac{d}{dt}(ln\,n), \label{Cont1}
\end{equation}
\begin{equation}
\frac{d\vec{v}}{dt}=-\frac{e}{m_i}\vec{\nabla}\phi+\omega_{ci}\vec{v}\times{\hat{z}}, \label{Momentum}
\end{equation}
where Eq. (\ref{Cont1}) denotes the continuity equation corresponding to the number density conservation of ions, and Eq. (\ref{Momentum}) denotes the Lorentz equation of motion for the cold ion fluid in an electrostatic field $\vec{E}=-\vec{\nabla}\phi$. Here $\vec{v}$ is the ion fluid velocity, $n$ is the ion number density, $c_s$ is the speed of sound, $\phi$ is the electrostatic potential and $\omega_{ci}$ is the ion cyclotron frequency. According to the quasi-neutrality condition, we have $n \approx n_e$ where $n_e$ denotes the electron number density. In our work, the electrons are assumed to obey the Boltzmann distribution:
\begin{equation}
n \approx n_e=n_{eo}exp(\frac{e\phi}{k_BT_e}). \label{Boltzmann}
\end{equation}
 We further assume that the waves propagate nearly perpendicular to the magnetic field so that all the derivatives in the $z$-direction are neglected. This assumption further rules out any coupling to the ion acoustic wave. The equilibrium ion density gradient is considered to be along the $x$-axis with a density gradient scale-length given by 
\begin{equation}
{L_n}^{-1}=-
\frac{d}{dx}[ln\, n_0(x)], \label{Ln}
\end{equation}
where $L_n$ denotes the density gradient scale-length and $n_0(x)$ denotes the equilibrium ion density. Substituting Eq. (\ref{Boltzmann}) in Eq. (\ref{Cont1}), we obtain
\begin{equation}
 \vec{\nabla}.\vec{v}=-[\frac{\partial \tilde{\phi}}{\partial t}+(\vec{v}.\vec{\nabla})ln \, n_0+(\vec{v}.\vec{\nabla})\tilde{\phi}] \approx \vec{\nabla}_{\perp}.\vec{v}_{\perp};\,\tilde{\phi}=\frac{e \phi}{k_B T_e}. \label{DeldotV}
\end{equation}
In writing the above Eq. \ref{DeldotV}, we have neglected the ion inertia in the direction of the external magnetic field, i.e. $\hat{z}$, assuming a pseudo-three-dimensional situation following \cite{Horton}. This type of pseudo-three-dimensional assumption is desirably consistent with the condition of the existence of the drift waves. In fact, this corresponds to the characteristic feature of the drift waves to have negligible axial propagations, which distinguishes them from the other kinds of waves present in the system having non-negligible axial propagations. This implies that $\vec{\nabla}.\vec{v} \approx \vec{\nabla}_{\perp}.\vec{v}_{\perp}=\frac{\partial v_x}{\partial x}+\frac{\partial v_y}{\partial y}$ as $\frac{\partial v_z}{\partial z}$ is negligible due to the weak axial propagations of the drift waves. In reality, the drift wave is basically a vortex mode. Therefore, it is reasonable to construct an equation for the vorticity $\vec{\Omega} =\vec{\nabla} \times \vec{v}$ by taking the curl of Eq. (\ref{Momentum}), as done in \cite{Horton}: 
\begin{equation}
\frac{d}{dt}(\Omega+\omega_{ci})+(\Omega+\omega_{ci})\vec{\nabla}_{\perp}.\vec{v}_{\perp}=0. \label{OmegaEq}
\end{equation}
In deriving the above Eq. (\ref{OmegaEq}), the following relations have been used in \cite{Horton}:
$$
\frac{d\vec{v}}{dt}= \frac{\partial \vec{v}}{\partial t}+(\vec{v}.\vec{\nabla})\vec{v}=\frac{\partial \vec{v}}{\partial t}+\frac{1}{2}{\nabla}v^2-\vec{v}\times\vec{\Omega}$$
\begin{equation}
\vec{\nabla}\times(\vec{v} \times \vec{\Omega})=-(\vec{\nabla}.\vec{v})\vec{\Omega}+(\vec{\Omega}.\vec{\nabla})\vec{v}-(\vec{v}.\vec{\nabla})\vec{\Omega}=-(\vec{\nabla}_{\perp}.\vec{v}_{\perp})\vec{\Omega}-(\vec{v}.\vec{\nabla})\vec{\Omega}. \label{Relation}
\end{equation}
After taking the divergence in both sides of Eq. (\ref{Momentum}), we get 
\begin{equation}
\frac{\partial}{\partial t}(\vec{\nabla}.\vec{v})+\frac{1}{2}{\nabla}^2v^2-\vec{\nabla}.(\vec{v}\times\vec{\Omega})=-\frac{e}{m_i} {\nabla}^2\phi+\vec{\nabla}.(\vec{v}\times{\vec{\omega}}_{ci}), \label{Rel2i}
\end{equation}
where we have used Eq. (\ref{Relation}). Now, after substituting Eq. (\ref{DeldotV}) in Eq. (\ref{Rel2i}), we get
\begin{equation}
\frac{\partial^2\tilde{\phi}}{\partial t^2}+\frac{\partial}{\partial t}(\vec{v}.\vec{\nabla})\tilde{\phi}-c^2_s\nabla^2\tilde{\phi}+\omega_{ci}\Omega_z-\frac{1}{L_n}\frac{\partial v_x}{\partial t}=\frac{1}{2}\nabla^2v^2-\Omega^2+\vec{v}.(\vec{\nabla}\times\vec{\Omega}), \label{PhiEq1}
\end{equation}
where Eq. (\ref{Ln}) has been used during the simplification process. Here $c_s$ denotes the sound speed that is given as: $c^2_s=\frac{k_BT_e}{m_i}$. Now, using Eq. (\ref{DeldotV}) in Eq. (\ref{OmegaEq}), we get
\begin{equation}
\frac{\partial \Omega_z}{\partial t}+(v_x\frac{\partial}{\partial x}+v_y\frac{\partial}{\partial y})\Omega_z-(\Omega_z+\omega_{ci})[\frac{\partial \tilde{\phi}}{\partial t}-\frac{v_x}{L_n}+(v_x\frac{\partial}{\partial x}+v_y\frac{\partial}{\partial y})\tilde{\phi}]=0,\label{Omegaz1}
\end{equation}
where $L_n$ is given by Eq. (\ref{Ln}). Here $\vec{\nabla}=\hat{x}\frac{\partial}{\partial x}+\hat{y}\frac{\partial}{\partial y}$. We assume that there is no propagation along the $x$-direction, i.e. $\frac{\partial}{\partial x}=0$. Here it should be noted that we have retained $L_n$ in our formulation whereas we have assumed that $\frac{\partial}{\partial x}=0$ as we intend to retain the density gradiemt scale length in $x$-direction and consider the propagation of the nonlinear drift wave in the $y$-direction, i.e. the other physical quantities excluding $L_n$ are assumed to have no dependence in the $x-$direction. Then, the following stationary frame transformation has been applied: 
\begin{equation}
\xi=y-vt;\,\frac{\partial}{\partial y}=\frac{d}{d \xi};\,\frac{\partial}{\partial t}=-v\frac{d}{d \xi}, \label{FramTr}
\end{equation}
to yield
\begin{equation}
-v\frac{d\Omega_z}{d\xi}+v_y\frac{d\Omega_z}{d\xi}-(\Omega_z+\omega_{ci})(-v\frac{d\tilde{\phi}}{d\xi}-\frac{v_x}{L_n}+v_y\frac{d\tilde{\phi}}{d\xi})=0, \label{OmgZFramTr}
\end{equation}
where the partial derivatives have been replaced by the standard derivatives as the dependent variables now depend only on the composite variable $\xi$. As the vorticity $\vec{\Omega}=\vec{\nabla}\times\vec{v}$, we have
\begin{equation}
\vec{\Omega}=(0,0,\Omega_z);\,\Omega_z=-\frac{\partial v_x}{\partial y}=-\frac{\partial v_x}{\partial \xi}, \label{OmgZ}
\end{equation}
where $\vec{\nabla}=(\frac{\partial}{\partial x},\frac{\partial}{\partial y},0)$, $\vec{v}=(v_x,v_y,0)$, and $\frac{\partial}{\partial x}=0$ have been used. Now Eq. (\ref{Momentum}) can be decomposed into two components:
\begin{equation}
\frac{\partial v_x}{\partial t}+v_y\frac{\partial v_x}{\partial y}=\omega_{ci} v_y, \label{MomtX}
\end{equation}
\begin{equation}
\frac{\partial v_y}{\partial t}+v_y\frac{\partial v_y}{\partial y}=-\frac{e}{m_i}\frac{\partial \phi}{\partial y}- \omega_{ci} v_x, \label{MomtY}
\end{equation}
where we have used $\frac{\partial}{\partial x}=0$. Eq. (\ref{MomtX}) again implies
\begin{equation}
v_y=\frac{\frac{\partial v_x}{\partial t}}{\omega_{ci}-\frac{\partial v_x}{\partial y}}. \label{Vy}
\end{equation}
Substituting Eqs. (\ref{OmgZ}) and (\ref{Vy}) in Eq. (\ref{OmgZFramTr}), we get the following equation after neglecting the third order terms.
\begin{equation}
\frac{d\tilde{\phi}}{d\xi}=\frac{\frac{d^2v_x}{d \xi^2}-\frac{2}{vL_n}v_x\frac{dv_x}{d\xi}+\frac{\omega_{ci}}{vL_n}v_x}{\frac{dv_x}{d\xi}-\omega_{ci}}. \label{DphiDxi}
\end{equation}
Now, following the same procedure, by putting $\frac{\partial}{\partial x}=0$ and using the frame transformation Eq. (\ref{FramTr}) in Eq. (\ref{PhiEq1}), we get
\begin{equation}
(v^2-c^2_s)\frac{d^2\tilde{\phi}}{d \xi^2}-(\omega_{ci}-\frac{v}{L_n})\frac{dv_x}{d\xi}=\frac{d}{d\xi}[v_y(v\frac{d\phi}{d\xi}+\frac{dv_y}{d\xi})], \label{D2phiDxi2Eq}
\end{equation}



where we have used Eqs. (\ref{OmgZ}) and (\ref{Vy}) during the simplification process. Following some rigorous calculations after substituting $\frac{d\tilde{\phi}}{d\xi}$ from Eq. (\ref{DphiDxi}) in the above Eq. (\ref{D2phiDxi2Eq}), we get
\begin{equation}
(v^2+c^2_s)\frac{d v_x}{d \xi}\frac{d^3v_x}{d\xi^3}-\omega_{ci}(2\omega_{ci}-\frac{3c^2_s}{vL_n}){(\frac{dv_x}{d\xi})}^2+\omega_{ci}(v^2-c^2_s)\frac{d^3v_x}{d\xi^3}+(3v^2-c^2_s){(\frac{d^2v_x}{d\xi^2})}^2+\frac{\omega_{ci}c^2_s}{vL_n} v_x\frac{d^2 v_x}{d\xi^2}+\omega^2_{ci}(\omega_{ci}-\frac{c^2_s}{vL_n})\frac{dv_x}{d\xi}=0,
\label{D2phiDxi2Eq1}
\end{equation}
where we have neglected the third order terms as before during the derivation of the above Eq. (\ref{D2phiDxi2Eq1}). In order to rewrite the above Eq. (\ref{D2phiDxi2Eq1}) in the standard dimensionless form, we apply the following normalization: $\bar{v}=\frac{v}{c_s};\,\bar{v_x}=\frac{v_x}{c_s};\,\bar{\xi}=\frac{\xi}{\rho_i};\,\bar{L_n}=\frac{L_n}{\rho_i}$, where $\rho_i$ is the ion Larmor radius given as: $\rho_{i}=\frac{c_s}{\omega_{ci}}$. Then, Eq. (\ref{D2phiDxi2Eq1}) becomes
\begin{equation}
(\bar{v}^2+1)\frac{d \bar{v}_x}{d \bar{\xi}}\frac{d^3\bar{v}_x}{d\bar{\xi}^3}+(\frac{3}{\bar{v}{\bar{L}_n}}-2){(\frac{d\bar{v}_x}{d\bar{\xi}})}^2+(\bar{v}^2-1)\frac{d^3\bar{v}_x}{d\bar{\xi}^3}+(3\bar{v}^2-1){(\frac{d^2\bar{v}_x}{d\bar{\xi}^2})}^2+\frac{1}{\bar{v}\bar{L}_n} \bar{v}_x\frac{d^2 \bar{v_x}}{d\bar{\xi}^2}+(1-\frac{1}{\bar{v}\bar{L}_n})\frac{d\bar{v}_x}{d\bar{\xi}}=0,
\label{Dimnlessfrm}
\end{equation}
In compact form, the above Eq. (\ref{Dimnlessfrm}) can be written as:
\begin{equation}
A\frac{d v_x}{d \xi}\frac{d^3v_x}{d\xi^3}+B{(\frac{dv_x}{d\xi})}^2+C\frac{d^3v_x}{d\xi^3}+D{(\frac{d^2v_x}{d\xi^2})}^2+Ev_x\frac{d^2v_x}{d\xi^2}+F\frac{dv_x}{d\xi}=0, \label{FinalNLEE}
\end{equation}
where $A, B, C, D, E$ and $F$ are the coefficients of different terms in Eq. (\ref{Dimnlessfrm}). These coefficients $A, B, C, D, E$ and $F$ in the above Eq. (\ref{FinalNLEE}) are given by
\begin{equation}
A=\bar{v}^2+1;\,\,B=\frac{3}{\bar{v}\bar{L}_n}-2;\,C=\bar{v}^2-1;\,D=3\bar{v}^2-1;\,E=\frac{1}{\bar{v}\bar{L}_n};\,F=1-\frac{1}{\bar{v}\bar{L}_n}. \label{abcdef}
\end{equation} 

For simplicity of further calculations, we have removed bar from the variables in Eq. (\ref{FinalNLEE}). This Eq. (\ref{FinalNLEE}) is the final novel nonlinear evolution equation in our system. This Eq. (\ref{FinalNLEE}) is a third order equation having some resemblances with the Jerk equation \cite{Wharton}. We know that Jerk equations can satisfy both chaotic and regular structures from the nonlinear dynamical point of view \cite{Wharton}. One important feature of Eq. (\ref{FinalNLEE}) governing the dynamics of the high frequency electrostatic drift waves is that the third and sixth terms in the left hand side can be considered as the terms similar to the stationary HM equation which explains the dynamics of the conventional low frequency electrostatic drift waves in $(1+1)$ dimensions. The remaining terms in the left hand side of Eq. (\ref{FinalNLEE}), which are quadratic in nature, are the terms arising due to the contributions of the nonlinear high frequency electrostatic drift waves. Therefore, Eq. (\ref{FinalNLEE}) can justifiably be regarded as the nonlinear high frequency generalization of the stationary HM equation in $(1+1)$ dimensions. This is also explained in appendix A from another perspective.

In this context, it should be noted that various types of differential equations have been explored abundantly across many scientific disciplines \cite{Li2006th, Jhangeer, MLi, Akinyemi, Nisar} as these are extremely useful and of current research interest. The travelling periodic, quasiperiodic and chaotic structures of perturbed Fokas-Lenells model (p-FLM), which is frequently reported in the field of nonlinear optical fibres, have been analyzed in \cite{Jhangeer} using the extended $(\frac{G'}{G^2})$-expansion technique. The p-FLM equation is a completely integrable generalization of the NLSE and is also the first negative member of the integrable hierarchy associated with the derivative NLSE. The phase portrait analysis has been performed by Jhangeer et al. \cite{Jhangeer} accompanied with sensitivity analysis in the parameter space after converting the model into a planar dynamical system with the help of Galilean transformation. The modulation instability along with multi-core directional couplers associated with optical solitons have been explored in \cite{Abbagari} where a new sub-ODE method is employed to extract the solutions. The dynamical study of longitudinal bud equation corresponding to a magneto-electro-elastic (MEE) circular rod has been performed by Khater et al. \cite{Khater} by deriving different solitary wave solutions with the implementation of the generalized Riccati equation mapping method and the generalized Kudryashov method.

Akinyemi et al. \cite{Akinyemi} have studied the fifth order fractional Schrodinger equation with Caputo derivative using the q-homotopy analysis transform method (q-HATM) to derive certain exact as well as approximate solutions. This method can be regarded as a combined form of the homotopy analysis method (HAM) and Laplace transform, and is very much useful for exploring various complex nonlinear models of fractional type. Nisar et al. \cite{Nisar} have recently investigated the generalized resonant nonlinear Schrodinger equation (NLSE) in order to explore some novel solutions not reported earlier. The NLSE has extensive applications in optical fibres, plasmas, Bose-Einstein condensates, superconductivity, quantum mechanics, mathematical biosciences including many other disciplines. In their work \cite{Nisar}, these novel solutions have been derived using the Bernoulli sub-ODE method (BSODEM) and the $(\frac{G'}{G})$-expansion method. The smooth and non-smooth solitary, kink and periodic solutions of an integrable higher order Korteweg-de Vries (KdV) type equation have been studied by Li et al. \cite{Li2006th} using the theory of planar dynamical systems. In \cite{MLi}, a generalized fourth-order dispersive nonlinear Schrödinger equation frequently encountered in Heisenberg spin chain with twist interaction has been explored. In particular, based on the theory of planar dynamical systems, various travelling wave solutions have been derived leading to bifurcations of the phase portraits \cite{MLi}.

For exploring the regular non-chaotic solutions of the nonlinear evolution equation Eq. (\ref{FinalNLEE}) in analogy with the solutions of the HM equation \cite{Hasegawa, Horton}, we decompose the left hand side of Eq. (\ref{FinalNLEE}) into two parts:
\begin{equation}
[A\frac{d}{d \xi}(\frac{dv_x}{d\xi}\frac{d^2v_x}{d\xi^2})+B\frac{d}{d\xi}(v_x\frac{dv_x}{d\xi})+C\frac{d^3v_x}{d\xi^3}+F\frac{dv_x}{d\xi}]+[(D-A){(\frac{d^2v_x}{d\xi^2})}^2+(E-B)v_x\frac{d^2v_x}{d\xi^2}]=0, \label{FinalNLEEdec}
\end{equation}
where the expressions in the two square brackets in the left hand sides of the above Eq. (\ref{FinalNLEEdec}) stand for the two parts. Due to such a particular decomposition, Eq. (\ref{FinalNLEE}) can be efficiently compared with the HM equation. The reason for this decomposition of the left hand side in Eq. (\ref{FinalNLEE}) into these two parts in Eq. (\ref{FinalNLEEdec}) can be more precisely understood later as explained in the appendix A; where detailed comparisons between Eq. (\ref{FinalNLEE}) and HM equation have been provided. Then, there can be one possibility that the two expressions in the square brackets in the above Eq. (\ref{FinalNLEEdec}) vanish separately as demanded by the right hand side of the same Eq. (\ref{FinalNLEEdec}). This implies
\begin{equation}
A\frac{d}{d \xi}(\frac{dv_x}{d\xi}\frac{d^2v_x}{d\xi^2})+B\frac{d}{d\xi}(v_x\frac{dv_x}{d\xi})+C\frac{d^3v_x}{d\xi^3}+F\frac{dv_x}{d\xi}=0, \label{FinalNLEEdeca}
\end{equation}
\begin{equation}
(D-A)\frac{d^2v_x}{d\xi^2}+(E-B)v_x=0. \label{FinalNLEEdecb}
\end{equation} 
Then, by integrating both sides of the nonlinear evolution Eq. (\ref{FinalNLEEdeca}) with respect to $\xi$ once, we get
\begin{equation}
A\frac{d v_x}{d \xi}\frac{d^2v_x}{d\xi^2}+Bv_x\frac{dv_x}{d\xi}+C\frac{d^2v_x}{d\xi^2}+Fv_x=K, \label{1FinalNLEE}
\end{equation}
where $K$ is the constant of integration. For simplicity, we assume that $K=0$. Then the above Eq. (\ref{1FinalNLEE}) can be rewritten as
\begin{equation}
\frac{d^2v_x}{d\xi^2}+f(\frac{dv_x}{d\xi})v_x=0;\,f(\frac{dv_x}{d\xi})=\frac{B\frac{dv_x}{d\xi}+F}{A\frac{dv_x}{d\xi}+C}. \label{fFinalNLEE1}
\end{equation}
Again, Eq. (\ref{FinalNLEEdecb}) implies

\begin{equation}
\frac{d^2v_x}{d\xi^2}+\frac{B-E}{A-D}v_x=0, \label{fFinalNLEE2}
\end{equation}

Therefore, the third order nonlinear evolution Eq. (\ref{FinalNLEE}) is equivalent to two reduced second order Eqs. (\ref{fFinalNLEE1}) and (\ref{fFinalNLEE2}). This further implies that the intersection of the solutions of these two reduced Eqs. (\ref{fFinalNLEE1}) and (\ref{fFinalNLEE2}) can potentially represent the desired solutions of the original third order nonlinear evolution Eq. (\ref{FinalNLEE}); but, these particular solutions belong to specific solutions and are not the general solution of Eq. (\ref{FinalNLEE}). The general solution of Eq. (\ref{FinalNLEE}) will be a superset of these particular solutions given by the intersections of the solutions of Eqs. (\ref{fFinalNLEE1}) and (\ref{fFinalNLEE2}). In this context, it should be noted that the general solution of the original third order novel nonlinear evolution Eq. (\ref{FinalNLEE}) is planned to be explored in future whereas, in the present work, we only explore the particular solutions given by the intersections of the solutions of the reduced Eqs. (\ref{fFinalNLEE1}) and (\ref{fFinalNLEE2}). Such methods of solving a higher order differential equation by reducing it to equivalent lower order differential equations under certain conditions have been reported extensively by different researchers \cite{Li2006th, MLi}. The similarities between the derived novel nonlinear evolution Eq. (\ref{FinalNLEE}) for the recently observed high frequency electrostatic drift waves and the well-known HM equation for the conventional low frequency electrostatic drift waves are discussed in Appendix A.

 
\section{Detailed study of linear dispersion relation} \label{DR}
\begin{figure}[hbt!]
\centering
\includegraphics[width=18cm]{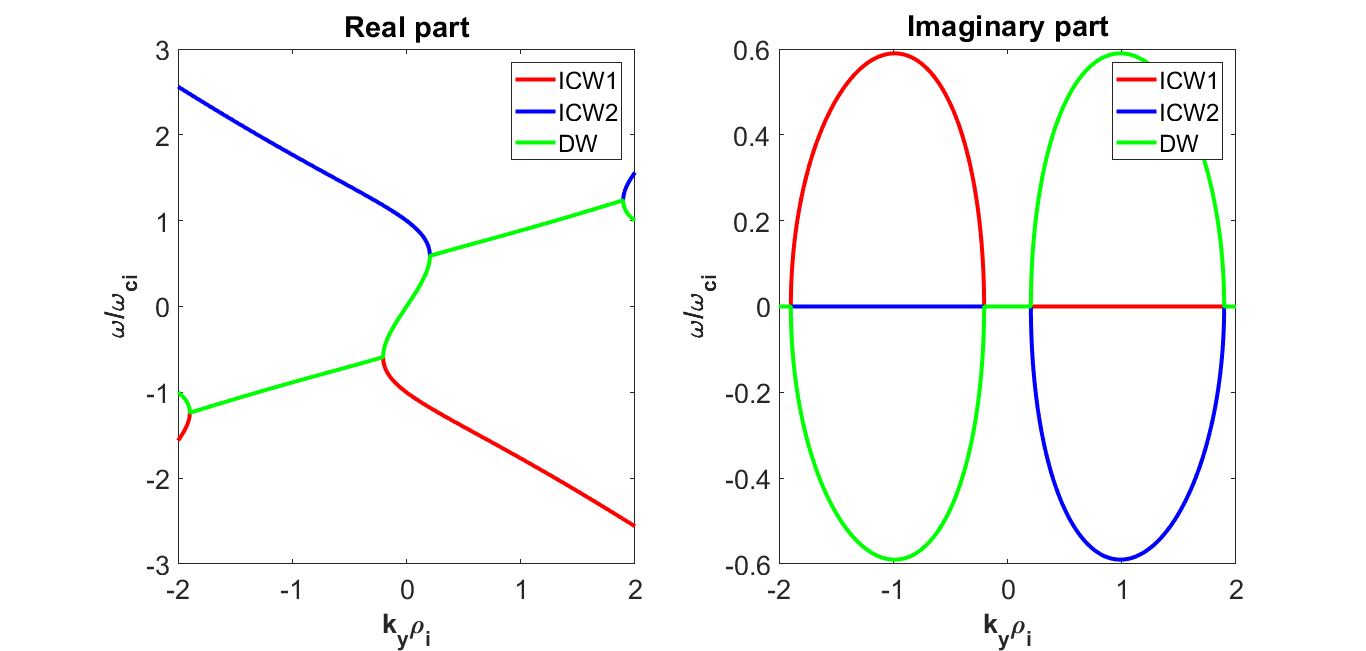}
\caption{Variations of the real and imaginary parts of the three exact roots of the dispersion relation Eq. (\ref{DispDL}) against $L_n$ for $k_y=0.56\, cm^{-1}$ and $\rho_i=1.8\,cm$ which are the same parameter values as reported in \cite{Ghosh2015} in the context of the high frequency electrostatic drift waves} \label{ExR1}
\end{figure}

In this section, we are going to analyze the linear dispersion relation for both the conventional low frequency and the recently observed high frequency electrostatic drift waves using the plasma fluid model equations given in Sec. \ref{Der}. In particular, for obtaining the linear dispersion relation governing the dynamics of the high frequency electrostatic drift waves in our system, we first linearize Eqs. (\ref{PhiEq1}) and (\ref{Omegaz1}) as:
\begin{equation}
\frac{\partial^2\tilde{\phi}}{\partial t^2}-c^2_s \nabla ^2 \tilde{\phi}+\omega_{ci}\Omega_z-\frac{1}{L_n}\frac{\partial v_x}{\partial t}=0, \label{PhiEq1L}
\end{equation}
\begin{equation}
\frac{\partial \Omega_z}{\partial t}-\omega_{ci}(\frac{\partial \tilde{\phi}}{\partial t})+\frac{\omega_{ci}}{L_n}v_x=0. \label{Omegaz1L}
\end{equation}
As we have assumed earlier that there is no propagation along the $x$-direction, i.e. $\frac{\partial}{\partial x}=0$, this implies that $\vec{\nabla}=\hat{y}\frac{\partial}{\partial y}$. Putting this value of $\vec{\nabla}$ and $\Omega_z$ from Eq. (\ref{OmgZ}) in Eqs. (\ref{PhiEq1L}) and (\ref{Omegaz1L}), we get
\begin{equation}
\frac{\partial^2\tilde{\phi}}{\partial t^2}-c^2_s\frac{\partial^2\tilde{\phi}}{\partial y^2}-\omega_{ci}\frac{\partial v_x}{\partial y}-\frac{1}{L_n}\frac{\partial v_x}{\partial t}=0, \label{PhiEq1L1}
\end{equation}
\begin{equation}
\frac{\partial^2 v_x}{\partial t \partial y}+\omega_{ci}\frac{\partial \tilde{\phi}}{\partial t}-\frac{\omega_{ci}}{L_n}v_x=0. \label{Omegaz1L1}
\end{equation}
We suppose that $\tilde{\phi}$ and $v_x$ are proportional to $exp[i(k_y y-\omega t)]$ implying that $\frac{\partial}{\partial t}=-i\omega $ and $\frac{\partial}{\partial y}=ik_y$. Substituting these values in Eqs. (\ref{PhiEq1L1}) and (\ref{Omegaz1L1}), and simplifying further to eleminate $\tilde{\phi}$ and $v_x$, we arrive at the following dispersion relation:
\begin{equation}
{\omega}^3-({{\omega}_{ci}}^2+k^2_y{c_s}^2)\omega+\frac{k_y}{L_n}{c_s}^2{\omega}_{ci}=0, \label{cubic_dr}
\end{equation}
which represents a cubic dispersion relation implying three roots depending upon the values of $\omega_{ci},\,k_y,\,c_s$ and $L_n$. A similar cubic dispersion relation having complex coefficients has also been derived by Ghosh et al. \cite{Ghosh2015,Ghosh2017} in the context of modelling their experimental results on the excitations of the high frequency electrostatic drift waves in the last decade. In dimensionless form the above dispersion relation Eq. (\ref{cubic_dr}) can be written as
\begin{equation}
\bar{\omega}^3-(1+\bar{k}^2_y)\bar{\omega}+\bar{k}_yN=0, \label{DispDL}
\end{equation}
where $\bar{\omega}=\frac{\omega}{\omega_{ci}},\,\bar{k}_y=k_y\rho_i$ and $N=\frac{\rho_i}{L_n}$. We know that the above cubic dispersion relation Eq. (\ref{DispDL}) has three roots. The typical variations of the real and imaginary parts of these three exact roots of Eq. (\ref{DispDL}) are shown in Fig. \ref{ExR1} against $k_y\rho_i$. These three roots correspond to the drift wave specified by green colour and two branches of the ion cyclotron wave specified by red and blue colours in the plot. Both the real and imaginary parts of the normalized mode frequency $\frac{\omega}{\omega_{ci}}$ have been plotted in this figure for inquiring the growth rates of these modes where the positive values of the imaginary part correspond to the growth rates, the vanishing value corresponds to the steady wave modes without any growth or decay, and the negative values of the imaginary part correspond to the decay rates. From this Fig. \ref{ExR1}, it can be seen that the region corresponding to the excitations of the high frequency electrostatic drift waves has positive values of the imaginary $\omega/\omega_{ci}$ which implies that these wave modes are excited in our system. In addition, it has to be noted that the positive growth rates of these novel high frequency electrostatic drift waves imply the presence of instabilities of these waves. The instabilities of these novel high frequency electrostatic drift waves are driven in our system solely by the density gradients instead of collisional mechanisms as reported in \cite{Ghosh2015, Ghosh2017}. The density gradients represent the sources of the free energy which drives these drift wave modes unstable. This is a very novel phenomenon which has not been reported in case of the high frequency electrostatic drift waves till now as far as our knowledge goes.

\begin{figure}[hbt!]
\centering
\includegraphics[width=18cm]{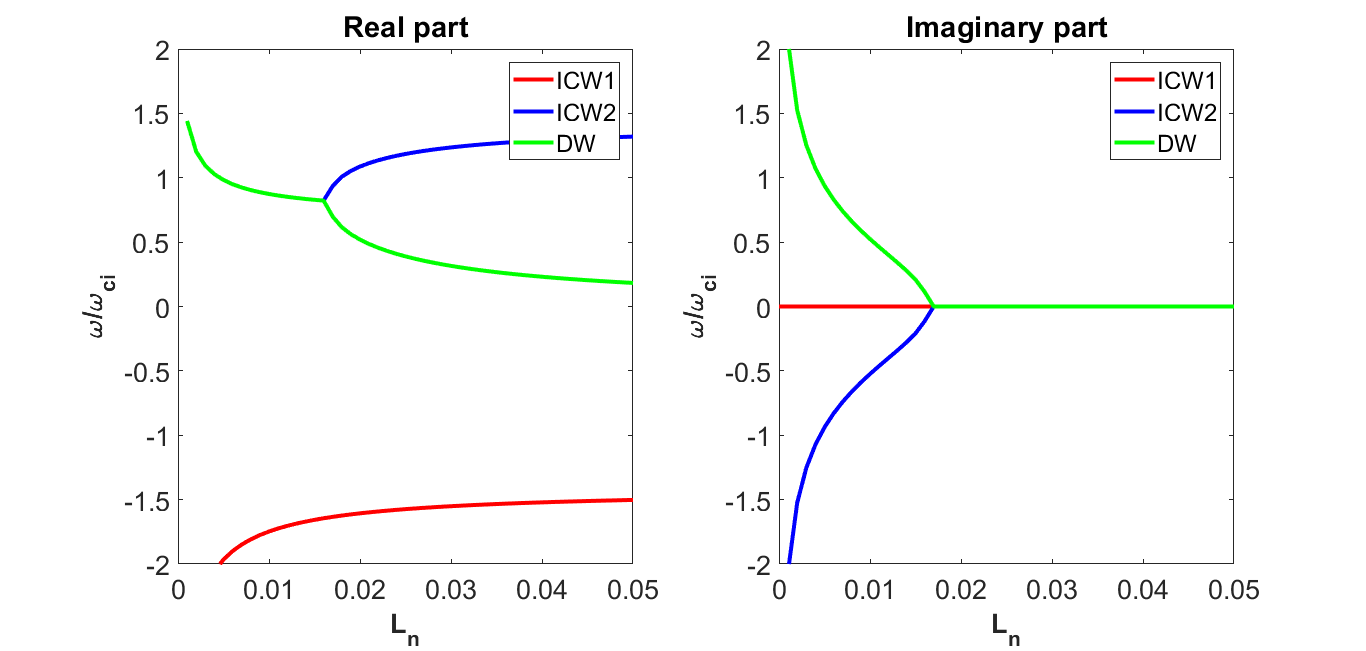}
\caption{Variations of the real and imaginary parts of the three exact roots of the dispersion relation Eq. (\ref{DispDL}) against $k_y\rho_i$ for $N=2$} \label{ExR2}
\end{figure}

 Similarly, the excitations of both the low frequency electrostatic drift and ion cyclotron waves can be verified from this Fig. \ref{ExR1}. Therefore, only the wave modes corresponding to either the vanishing or positive values of the imaginary part are to be excited. In order to ascertain the drift wave nature of one of these three roots as specified by the green colour in Fig. \ref{ExR1}, we have also analysed the behaviours of these three roots against the variations of the density gradient scale length $L_n$ in Fig. \ref{ExR2}. It can be seen from this figure that the exact root of the dispersion relation Eq. (\ref{DispDL}) corresponding to the drift wave, as specified by green colour in the plot, varies significantly in both the limits $\omega \approx \omega_{ci}$ and $\omega <<\omega_{ci}$; which justifies its drift wave nature. Therefore, in order to establish this fact further, we now proceed to analyse the cubic dispersion relation Eq. (\ref{cubic_dr}) in different limits of the mode frequency $\omega$ in comparison with the ion cyclotron frequency $\omega_{ci}$ as given in Appendix B.

\section{Fixed points and bifurcations with corresponding phase portraits for the reduced planar dynamical system Eq. (\ref{fFinalNLEE1})} \label{PPBifur}
The reduced second order nonlinear Eq. (\ref{fFinalNLEE1}) derived in Sec \ref{Der}  represents a typical example of the very well-known planar dynamical system; which can be characterized by the qualitative theory of differential equations on the phase plane. Various types of planar dynamical systems have been extensively investigated by many groups of researchers \cite{Wang,Li2020} in different contexts of applied mathematics with predominance among other related disciplines. In our system Eq. (\ref{fFinalNLEE1}), the equivalent planar dynamical system can be written as
\begin{equation}
\frac{d v_x}{d \xi}=y;\,\frac{d y}{d \xi}=-\frac{By+F}{Ay+C}v_x.\label{Phaseplane}
\end{equation}
From the above planar dynamical system Eq. (\ref{Phaseplane}), it is clear that there is only one singular line as follows:
\begin{equation}
Ay+C=0;\,or\,y=\frac{-C}{A}. \label{SingLine}
\end{equation}
The planar dynamical system Eq. (\ref{Phaseplane}) is a singular system due to the presence of a singular line given in Eq. (\ref{SingLine}). According to the theory of planar dynamical systems, the associated regular system of the planar dynamical system Eq. (\ref{Phaseplane}) is given by
\begin{equation}
\frac{d v_x}{d \eta}=y(Ay+C);\,\frac{d y}{d \eta}=-(By+F)v_x,\label{Phaseplane1}
\end{equation}
which is valid except on the singular line $y=\frac{-C}{A}$. In the above Eq. (\ref{Phaseplane1}), the new variable $\eta$ is defined as:
\begin{equation}
d\xi=(Ay+C)d\eta. \label{eta}
\end{equation}
The first integral of the regular system Eq. (\ref{Phaseplane1}) can be calculated as:
\begin{equation}
H(v_x,y)=\frac{v^2_x}{2}+\frac{A}{2B}y^2+\frac{BC-AF}{B^2}y+\frac{AF-BC}{B^3}F\,ln(y+\frac{F}{B})=constant\, (let\, h). \label{Ham}
\end{equation}
where $H(v_x,y)$ is a new function representing the first integral of the planar dynamical system Eq. (\ref{Phaseplane1}). It should be mentioned here that the first integral and equivalently the corresponding phase portraits of the singular system Eq. (\ref{Phaseplane}) is the same as that of the corresponding regular system Eq. (\ref{Phaseplane1}) according to the theory of planar dynamical systems. As the first integral $H(v_x,y)$ is becoming a constant on the reght hand side as expressed in Eq. (\ref{Ham}), it can equivalently be referrred to as the Hamiltonian function of the system considered. This is analogous to the fact that the Hamiltonian of a conservative system is constant. 
Now we are going to find the equilibrium or fixed points of the planar dynamical system Eq. (\ref{Phaseplane1}). At the fixed points, $\frac{d v_x}{d \eta}$ and $\frac{d y}{d \eta}$ in Eq. (\ref{Phaseplane1}) vanish, thereby, implying that the fixed points (FPs) of Eq. (\ref{Phaseplane1}) are as follows:
\begin{equation}
(v_x,y)=(0,0);\,(v_x,y)=(0,-\frac{C}{A});\,(v_x,y)=(a,-\frac{F}{B}), \label{FPs}
\end{equation}
where $a$ is any constant (preferably non-zero). The FP $(v_x,y)=(a,-\frac{F}{B})$ is subject to the condition: 
\begin{equation}
AF=BC. \label{FP3cond}
\end{equation}
For further analysis in the framework of the theory of planar dynamical systems \cite{Li2020,Wang}, we have to determine the Jacobian matrices at the fixed points given in Eq. (\ref{FPs}). In general form, it is well-known that the Jacobian matrix \cite{Strogatz} corresponding to the regular system Eq. (\ref{Phaseplane1}) is given by
\begin{equation}
J(v_x,y)=\begin{bmatrix}
   \frac{\partial}{\partial v_x}(\frac{dv_x}{d\eta}) & \frac{\partial}{\partial y}(\frac{dv_x}{d\eta})\\
   \frac{\partial}{\partial v_x}(\frac{dy}{d\eta}) & \frac{\partial}{\partial y}(\frac{dy}{d\eta})
\end{bmatrix}. \label{J1}
\end{equation}
Evaluating the derivatives in the above matrix Eq. (\ref{J1}) using Eq. (\ref{Phaseplane1}), we get
\begin{equation}
J(v_x,y)=\begin{bmatrix}
   0 & 2Ay+C\\
   -(By+F) &   -Bv_x 
\end{bmatrix}
.\label{J2} 
\end{equation}
At the fixed points Eq. (\ref{FPs}), the Jacobian matrix $J(v_x,y)$ becomes
\begin{equation}
J(0,0)=\begin{bmatrix}
   0 & C\\
   -F & 0
\end{bmatrix}
,\label{Jfp1} 
\end{equation}

\begin{equation}
J(0,-\frac{C}{A})=\begin{bmatrix}
   0 & -C\\
   \frac{BC}{A}-F & 0
\end{bmatrix}
,\label{Jfp2} 
\end{equation}

\begin{equation}
J(a,-\frac{F}{B})=\begin{bmatrix}
   0 & -\frac{2AF}{B}+C\\
   0 & -Ba
\end{bmatrix}
.\label{Jfp3} 
\end{equation}
It can be easily seen that the traces of the above Jacobian matrices Eqs. (\ref{Jfp1}) and (\ref{Jfp2}) are vanishing whereas that of the Jacobian matrix Eq. (\ref{Jfp3}) is non-vanishing, i.e.
\begin{equation}
T(0,0)=Tr \,J(0,0)=0;\,T(0,-\frac{C}{A})=Tr\, J(0,-\frac{C}{A})=0;\,T(a,-\frac{F}{B})=Tr\, J(a,-\frac{F}{B})=-Ba, \label{Trace}
\end{equation}
where $T(0,0),\,T(0,-\frac{C}{A})$ and $T(a,-\frac{F}{B})$ denote the traces of the Jacobian matrices at the fixed points $(0,0),\,(0,-\frac{C}{A})$ and $(a,-\frac{F}{B})$ respectively. Similarly, the determinants of the Jacobian matrix $J$ at the fixed points are given by
\begin{equation}
M(0,0)=det J(0,0)=CF;\,M(0,-\frac{C}{A})=det J(0,-\frac{C}{A})=C(\frac{BC}{A}-F);\,M(a,-\frac{F}{B})=det J(a,-\frac{F}{B})=0, \label{M} 
\end{equation}
where $M(0,0),\,M(0,-\frac{C}{A})$ and $M(a,-\frac{F}{B})$ denote the determinants of the Jacobian matrices at the fixed points $(0,0),\,(0,-\frac{C}{A})$ and $(a,-\frac{F}{B})$ respectively. According to the theory of planar dynamical systems, it is well-known that for a fixed point of a planar dynamical system, if the determinant of the Jacobian matrix $M$ is negative, then the fixed point is a saddle point; if $M$ is positive, then the fixed point is a center point; if $M$ is vanishing and the Poincare index of the fixed point is also vanishing, then the fixed point is a cusp \cite{Strogatz}. From the values of traces and determinants given in Eqs. (\ref{Trace}) and (\ref{M}), it is clear that the fixed point $(0,0)$ is a center if $CF$ is positive and a saddle point if $CF$ is negative; likewise, the fixed point $(0,-\frac{C}{A})$ is a center if $C(\frac{BC}{A}-F)$ is positive and a saddle point if $C(\frac{BC}{A}-F)$ is negative. On the other hand, the fixed point $(a,-\frac{F}{B})$ is a non-isolated fixed point, and can also be a star, degenerate node or cusp as well depending upon the value of $Ba$. Now we are going to discuss the bifurcations of the fixed points, using the level curves or phase portraits explicitly for each fixed point, in the following subsections accompanied with the corresponding bifurcation curves in the parameter space.
\subsection{$FP1: (v_x,y)=(0,0)$} 
In this case, for the fixed point $(v_x,y)=(0,0)$, it is clear from Eqs. (\ref{Trace}) and (\ref{M}) that the trace of the Jacobian matrix $T(0,0)$ is vanishing whereas the determinant of the Jacobian matrix $M(0,0)$ is $CF$. Therefore, the nature of the fixed point $(0,0)$ is to be determined by the signs of $CF$ according to the theory of planar dynamical systems. In this case, the bifurcation curves are given by $C=0$ and $F=0$, i.e. $C-axis$ and $F-axis$ on the $CF-$parametric plane.

\begin{figure}[hbt!]
\centering
\includegraphics[width=18cm]{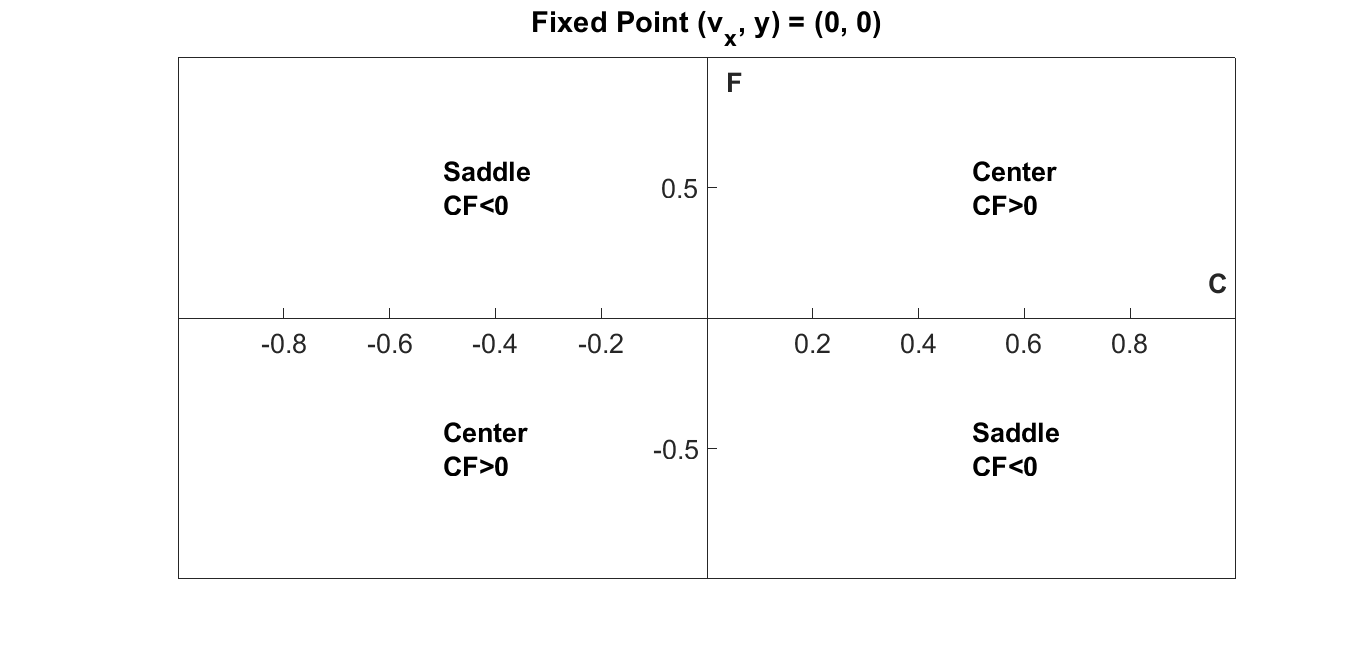}
\caption{Bifurcation curves corresponding to the fixed point $(0,0)$ on $CF-$parametric plane characterizing the nature of the fixed point $(0,0)$ as the parameters $C$ and $F$ are varied; $C=0$, i.e. $F-$axis and $F=0$, i.e. $C-$axis represent the bifurcation curves.} \label{BCOne}
\end{figure}

\begin{figure}
\includegraphics[width=5.6cm]{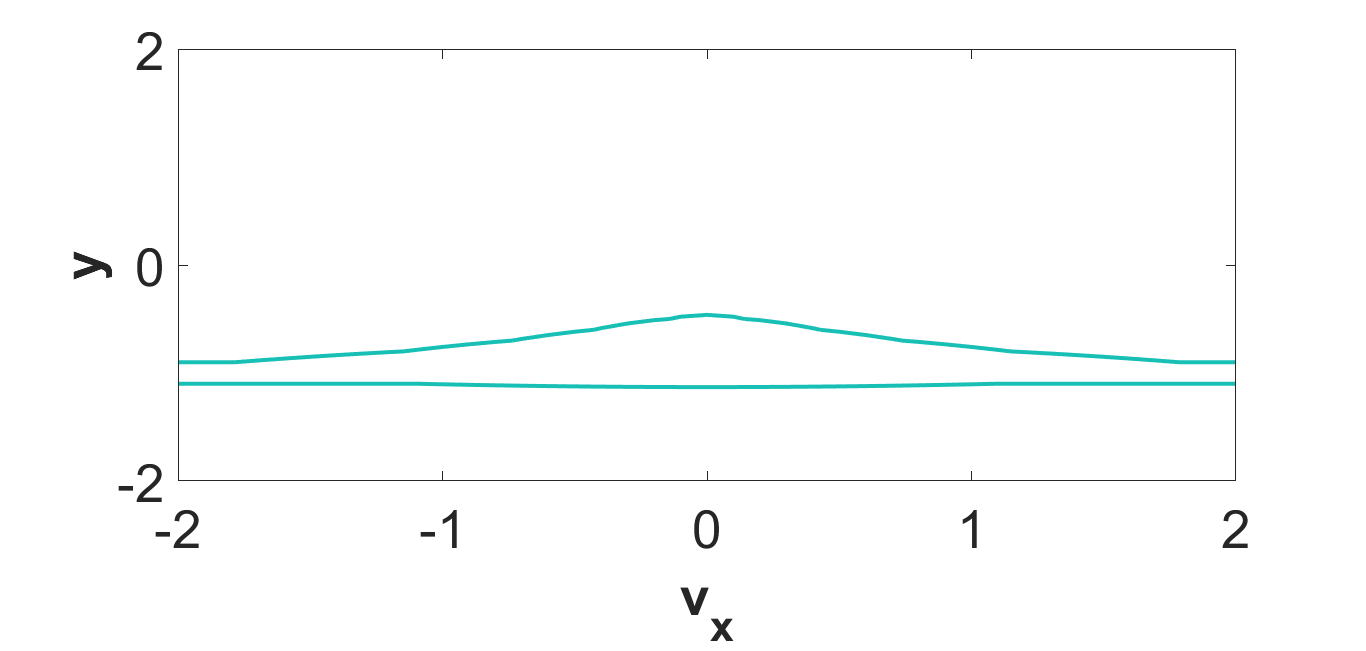}
 \ \ \ \
\includegraphics[width=5.6cm]{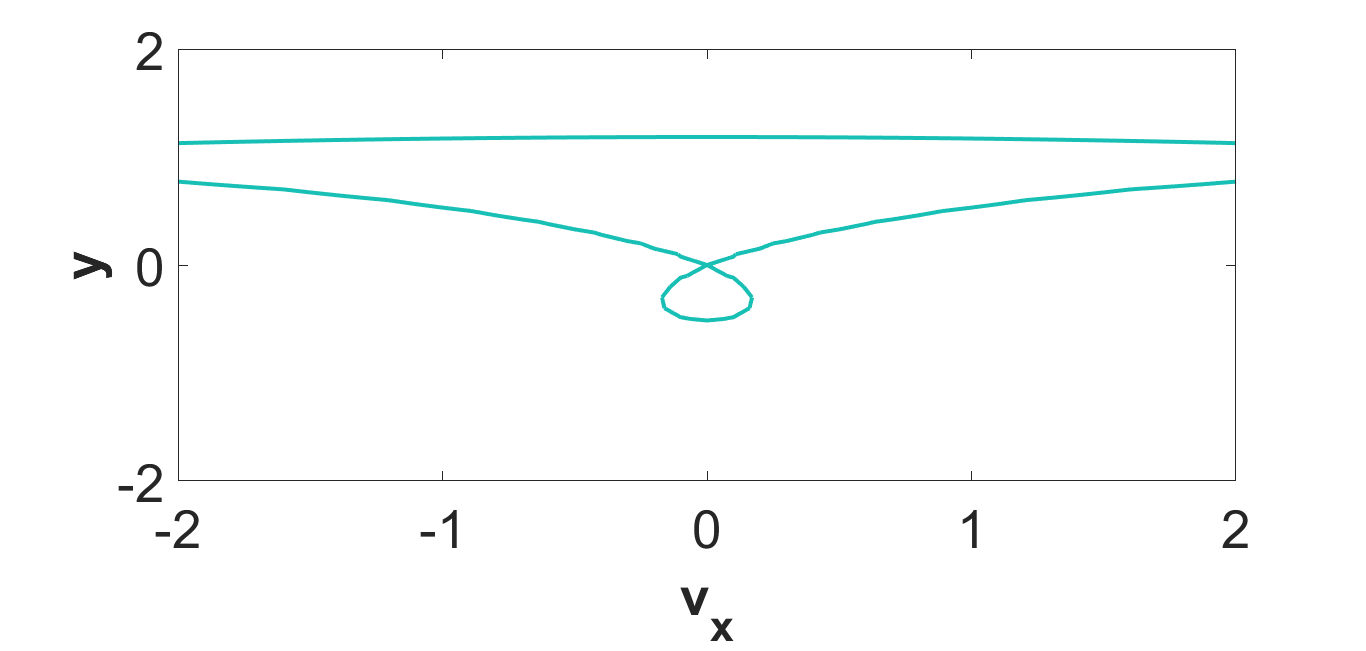}
 \ \ \ \
\includegraphics[width=5.6cm]{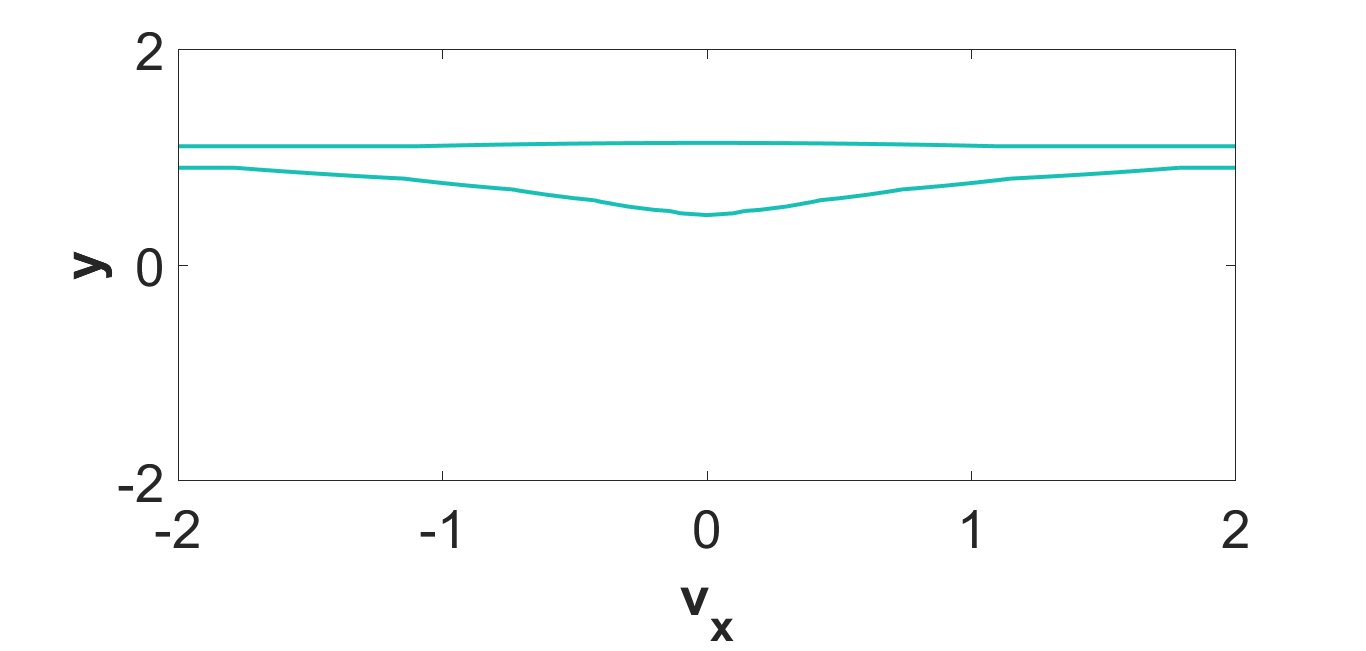}
(a) $A=3,\,B=1,\,C=1,\,F=1$
\qquad \qquad (b) $A=3,\,B=1,\,C=1,\,F=-1$
\qquad \qquad (c) $A=3,\,B=1,\,C=-1,\,F=-1$
\includegraphics[width=5.6cm]{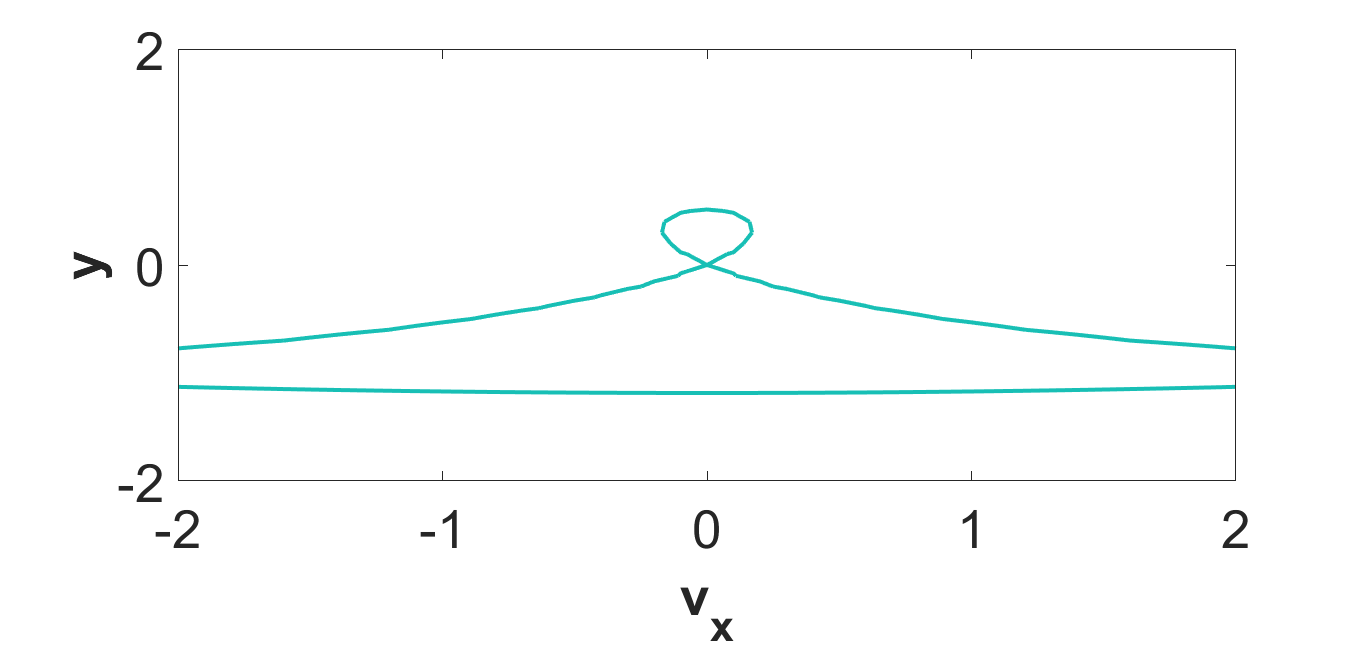}
 \ \ \ \
\includegraphics[width=5.6cm]{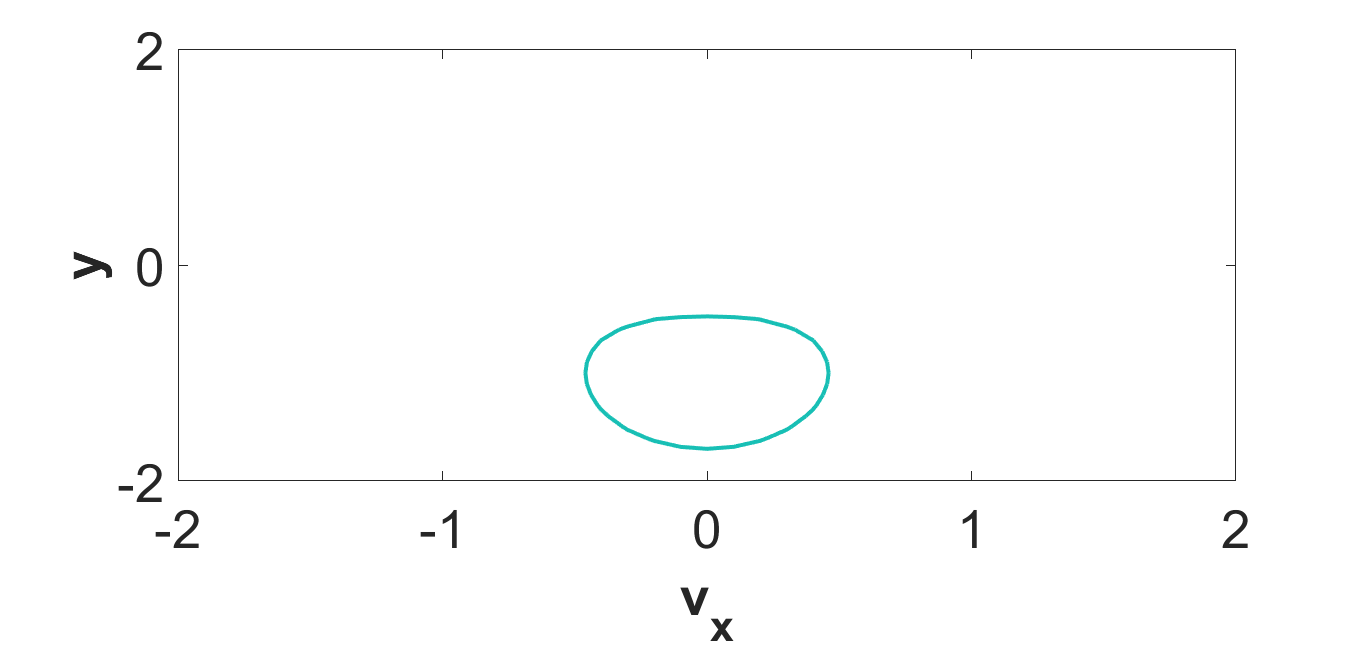}
 \ \ \ \
\includegraphics[width=5.6cm]{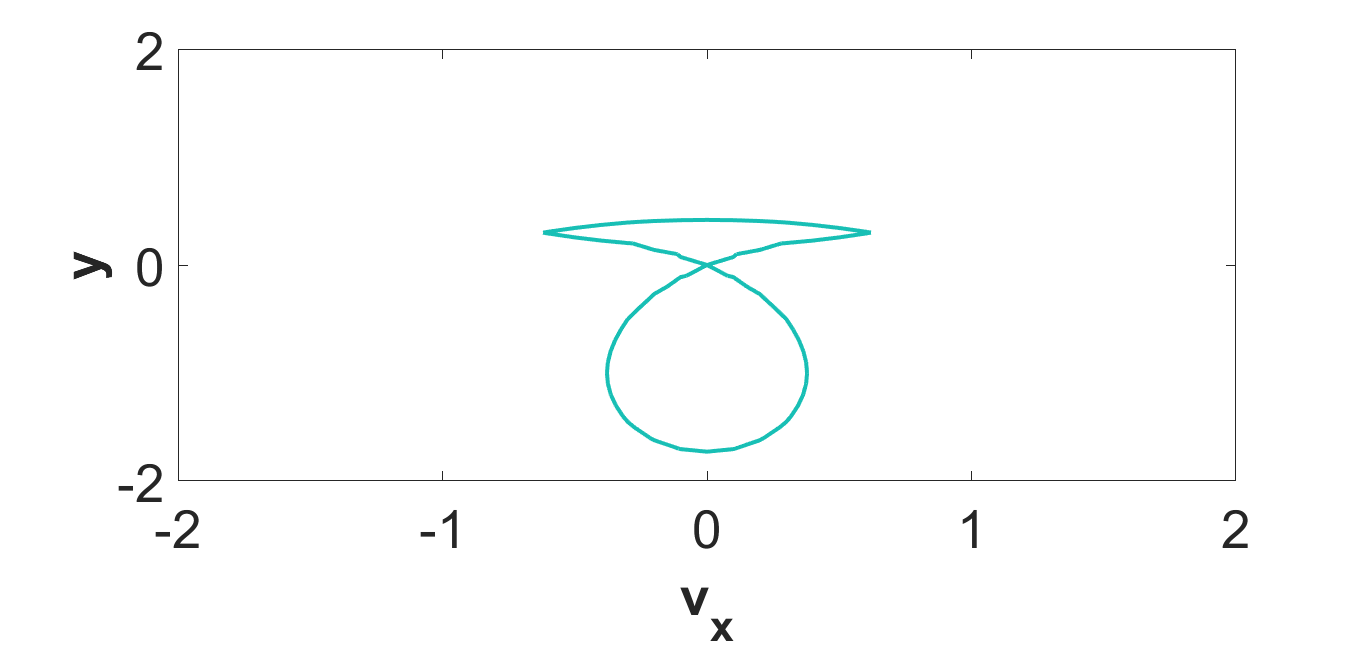}
(d) $A=3 \,B=1,\,C=-1 \,and \,F=1$
\quad \qquad (e) $A=1 \,B=3,\,C=1 \,and \,F=1$
\quad \qquad (f) $A=1 \,B=3,\,C=1 \,and \,F=-1$
 
\includegraphics[width=5.6cm]{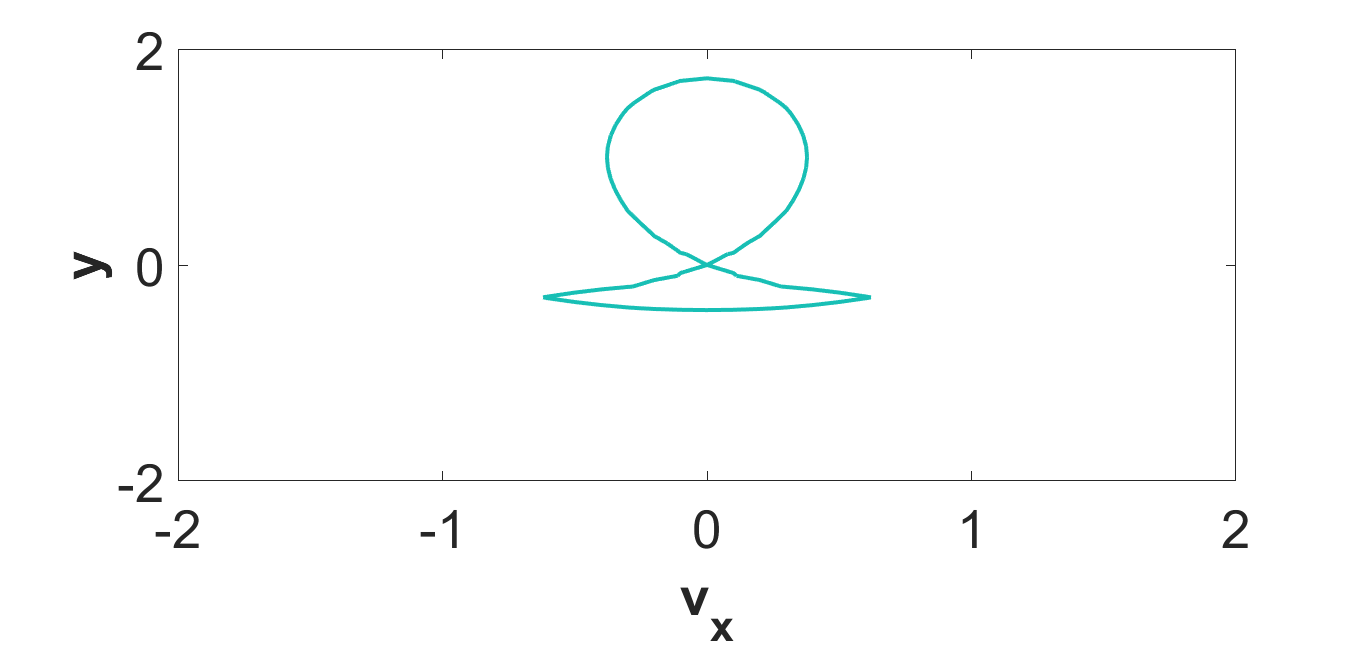}
\ \ \ \
\includegraphics[width=5.6cm]{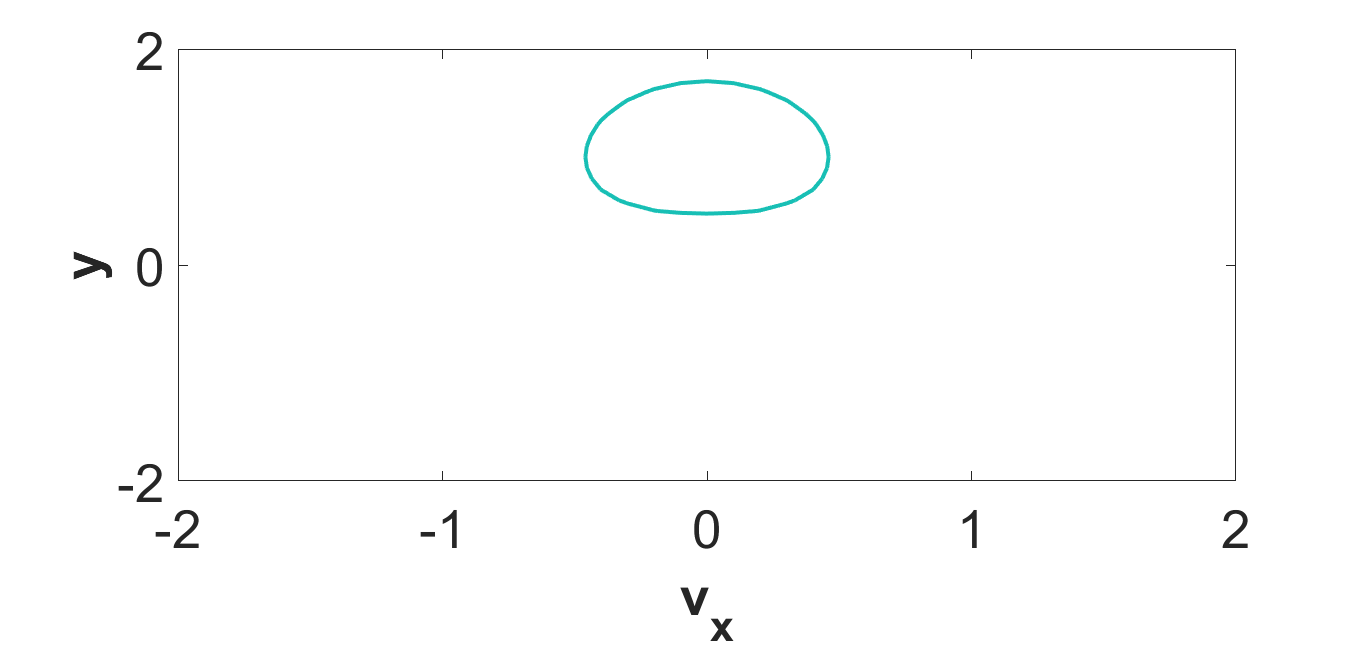}
 \ \ \ \
\includegraphics[width=5.6cm]{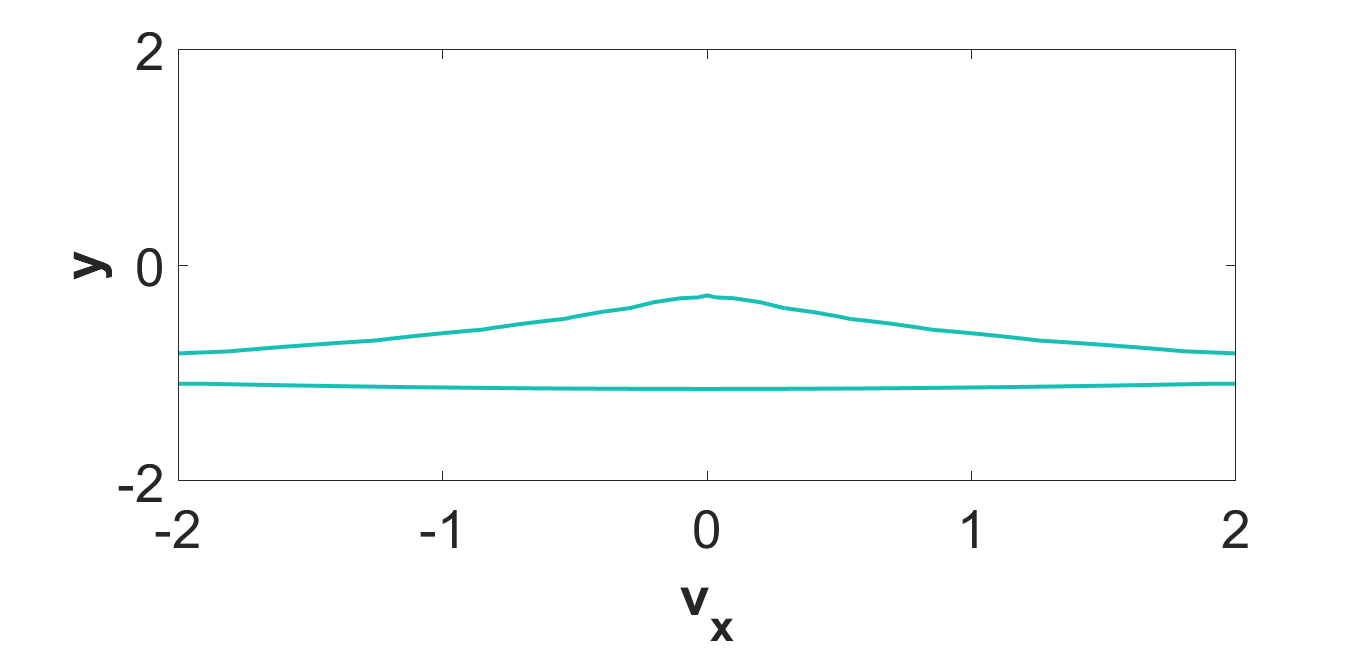}
(g) $A=1 \,B=3,\,C=-1 \,and \,F=1$
\qquad \qquad (h) $A=1 \,B=3,\,C=-1 \,and \,F=-1$
\quad \qquad (i) $A=5 \,B=1,\,C=1 \,and \,F=1$
\caption{Bifurcations of phase portraits about the fixed point $(0,0)$ in the corresponding parameter space} \label{BifPP1}
\end{figure}

The fixed point $(0,0)$ is a center provided the values of $(C,F)$ lie in the first or third quadrants; similarly, it is a saddle point provided the values of $(C,F)$ lie in the second or fourth quadrants. This peculiar behaviour has been shown in Fig. \ref{BCOne}. It should be noted that the center point is neutrally stable as the trace is vanishing which indicates the purely imaginary eigenvalues of the Jacobian matrix; similarly, the saddle point is semi-stable \cite{Strogatz}. The corresponding bifurcations of phase portraits of the planar dynamical system Eq. (\ref{Phaseplane1}) are shown in Fig. \ref{BifPP1}. The bifurcations of phase portraits Fig. \ref{BifPP1} are obtained using the Hamiltonian function Eq. (\ref{Ham}) at the fixed point $(0,0)$:
\begin{equation}
H(v_x,y)=\frac{v^2_x}{2}+\frac{A}{2B}y^2+\frac{BC-AF}{B^2}y+\frac{AF-BC}{B^3}F\,ln(y+\frac{F}{B})=H(0,0)=\frac{AF-BC}{B^3}F\,ln(\frac{F}{B}). \label{Ham00}
\end{equation}

In this type of bifurcation of phase portraits as shown in Fig. \ref{BifPP1}, the fixed point $(0,0)$ becomes a semi-stable saddle point from a neutrally stable center point and vice-versa as the parameters are varied.

\subsection{$FP2: (v_x,y)=(0,-\frac{C}{A})$} 
In this case, for the fixed point $(v_x,y)=(0,-\frac{C}{A})$, it is clear from Eqs. (\ref{Trace}) and (\ref{M}) that the trace of the Jacobian matrix $T(0,-\frac{C}{A})$ is vanishing whereas the determinant of the Jacobian matrix $M(0,-\frac{C}{A})$ is $C(\frac{BC}{A}-F)$. Therefore, the nature of the fixed point $(0,-\frac{C}{A})$ is to be determined by the signs of $(0,-\frac{C}{A})$ according to the theory of planar dynamical systems. In this case, the bifurcation curve is given by $F=(\frac{B}{A})C$ on the $CF-$parametric plane.

\begin{figure}
\includegraphics[width=18cm]{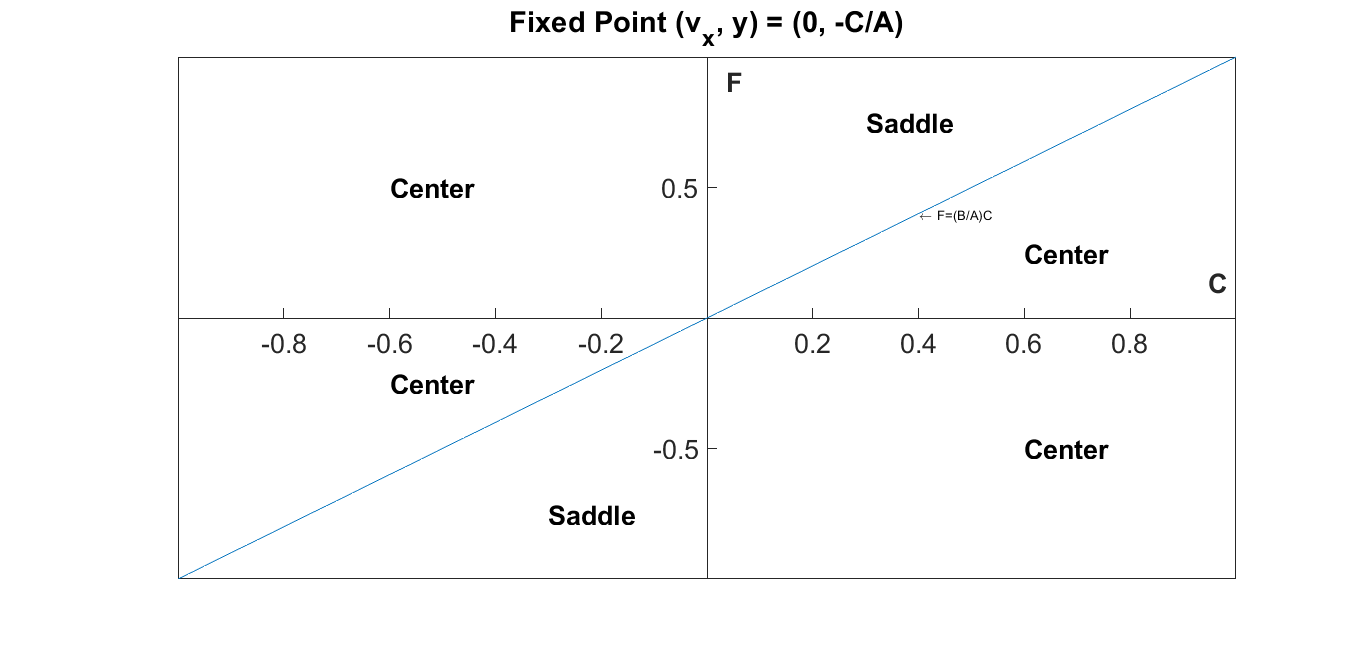}
\includegraphics[width=18cm]{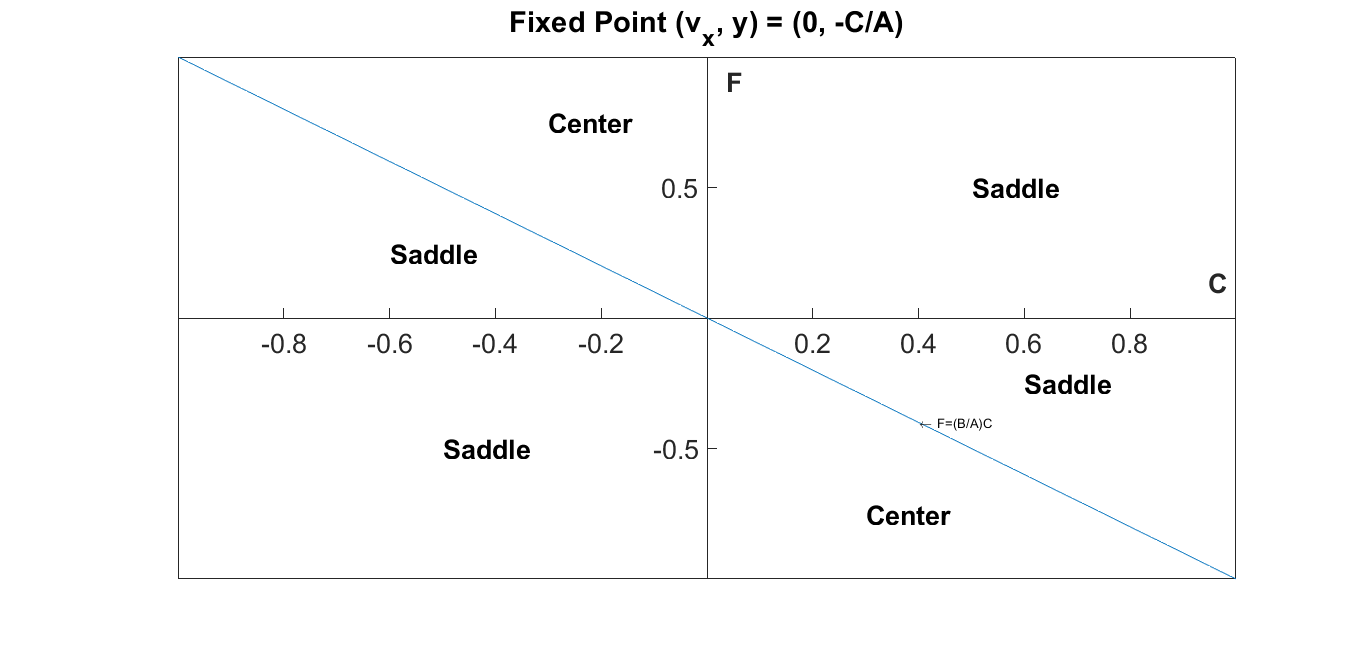}
 
\caption{Bifurcation curves for some arbitrarily chosen parameter values, i.e. $\frac{B}{A}=1$ and $\frac{B}{A}=-1$ for top and bottom figures respectively, corresponding to the FP $(v_x,y)=(0,-\frac{C}{A})$} \label{BCTwo}
\end{figure}

 Whether the fixed point $(0,-\frac{C}{A})$ is a saddle or center can be determined by observing different regions in Fig. \ref{BCTwo} on the $CF-$parametric plane as done in the previous case for the fixed point $(0,0)$. Different regions in Fig. \ref{BCTwo} are specified as saddles or centers for the values of the parameters $A$ and $B$ such that $\frac{B}{A}=\pm 1$. It should be noted that the center point is neutrally stable as the trace is vanishing which implies the existence of purely imaginary eigenvalues of the Jacobian matrix whereas the saddle point is semi-stable.
 
\begin{figure}
\includegraphics[width=5.6cm]{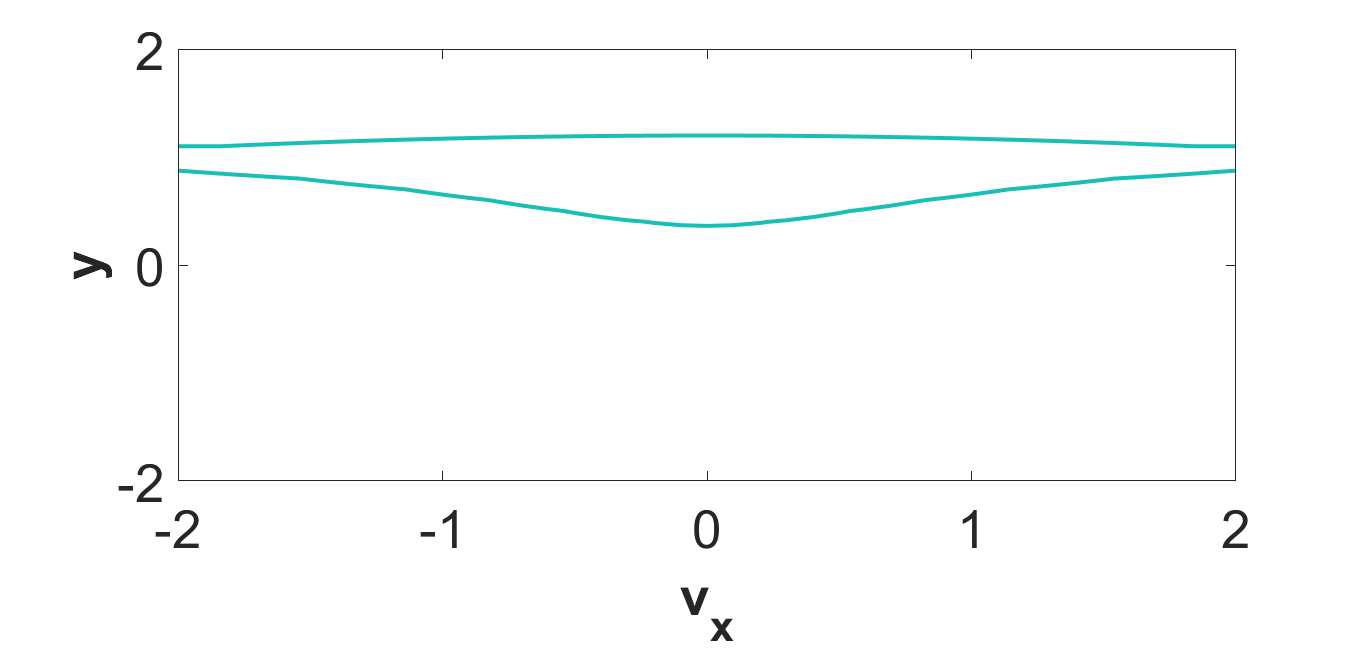}
 \ \ \ \
\includegraphics[width=5.6cm]{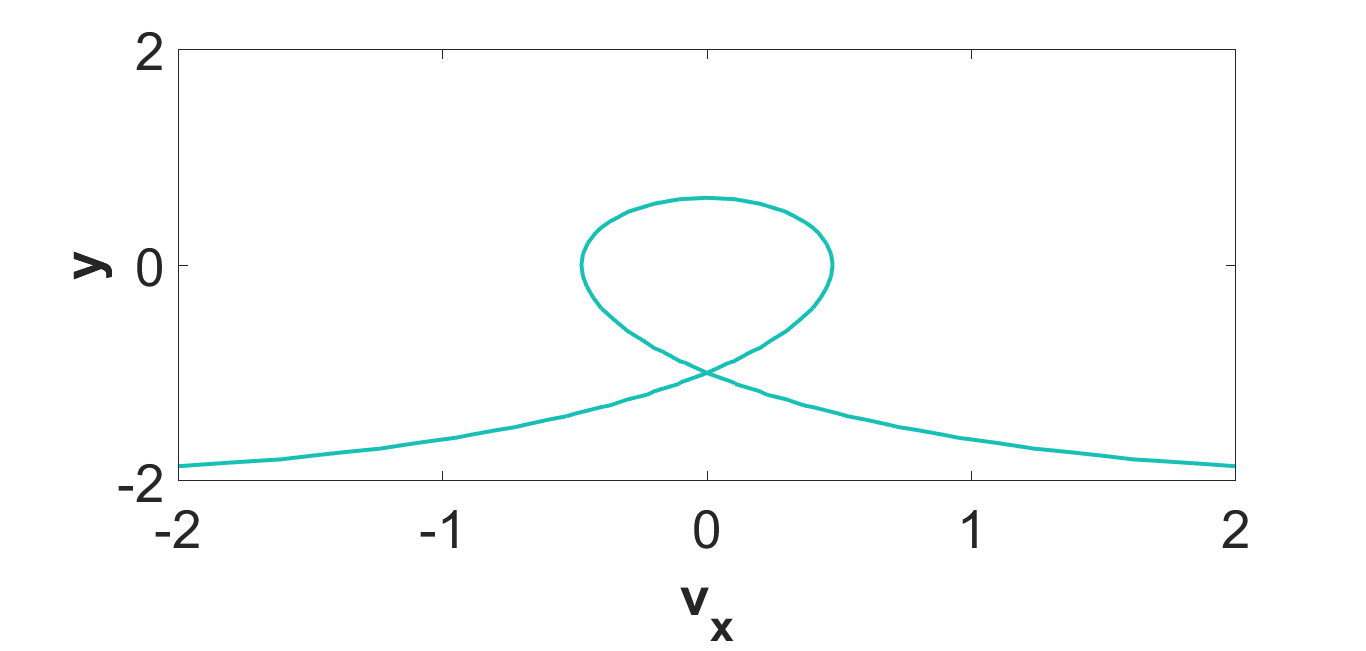}
 \ \ \ \
\includegraphics[width=5.6cm]{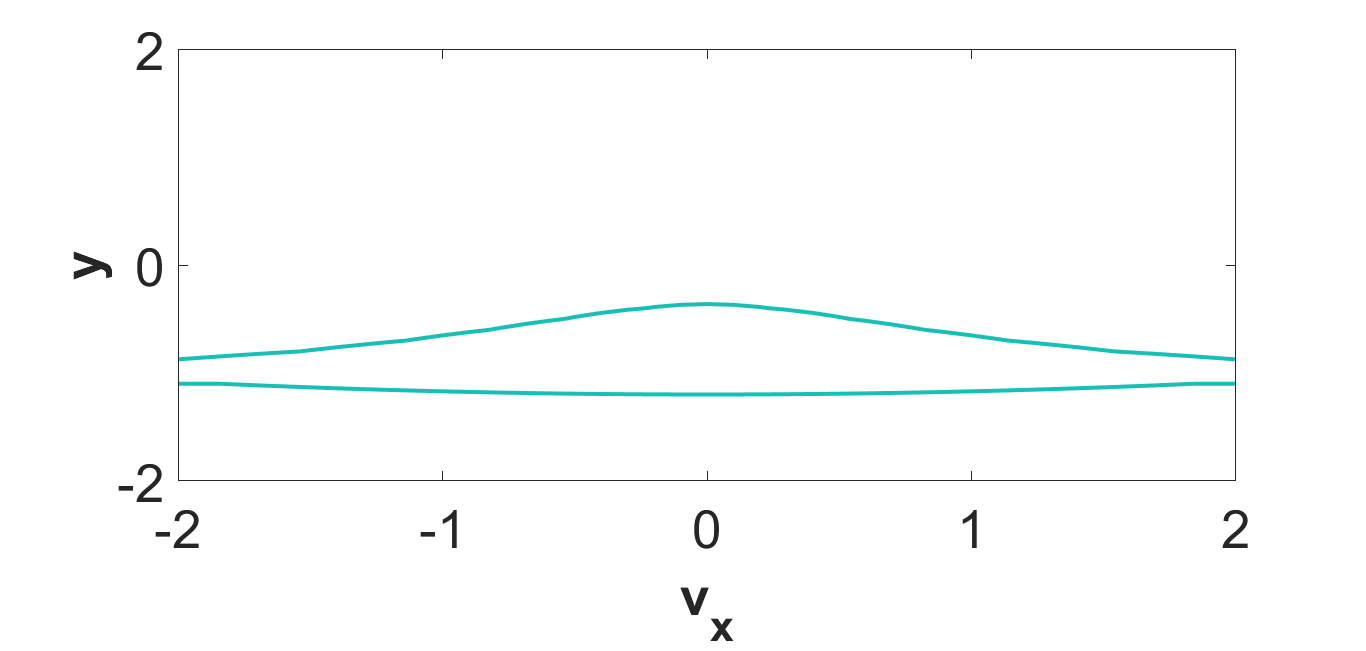}
(a) $A=1,\,B=1,\,C=1,\,F=-1$
\qquad \qquad (b) $A=1,\,B=1,\,C=1,\,F=2$
\qquad \qquad (c) $A=1,\,B=1,\,C=-1,\,F=1$
\includegraphics[width=5.6cm]{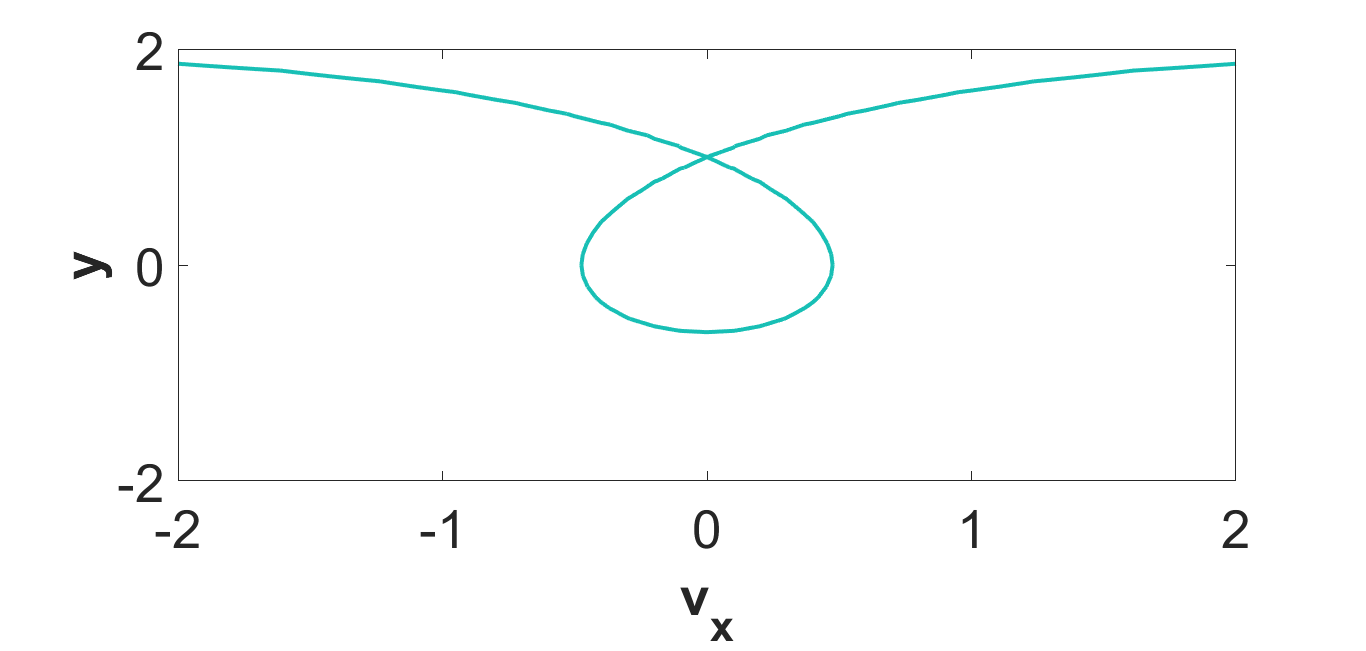}
 \ \ \ \
\includegraphics[width=5.6cm]{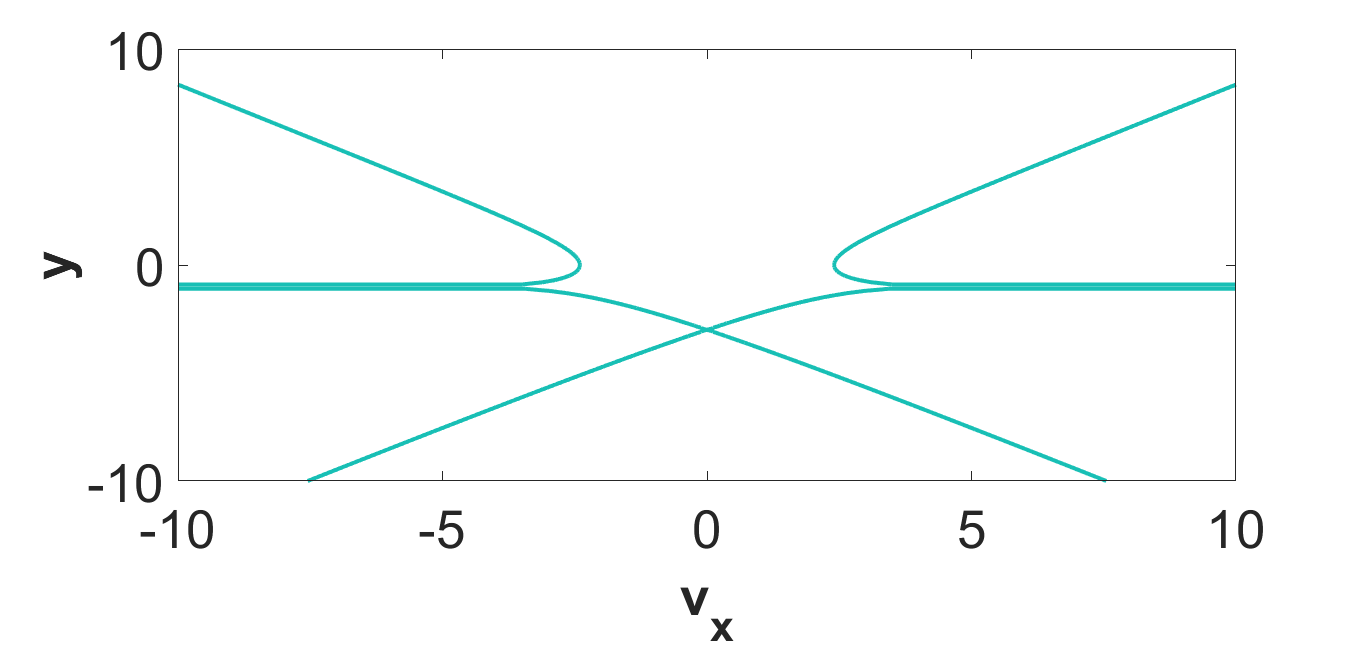}
 \ \ \ \
\includegraphics[width=5.6cm]{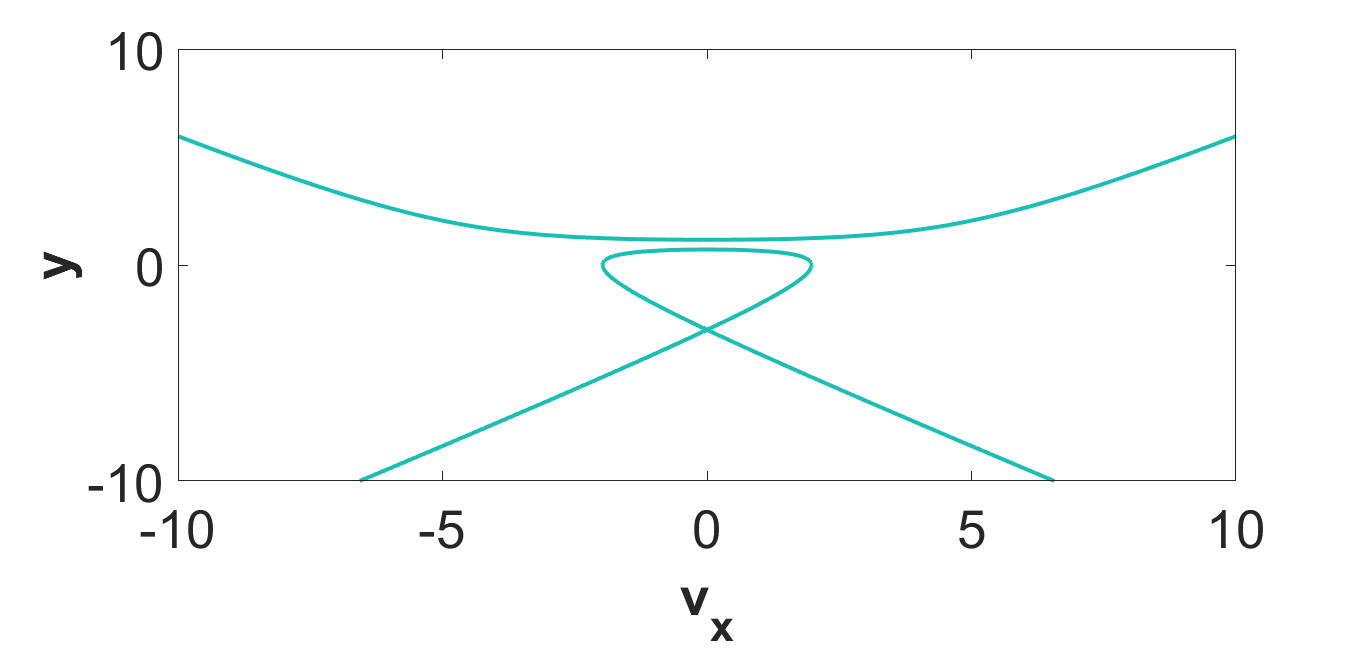}
(d) $A=1 \,B=1,\,C=-1 \,and \,F=-2$
\quad \qquad (e) $A=1 \,B=-1,\,C=3 \,and \,F=-1$
\quad \qquad (f) $A=1 \,B=-1,\,C=3 \,and \,F=1$
 
\includegraphics[width=5.6cm]{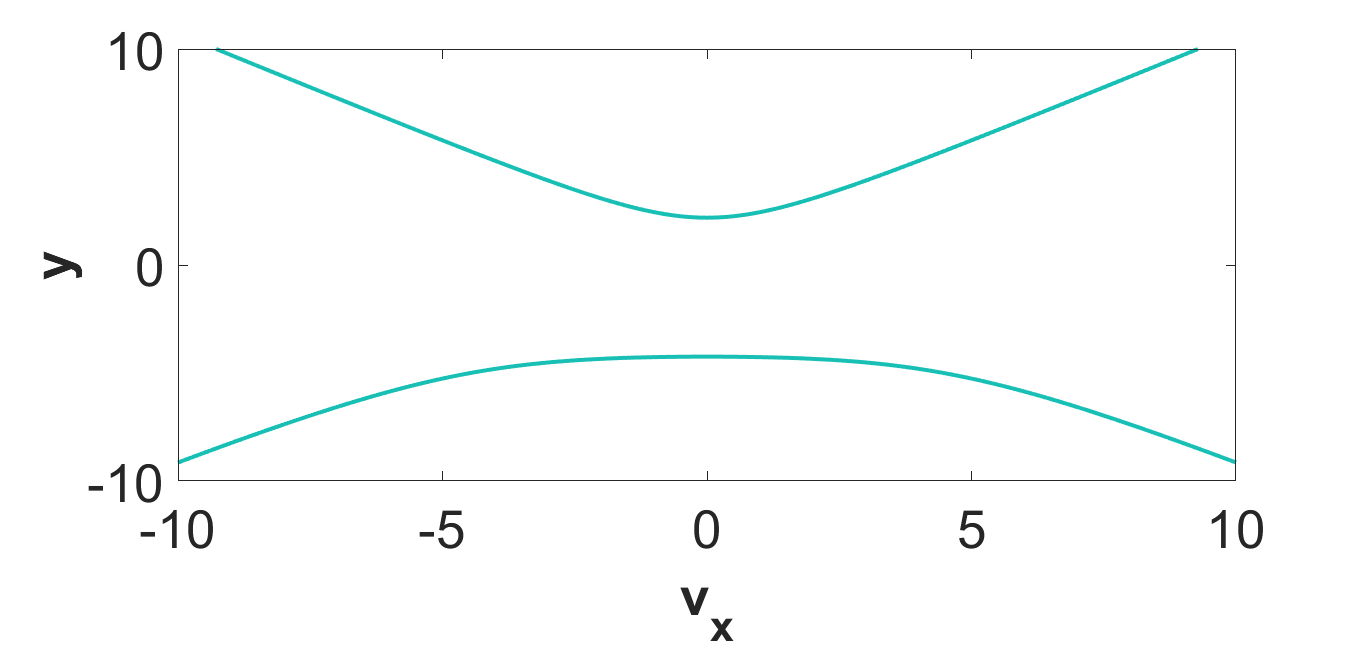}
\ \ \ \
\includegraphics[width=5.6cm]{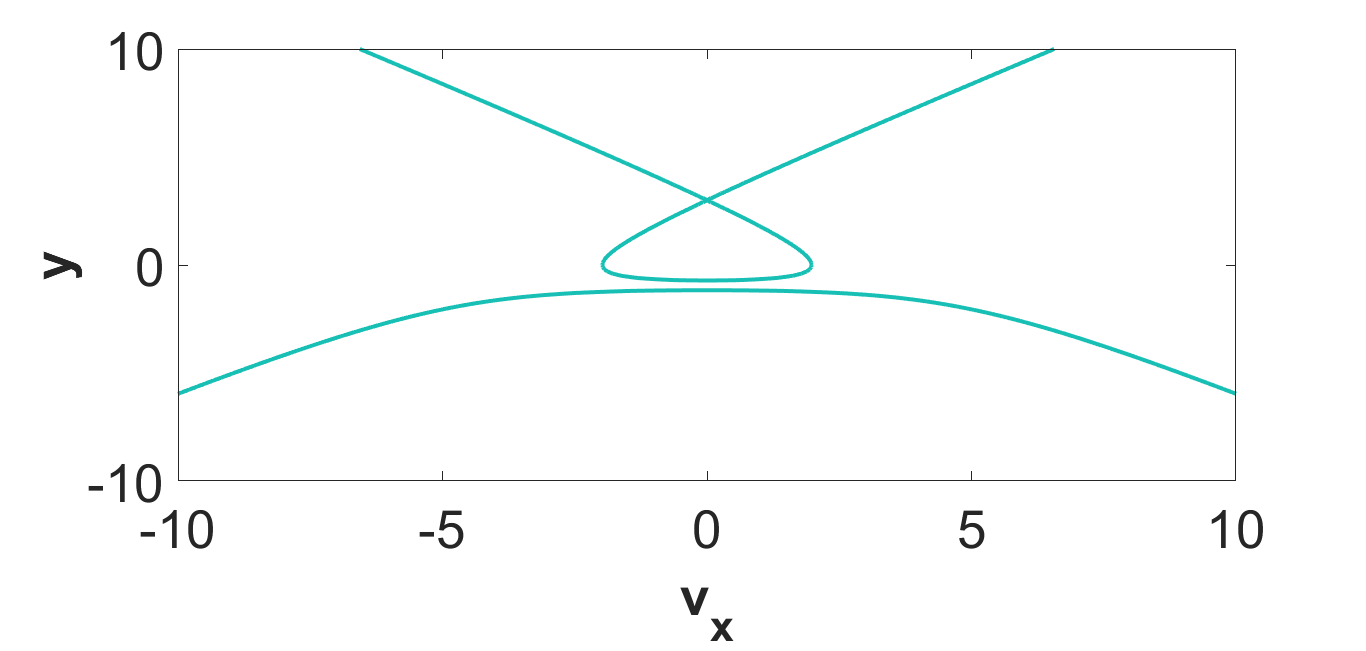}
 \ \ \ \
\includegraphics[width=5.6cm]{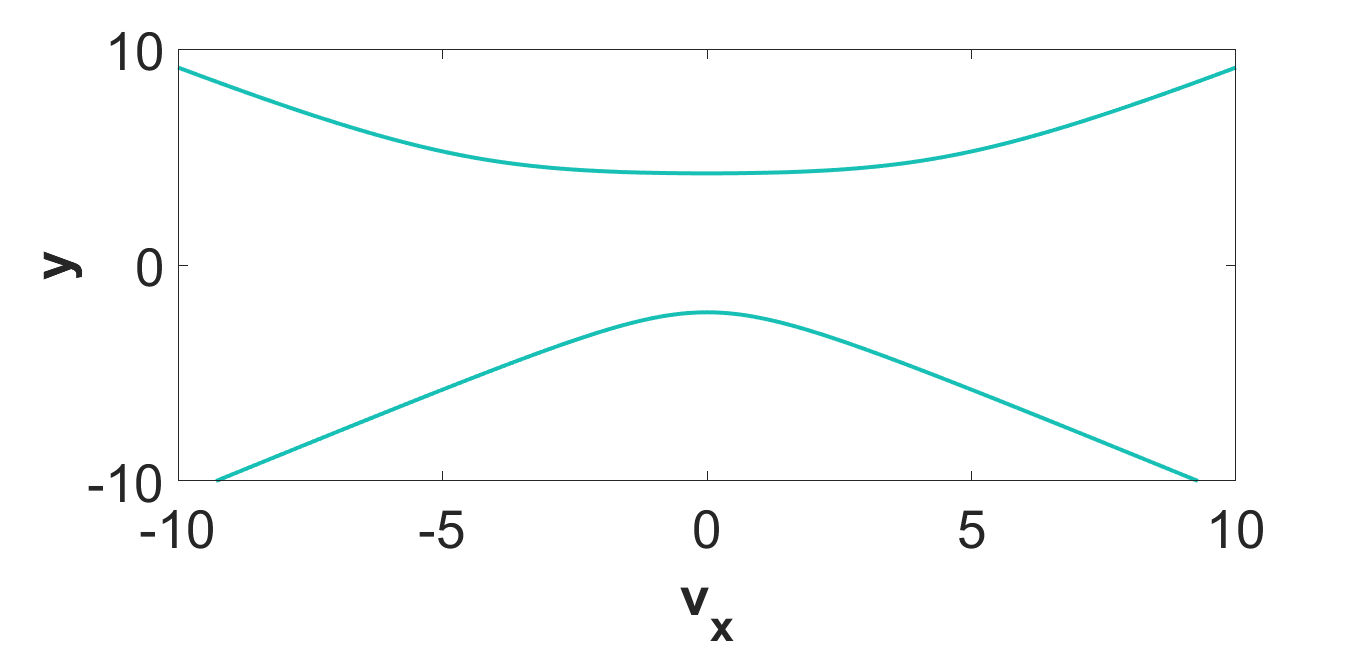}
(g) $A=1 \,B=-1,\,C=3 \,and \,F=-4$
\qquad (h) $A=1 \,B=-1,\,C=-3 \,and \,F=-1$
\qquad (i) $A=1 \,B=-1,\,C=-3 \,and \,F=4$
\caption{Bifurcations of phase portraits about the fixed point $(0,-\frac{C}{A})$ in the corresponding parameter space} \label{BifPP2} 
\end{figure}

 The corresponding bifurcations of phase portraits of the planar dynamical system Eq. (\ref{Phaseplane1}) are shown in Fig. \ref{BifPP2}. The bifurcations of phase portraits Fig. \ref{BifPP2} are obtained using the Hamiltonian function Eq. (\ref{Ham}) at the fixed point $(0,-\frac{C}{A})$:
$$
H(v_x,y)=\frac{v^2_x}{2}+\frac{A}{2B}y^2+\frac{BC-AF}{B^2}y+\frac{AF-BC}{B^3}F\,ln(y+\frac{F}{B})=H(0,-\frac{C}{A})=\frac{C^2}{2AB}+\frac{AF-BC}{AB^2}C+$$
\begin{equation}
\frac{AF-BC}{B^3}F\,ln(\frac{AF-BC}{AB}). \label{Ham0CA}
\end{equation}
In this type of bifurcation of phase portraits as shown in Fig. \ref{BifPP2}, the fixed point $(0,-\frac{C}{A})$ becomes a semi-stable saddle point from a neutrally stable center point and vice-versa as the parameters are varied.

\subsection{$FP3: (v_x,y)=(a,-\frac{F}{B})$} 
In this case, from Eqs. (\ref{Trace}) and (\ref{M}), it is clear that the trace of the Jacobian matrix is given as: $T(a,-\frac{F}{B})=-Ba$ whereas the determinant of the Jacobian matrix is vanishing. Therefore, according to the theory of planar dynamical systems, the fixed point  $(a,-\frac{F}{B})$ can represent a class of non-isolated fixed points or stars or degenerate nodes. This fixed point will be a cusp provided the Poincare index of the fixed point is zero \cite{Strogatz}.
 As, in this case, the determinant of the Jacobian matrix is vanishing, at least one of the eigenvalues is zero. This implies that the origin is not an isolated fixed point.


\section{Exact and approximate travelling wave solutions for the reduced planar dynamical system Eq. (\ref{fFinalNLEE1})} \label{Solution}
In order to get the solutions of the planar dynamical system Eq. (\ref{Phaseplane1}), which is the associated regular system of the planar dynamical system corresponding to Eq. (\ref{fFinalNLEE1}), we proceed by considering the values of the first integral or Hamiltonian function Eq. (\ref{Ham}) at the fixed points given in Eq. (\ref{FPs}) in the following manner. 
\subsection{FP1: $(v_x,y)=(0,0)$}
\begin{figure}[hbt!]
\centering
\includegraphics[width=20cm]{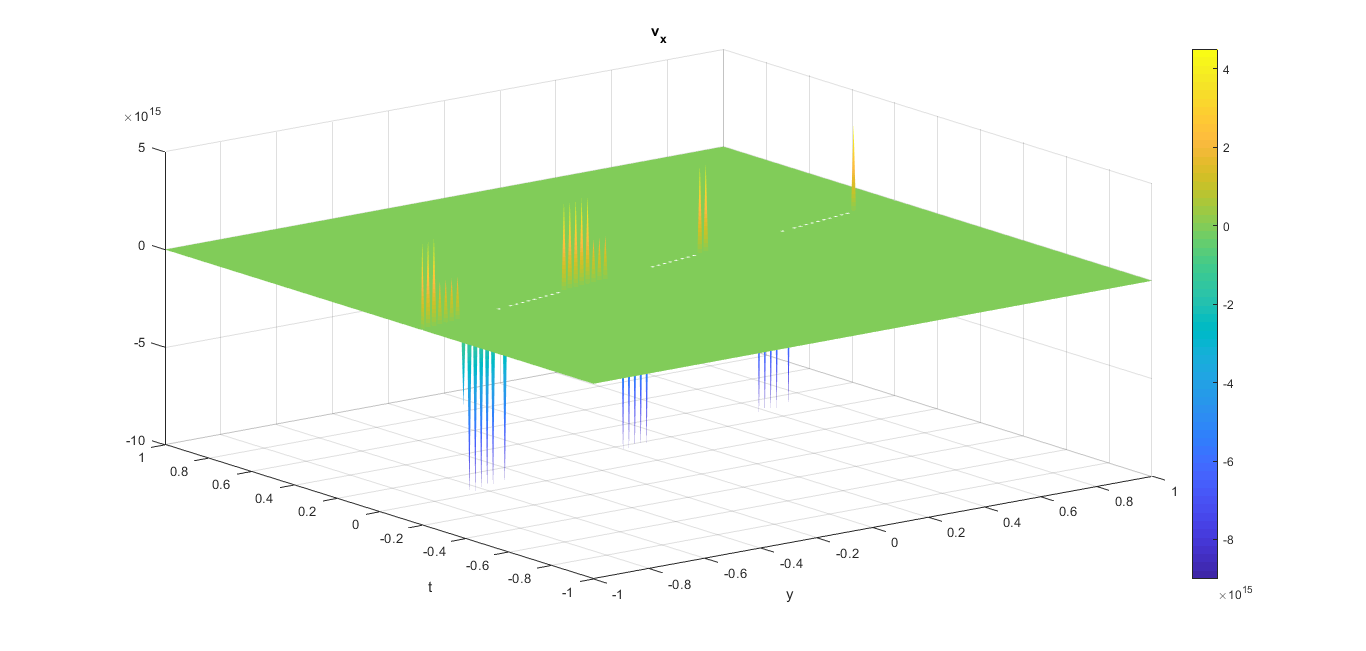}
\caption{Typical behaviour of the solution Eq. (\ref{solfp1}) for arbitrarily chosen parameter values $A=C=F=1,\,v=5$ and $k_1=0$.} \label{fp1so}
\end{figure}

\begin{figure}
\includegraphics[width=5.6cm]{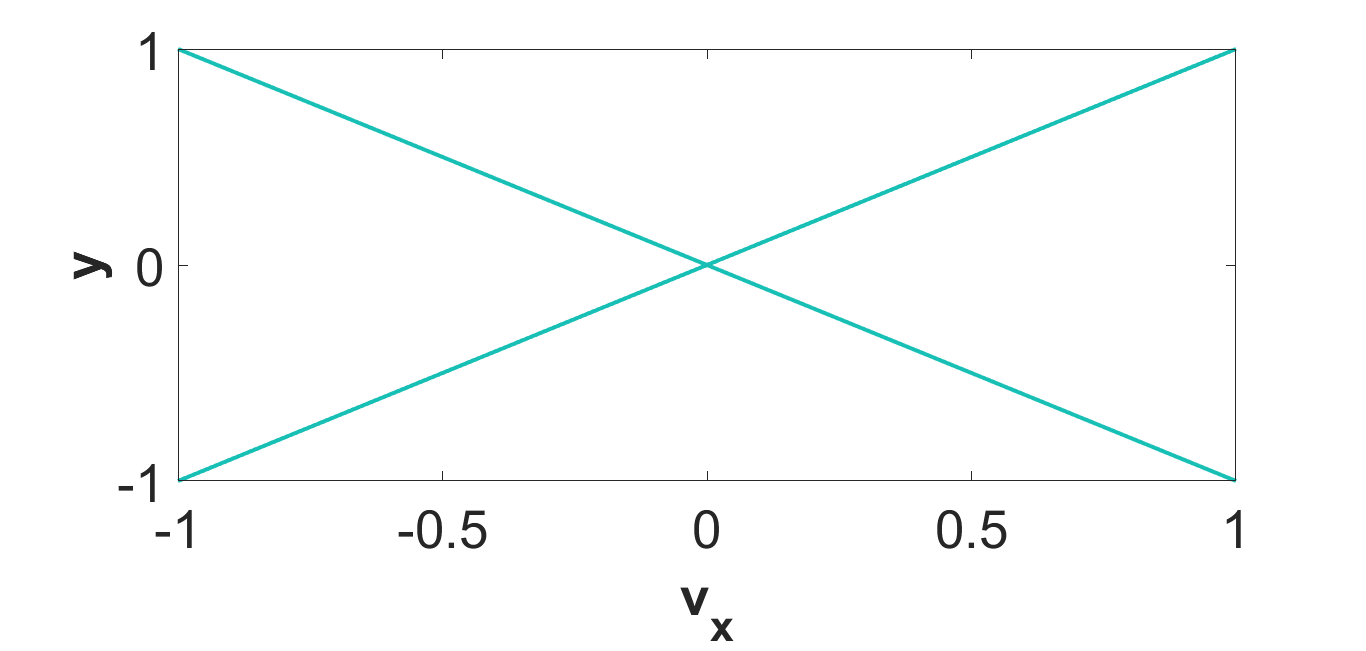}
 \ \ \ \
\includegraphics[width=5.6cm]{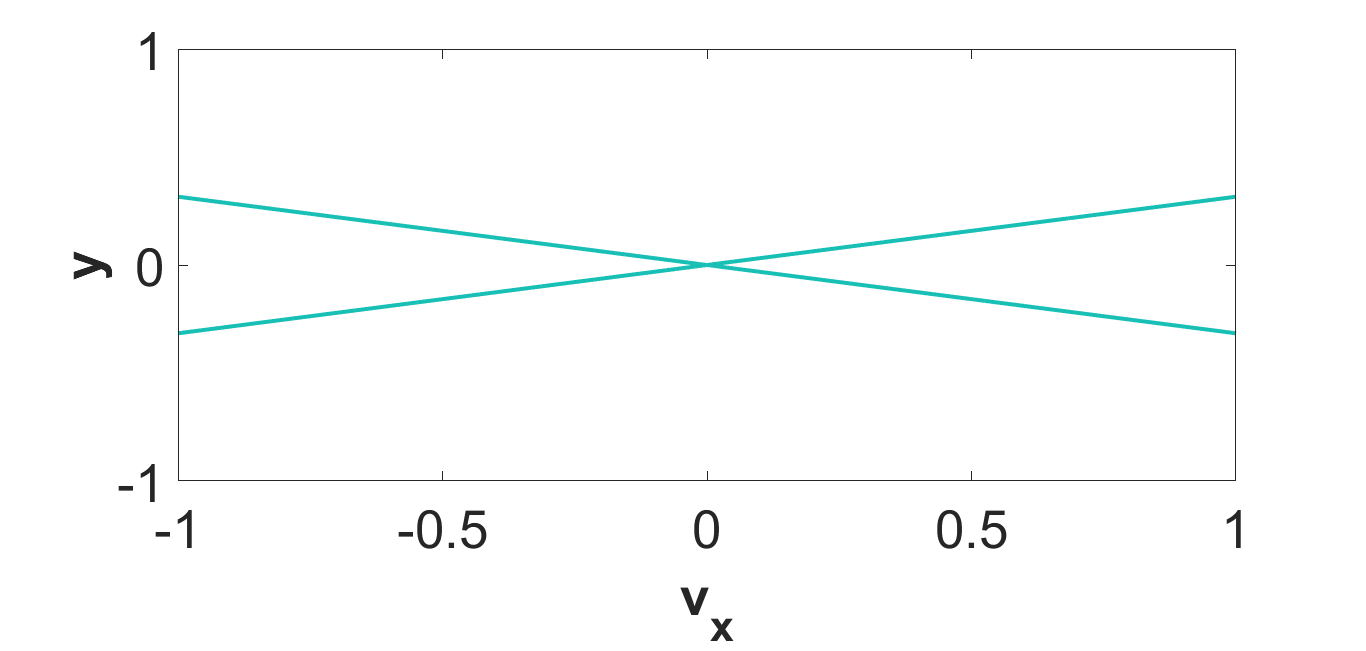}
 \ \ \ \
\includegraphics[width=5.6cm]{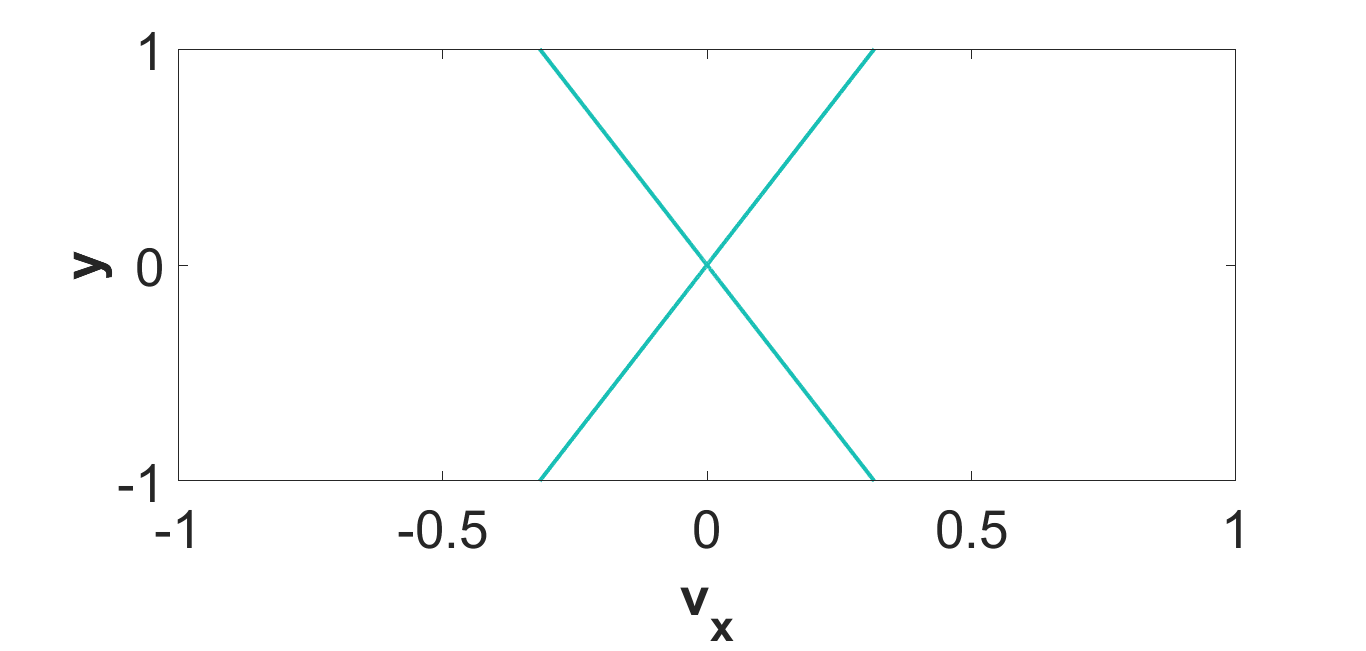}

(a) $C=1 \,and \,F=-1$
\quad \qquad \qquad \quad \qquad (b) $C=10 \,and \,F=-1$ 
\quad \qquad \qquad \quad \qquad (c) $C=-1 \,and \,F=10$
\caption{Level curves as represented by Eq. (\ref{lc1}) for some arbitrarily chosen parameter values corresponding to the FP $(v_x,y)=(0,0)$} \label{lc1fig}
\end{figure}

In this case, for finding the solution of Eq. (\ref{Phaseplane1}), we approximate the first integral Eq. (\ref{Ham}) as
\begin{equation}
H(v_x,y)=\frac{v^2_x}{2}+\frac{A}{2B}y^2+\frac{BC-AF}{B^2}y+\frac{AF-BC}{B^3}F[ln(\frac{F}{B})+\frac{B}{F}y-\frac{B^2}{2F^2}y^2], \label{Ham1app}
\end{equation} 
where we have used the following expansion for $ln(1+x)$.
\begin{equation}
ln(1+x)=x-\frac{x^2}{2}+\frac{x^3}{3}+...
\end{equation}
Now the value of the first integral Eq. (\ref{Ham1app}) at the FP1 $(v_x,y)=(0,0)$ is given by
\begin{equation}
H(0,0)=\frac{AF-BC}{B^3}F\,ln(\frac{F}{B})=h_1 \label{h1}
\end{equation}
where $h_1$ is a constant corresponding to the first integral Eq. (\ref{Ham}) for the FP1. Using Eqs. (\ref{Ham1app}) and (\ref{h1}), the locus of points on $(v_x,y)$ plane or phase plane satisfying the constant value $h_1$ is given by
\begin{equation}
\frac{v^2_x}{2}+\frac{A}{2B}y^2+\frac{BC-AF}{B^2}y+\frac{AF-BC}{B^3}F[ln(\frac{F}{B})+\frac{B}{F}y-\frac{B^2}{2F^2}y^2]=\frac{AF-BC}{B^3}F\,ln(\frac{F}{B}). \label{Locus1}
\end{equation}
After some simplifications in the above Eq. (\ref{Locus1}), we get
\begin{equation}
y^2=-\frac{F}{C} v^2_x. \label{y1}
\end{equation}
Now, from the first Eq. in the planar dynamical system (\ref{Phaseplane1}), we have
\begin{equation}
\frac{dv_x}{d\eta}=Ay^2+Cy. \label{Firstreg1}
\end{equation}
Substituting the value of $y$ in terms of $v_x$ from Eq. (\ref{y1}) in the above Eq. (\ref{Firstreg1}), we get
\begin{equation}
\int \frac{dv_x}{\sqrt{\frac{F}{C}}v_x(A \sqrt{\frac{F}{C}}v_x+C)}=\int d\eta. \label{aFP1}
\end{equation}
After evaluating the above integrations in Eq. (\ref{aFP1}), we obtain
\begin{equation}
v_x=\frac{C^{3/2}}{A\sqrt{F}}{[e^{-\sqrt{CF}(\eta +k_1)}-1]}^{-1}, \label{solfp1}
\end{equation}
where $k_1$ is the constant of integration. The pictorial representation of the above solution Eq. (\ref{solfp1}) is shown in Fig. \ref{fp1so} for some arbitrary values of the parameters. Therefore, in this case for the $FP1$, the solutions can be found at the level curves given by
\begin{equation}
y^2+\frac{F}{C} v^2_x=0, \label{lc1} 
\end{equation}
which is evident from Eqs. (\ref{y1}) and (\ref{solfp1}). The level curves represented by the above Eq. (\ref{lc1}) are shown in Fig. \ref{lc1fig} for different parameter values chosen arbitrarily.

\subsection{FP2: $(v_x,y)=(0,-\frac{C}{A})$}
\begin{figure}[hbt!]
\centering
\includegraphics[width=20cm]{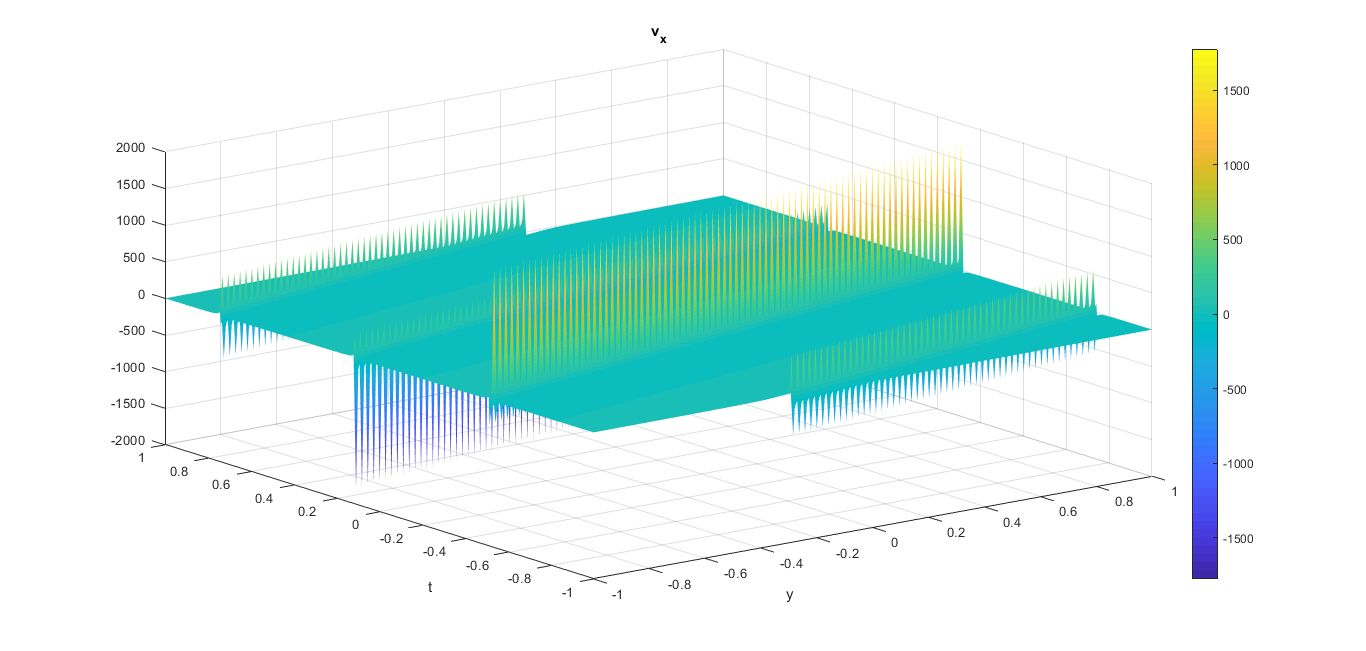}
\caption{Typical behaviour of the solution Eq. (\ref{solfp2}) for arbitrarily chosen parameter values $A=C=F=F'=1,\,\Lambda_1=\Lambda_2=1,\,v=5$ and $k_2=0$.} \label{fp2so}
\end{figure}

\begin{figure}
\includegraphics[width=5.6cm]{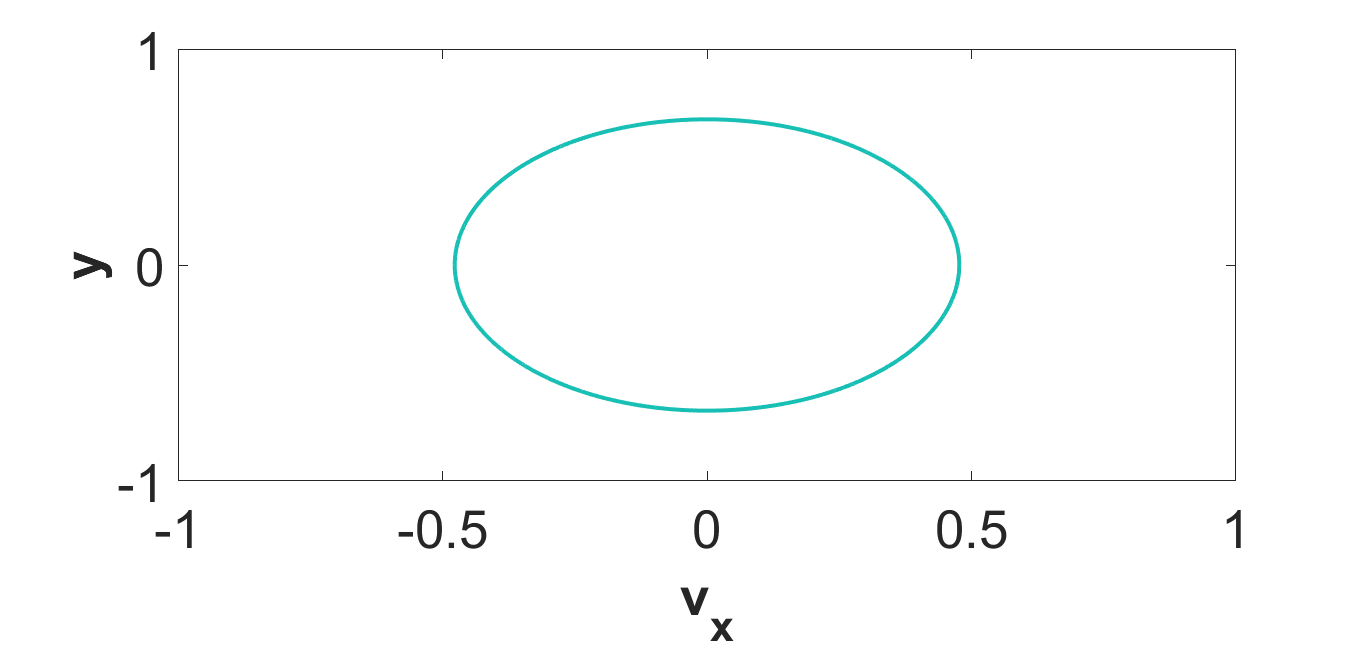}
 \ \ \ \
\includegraphics[width=5.6cm]{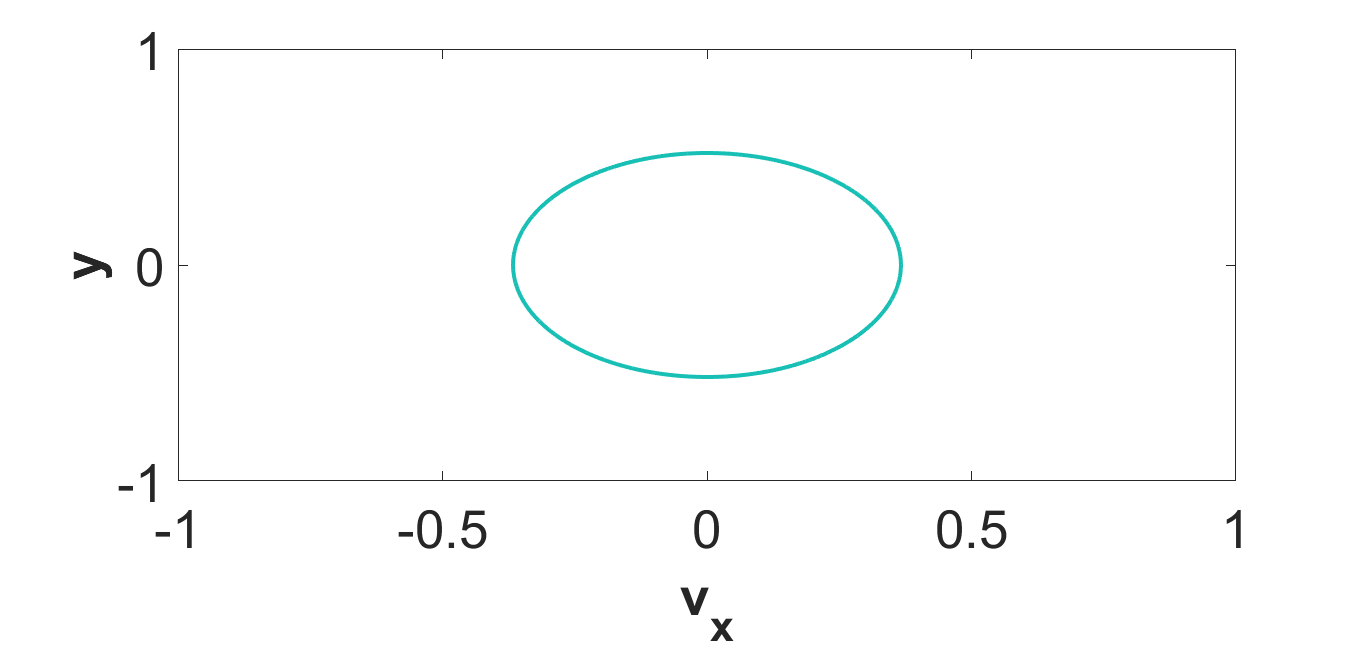}
 \ \ \ \
\includegraphics[width=5.6cm]{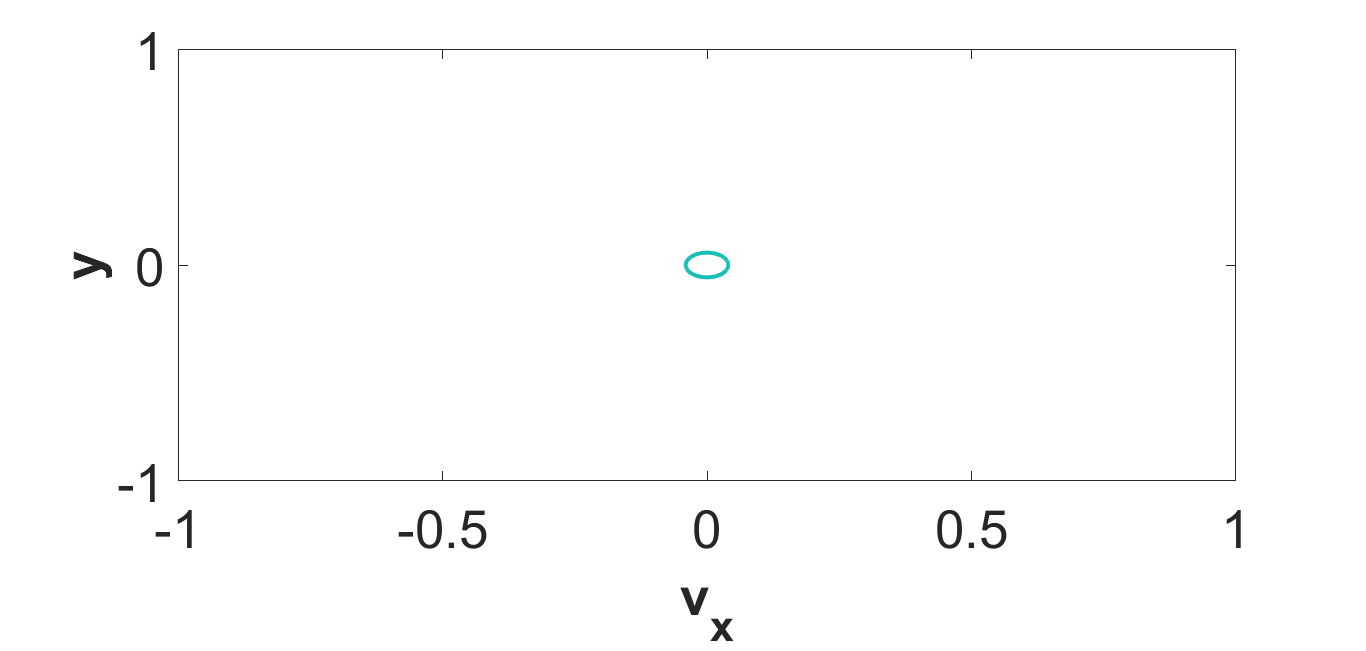}
(a) $A=1, \,B=1,\,C=1 \,and \,F=2$
\quad \qquad (b) $A=-1, \,B=1,\,C=1 \,and \,F=2$
\quad \qquad (c) $A=-10 \,B=1,\,C=1 \,and \,F=2$
\includegraphics[width=5.6cm]{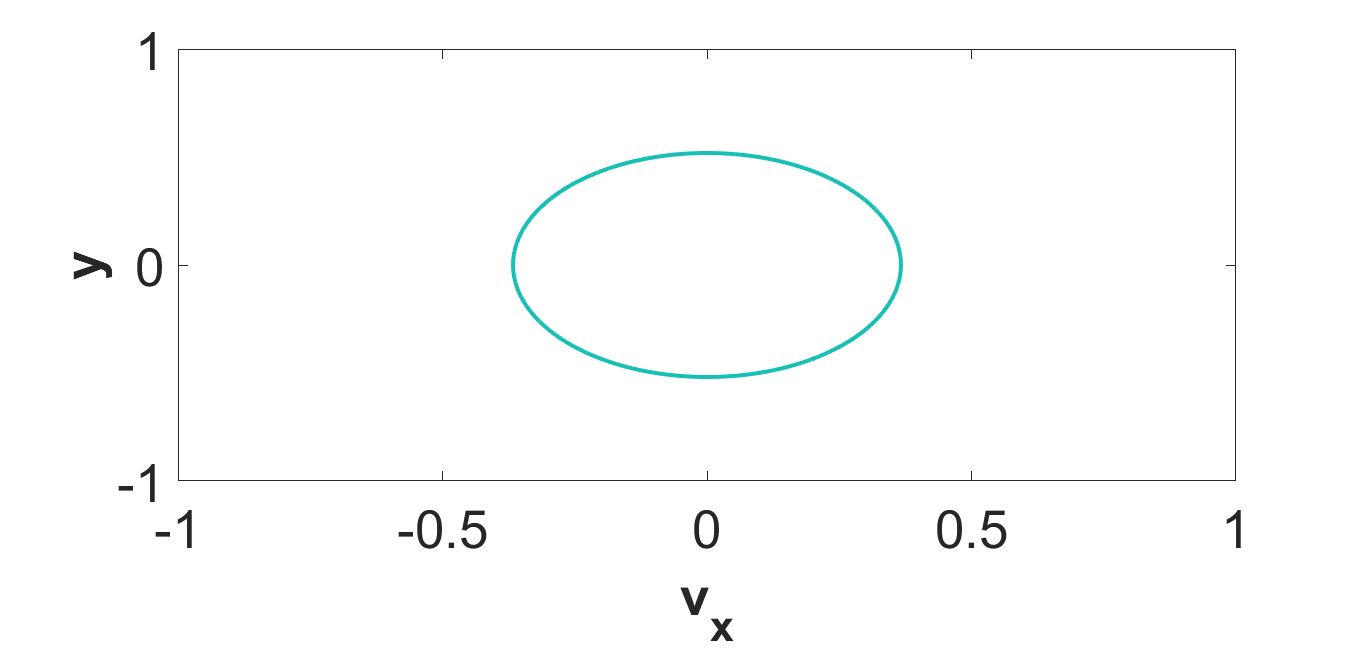}
 \ \ \ \
\includegraphics[width=5.6cm]{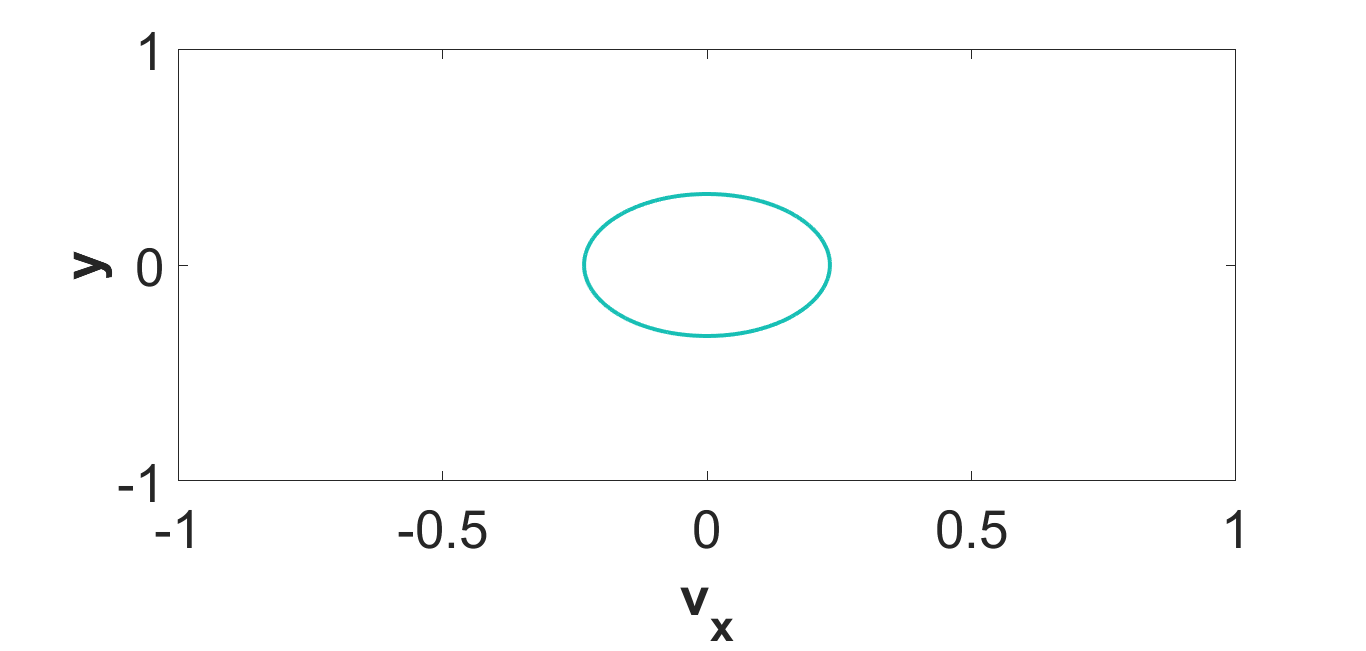}
 \ \ \ \
\includegraphics[width=5.6cm]{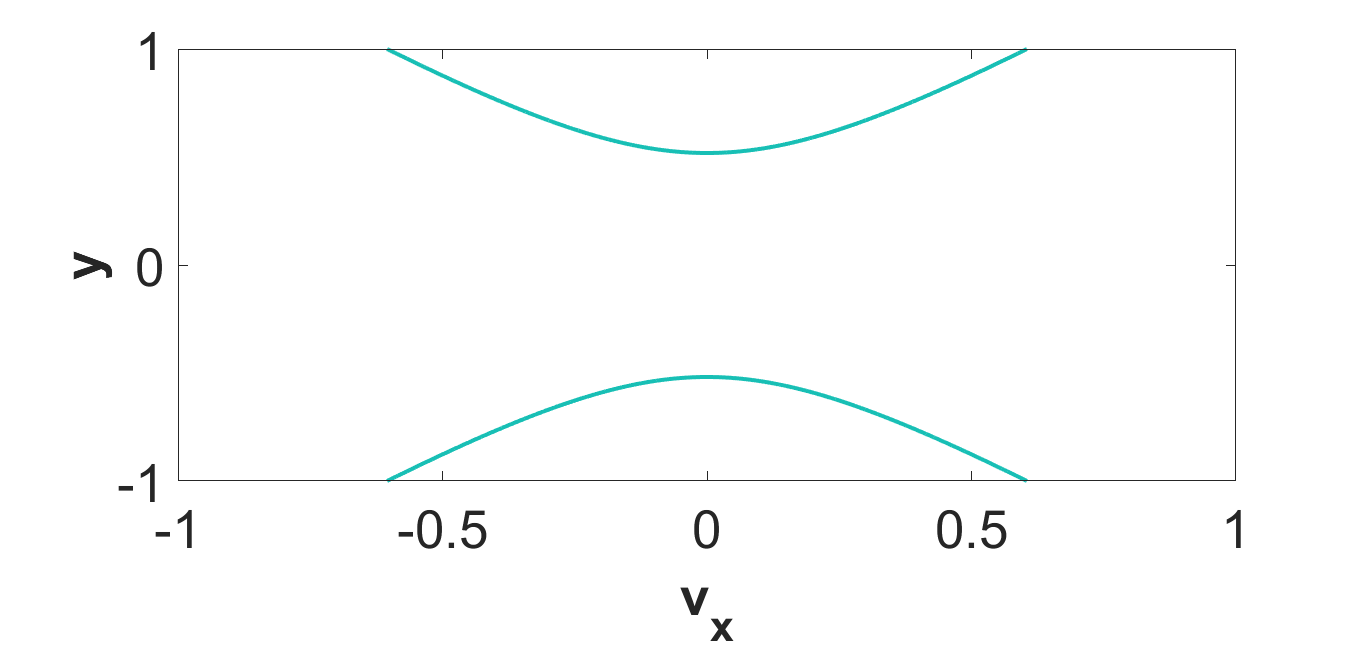}
(d) $A=1 \,B=-1,\,C=1 \,and \,F=2$
\quad \qquad (e) $A=1 \,B=-10,\,C=1 \,and \,F=2$
\quad \qquad (f) $A=1 \,B=1,\,C=-1 \,and \,F=2$
 
\includegraphics[width=5.6cm]{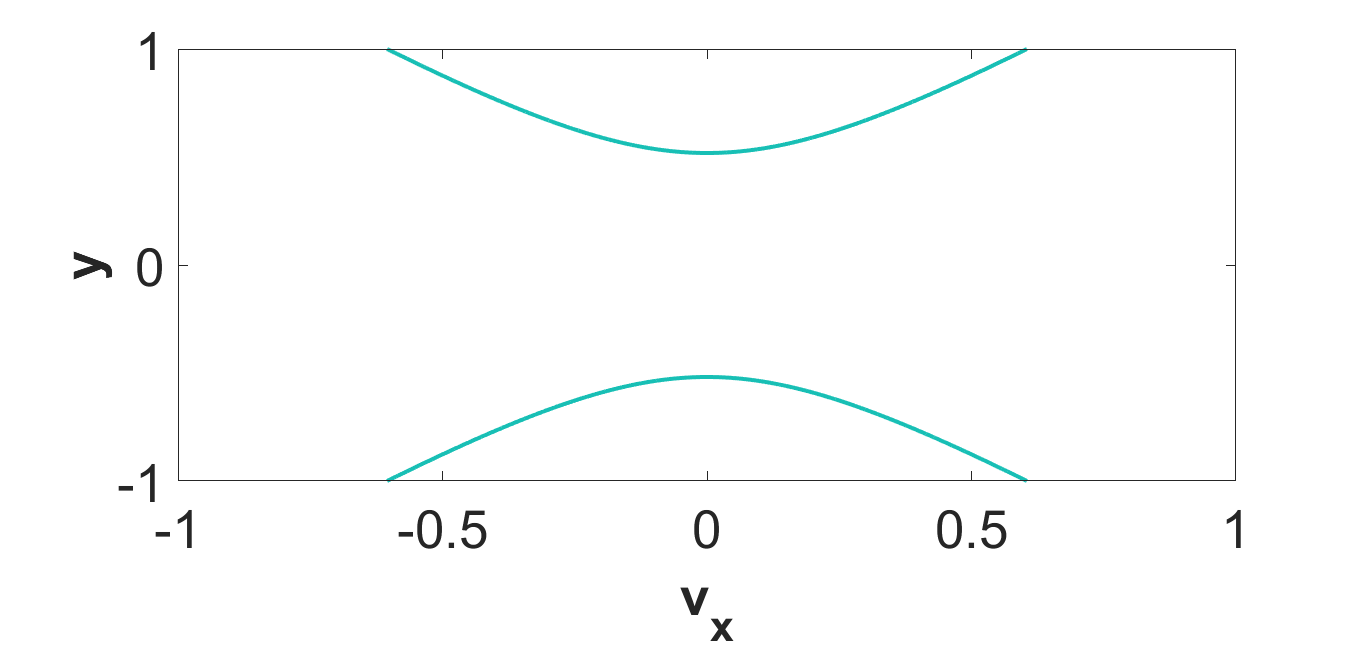}
 \ \ \ \
\includegraphics[width=5.6cm]{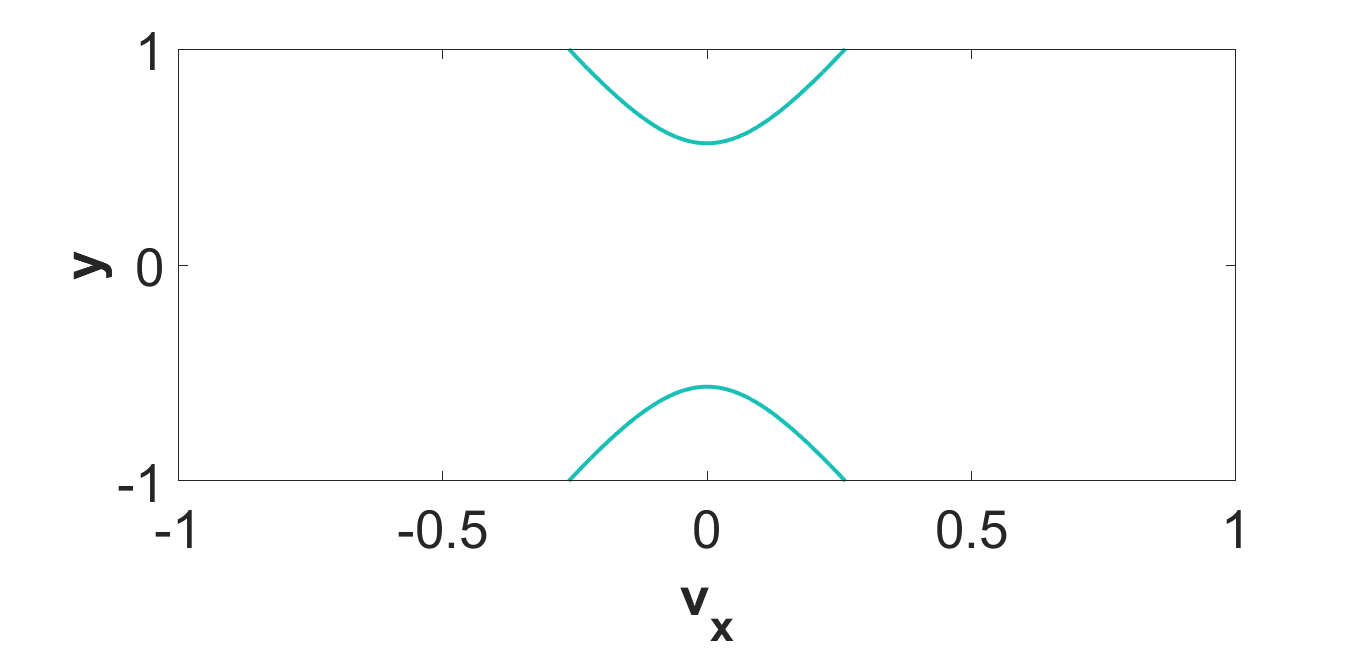}
 \ \ \ \
\includegraphics[width=5.6cm]{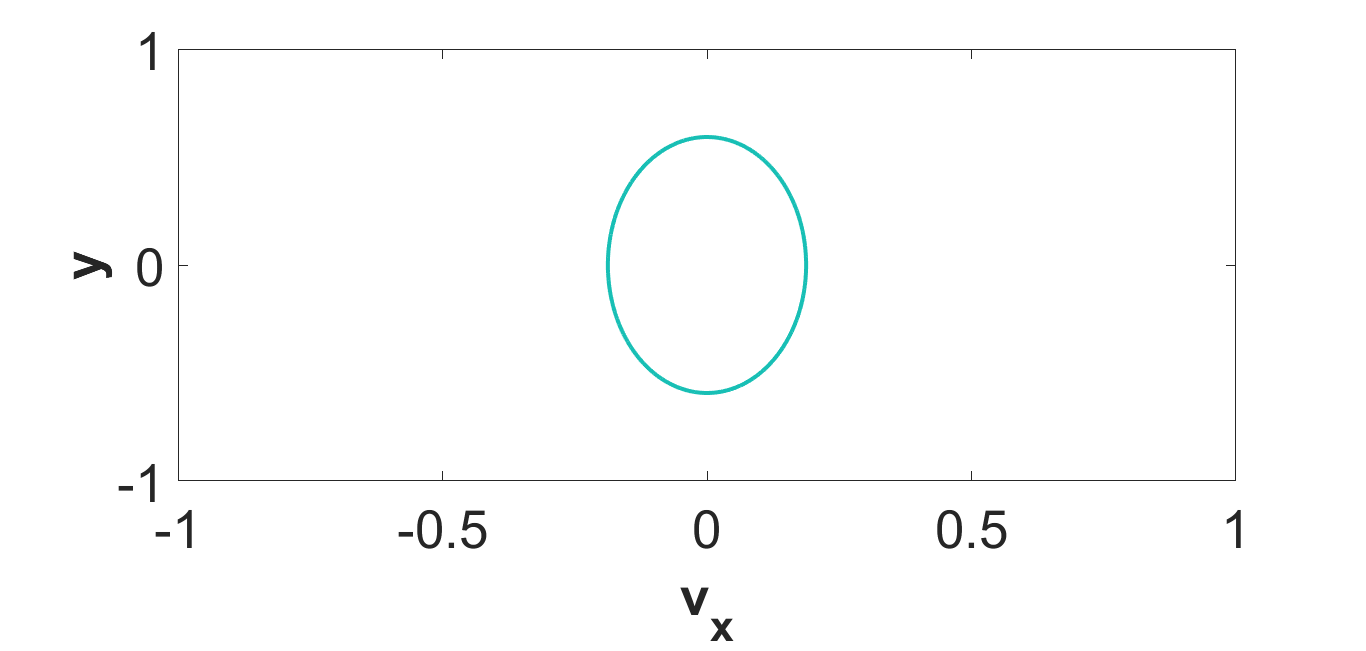}
(g) $A=1 \,B=1,\,C=1 \,and \,F=-2$
\quad \qquad (h) $A=1 \,B=1,\,C=1 \,and \,F=-10$
\quad \qquad (i) $A=1 \,B=1,\,C=1 \,and \,F=10$
\caption{Level curves as represented by Eq. (\ref{lc2}) for some arbitrarily chosen parameter values corresponding to the FP $(v_x,y)=(0,-\frac{C}{A})$} \label{lc2fig}
\end{figure}

As in the previous subsection, we use the approximate first integral Eq. (\ref{Ham1app}) in this subsection also. Now the value of the first integral Eq. (\ref{Ham1app}) at the FP2 $(v_x,y)=(0,-\frac{C}{A})$ is given by
\begin{equation}
H(0,-\frac{C}{A})=\frac{FC}{B^2}-\frac{C^2}{2AB}+\frac{AF-BC}{B^3}F \,ln(\frac{AF-BC}{AB})=h_2 \label{h2}
\end{equation}
where $h_2$ is a constant corresponding to the first integral Eq. (\ref{Ham}) for the FP2. Using Eqs. (\ref{Ham1app}) and (\ref{h2}), the locus of points on $(v_x,y)$ plane or phase plane satisfying the constant value $h_2$ is given by
\begin{equation}
\frac{v^2_x}{2}+\frac{A}{2B}y^2+\frac{BC-AF}{B^2}y+\frac{AF-BC}{B^3}F[ln(\frac{F}{B})+\frac{B}{F}y-\frac{B^2}{2F^2}y^2]=\frac{FC}{B^2}-\frac{C^2}{2AB}+\frac{AF-BC}{B^3}F \,ln(\frac{AF-BC}{AB}). \label{Locus2}
\end{equation}
This implies
\begin{equation}
y^2=-\frac{2F}{C}[\frac{v^2_x}{2}+\frac{AF-BC}{B^3}F\, ln(\frac{AF}{AF-BC})-\frac{FC}{B^2}+\frac{C^2}{2AB}]. \label{y2}
\end{equation}
From the first equation in the planar dynamical system Eq. (\ref{Phaseplane1}), we have
\begin{equation}
\frac{dv_x}{d\eta}=Ay^2+Cy. \label{Firstreg1}
\end{equation}
Substituting the value of $y$ in terms of $v_x$ from Eq. (\ref{y2}) in the above Eq. (\ref{Firstreg1}), we get
\begin{equation}
\int \frac{dv_x}{(-\frac{2AF}{C})[\frac{v^2_x}{2}+\frac{AF-BC}{B^3}F \,ln(\frac{AF}{AF-BC})-\frac{FC}{B^2}+\frac{C^2}{2AB}]+C\sqrt{(-\frac{2F}{C})[\frac{v^2_x}{2}+\frac{AF-BC}{B^3}F\, ln(\frac{AF}{AF-BC})-\frac{FC}{B^2}+\frac{C^2}{2AB}]}}=\int d\eta. \label{aFP2}
\end{equation}
After evaluating the above integrations in Eq. (\ref{aFP2}), we obtain
\begin{equation}
v_x=2\sqrt{2F'}\frac{\Lambda_1\Lambda_2tan[
\frac{\Lambda_1\Lambda_2\sqrt{CF}(\eta+k_2)}{2}]}{\Lambda^2_1-\Lambda^2_2 tan^2[
\frac{\Lambda_1\Lambda_2\sqrt{CF}(\eta+k_2)}{2}]},\label{solfp2}
\end{equation}
where $k_2$ is the constant of integration and the parameters $\Lambda_1,\,\Lambda_2$ and $F'$ are given by
$$ F'=\frac{AF-BC}{B^3}F\, ln(\frac{AF}{AF-BC})-\frac{FC}{B^2}+\frac{C^2}{2AB}$$
\begin{equation}
\Lambda_1=\sqrt{\sqrt{\frac{2FF'}{C^3}}A-1};\,\Lambda_2=\sqrt{\sqrt{\frac{2FF'}{C^3}}A+1}
\end{equation}
The above solution Eq. (\ref{solfp2}) is pictorially represented in Fig. \ref{fp2so} for some arbitrary values of parameters. Therefore, in this case for the FP2, the solutions can be found at the level curves given by
\begin{equation}
y^2+\frac{2F}{C}[\frac{v^2_x}{2}+\frac{AF-BC}{B^3}F\, ln(\frac{AF}{AF-BC})-\frac{FC}{B^2}+\frac{C^2}{2AB}]=0, \label{lc2}
\end{equation}
which is evident from Eqs. (\ref{y2}) and (\ref{solfp2}). These level curves are plotted in Fig. \ref{lc2fig}.


\subsection{FP3: $(v_x,y)=(a,-\frac{F}{B})$}
\begin{figure}[hbt!]
\centering
\includegraphics[width=20cm]{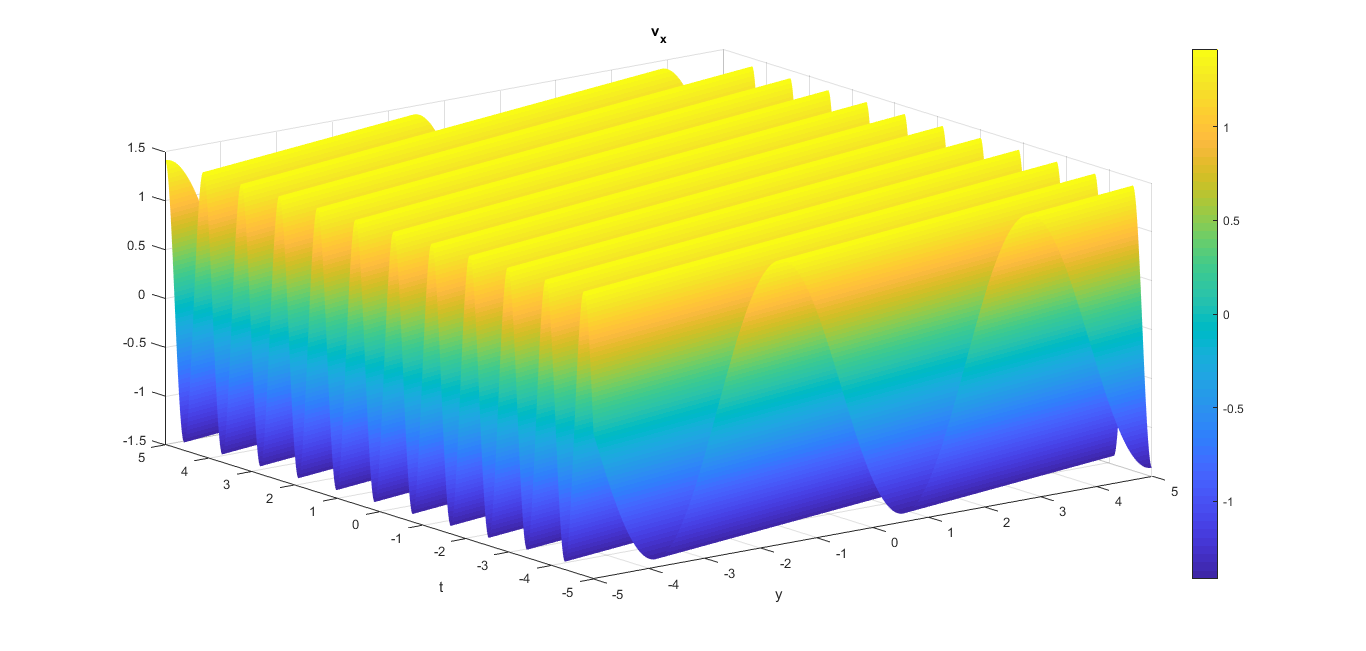}
\caption{Typical behaviour of the solution Eq. (\ref{solfp3}) for arbitrarily chosen parameter values $a=1,\,A=F=B=1,\,\lambda_1=\lambda_2=1,\,C=2,$ and $k_3=0$.} \label{fp3so}
\end{figure}

\begin{figure}
\includegraphics[width=5.6cm]{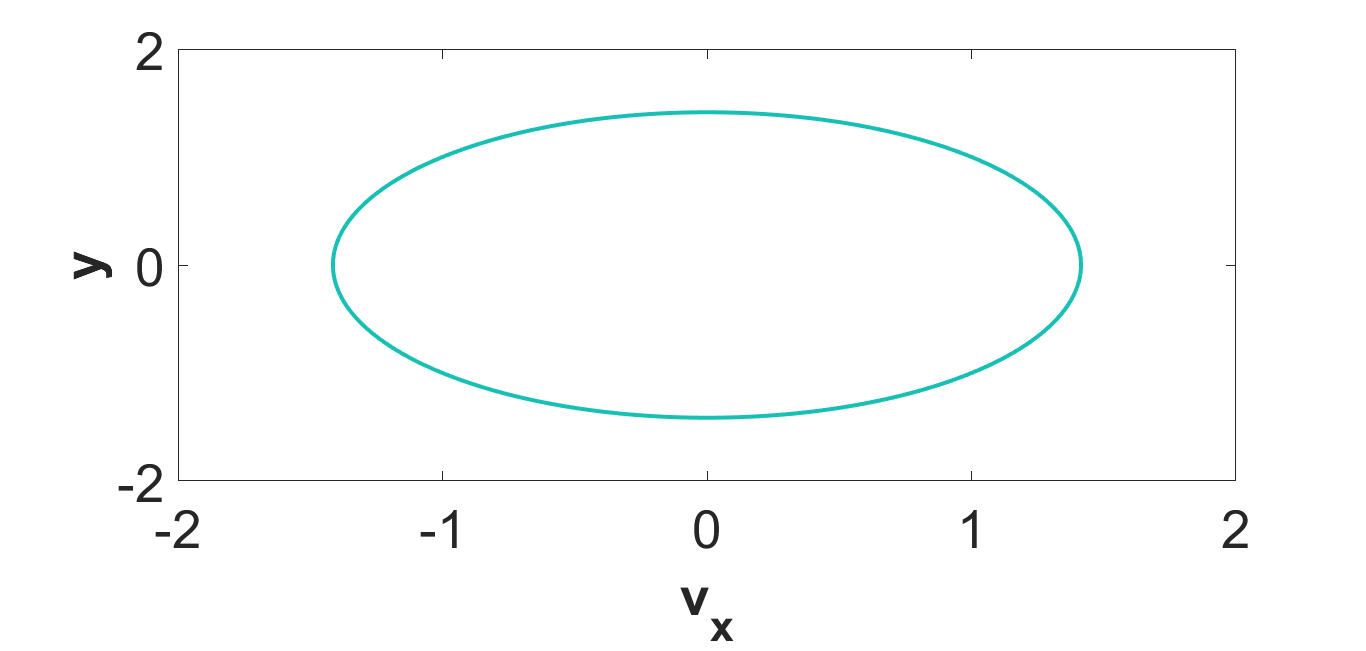}
 \ \ \ \
\includegraphics[width=5.6cm]{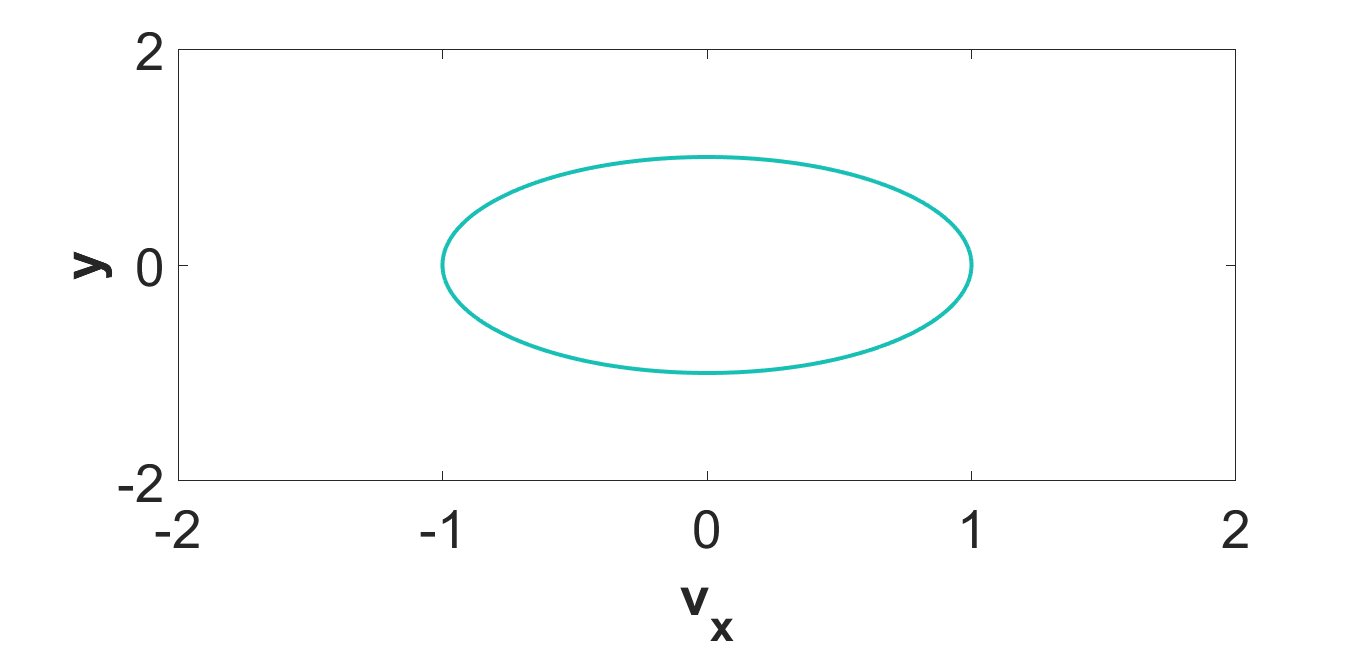}
 \ \ \ \
\includegraphics[width=5.6cm]{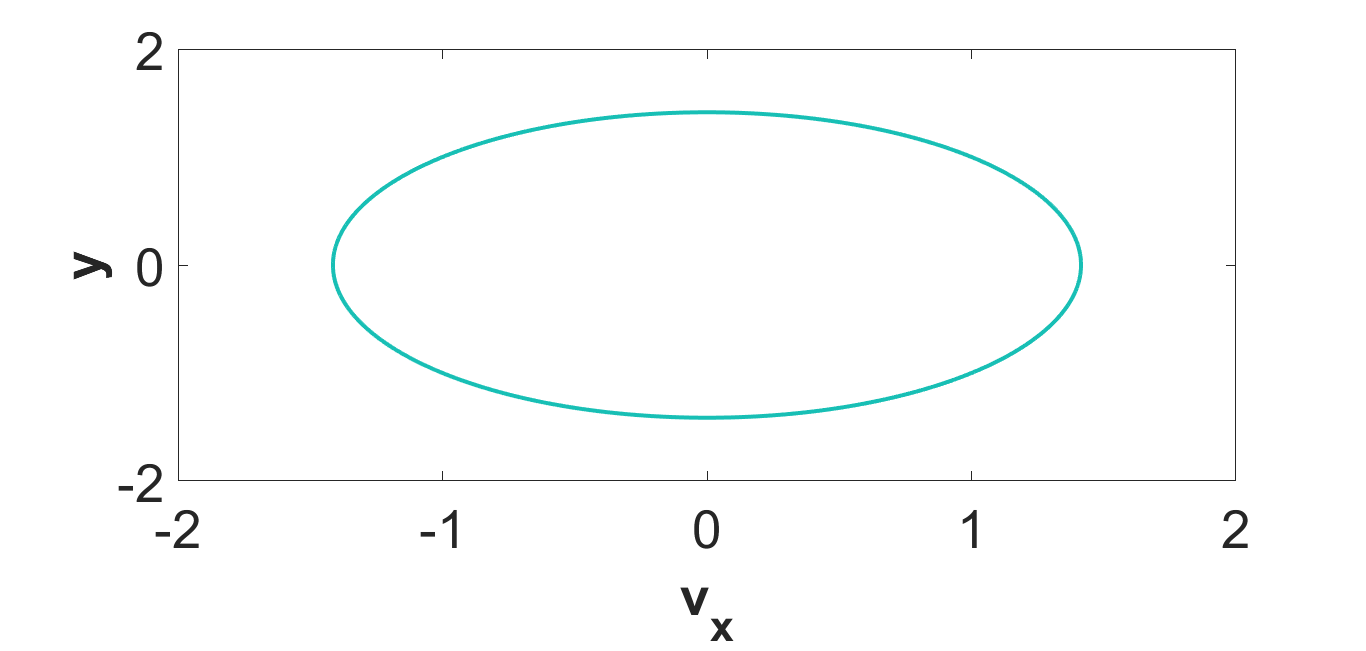}
(a) $a=1, \,A=1,\,B=1 \,and \,F=1$
\quad \qquad (b) $a=0, \,A=1,\,B=1 \,and \,F=1$
\quad \qquad (c) $a=-1 \,A=1,\,B=1 \,and \,F=1$
\includegraphics[width=5.6cm]{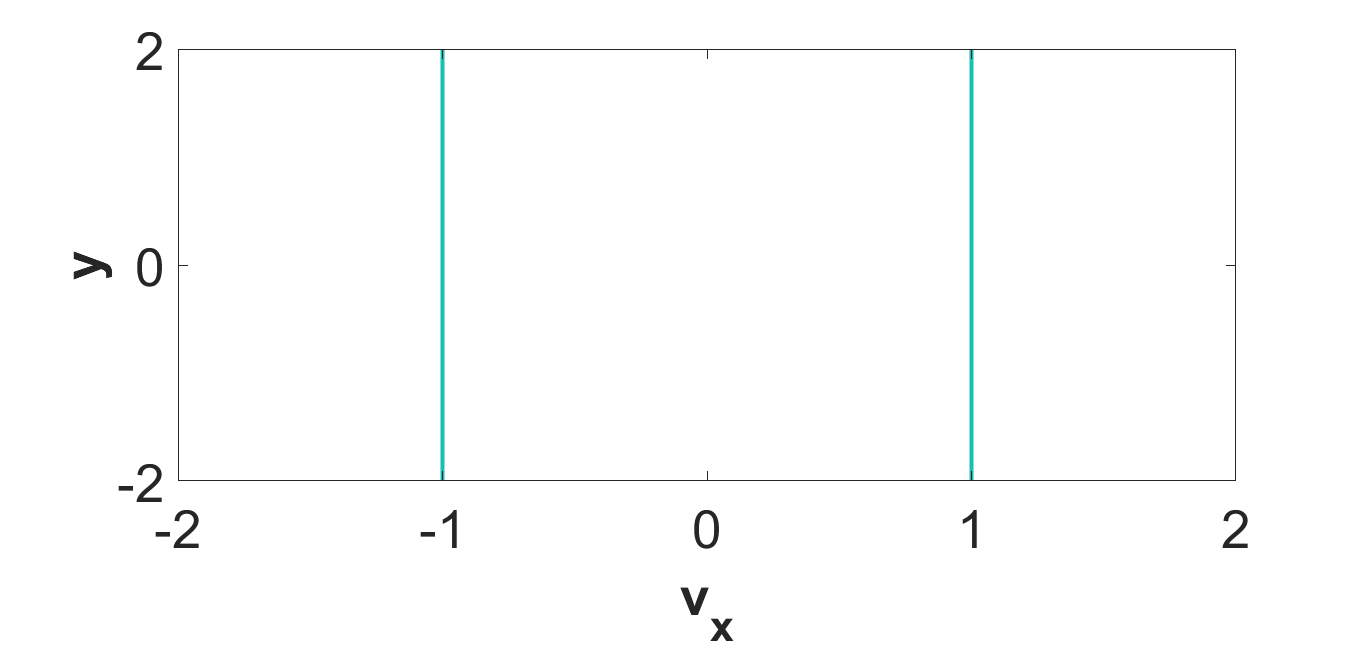}
 \ \ \ \
\includegraphics[width=5.6cm]{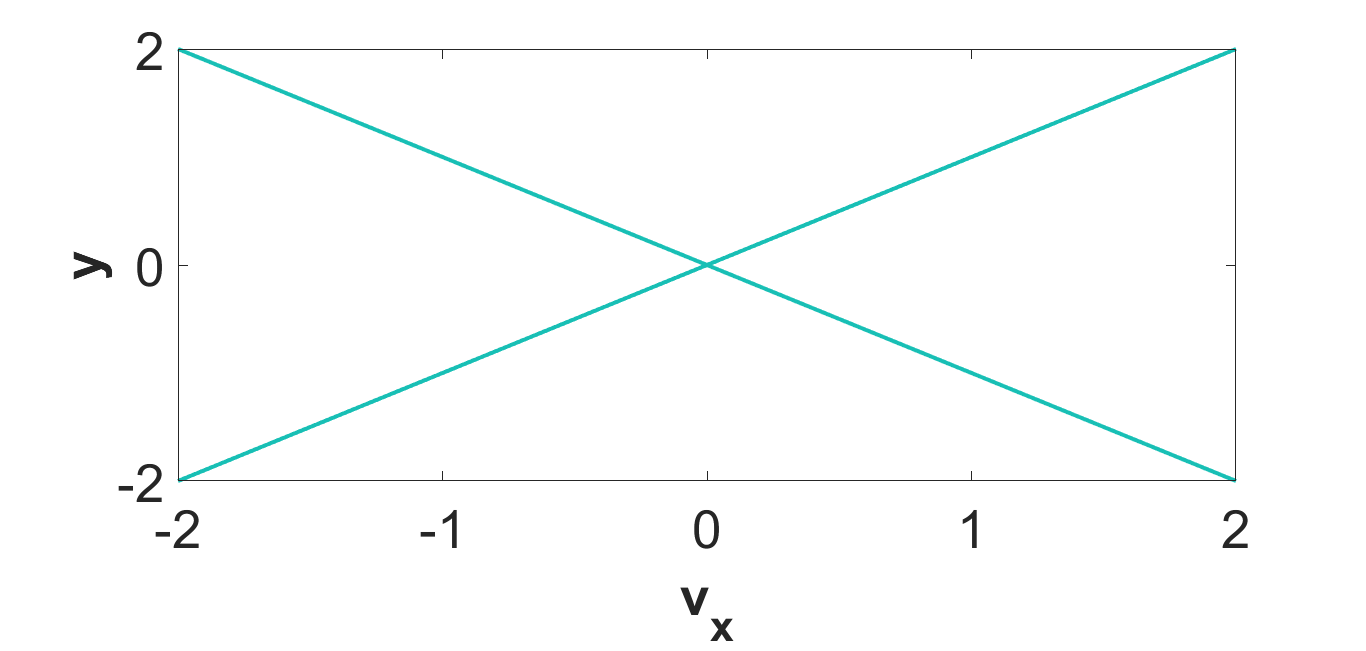}
 \ \ \ \
\includegraphics[width=5.6cm]{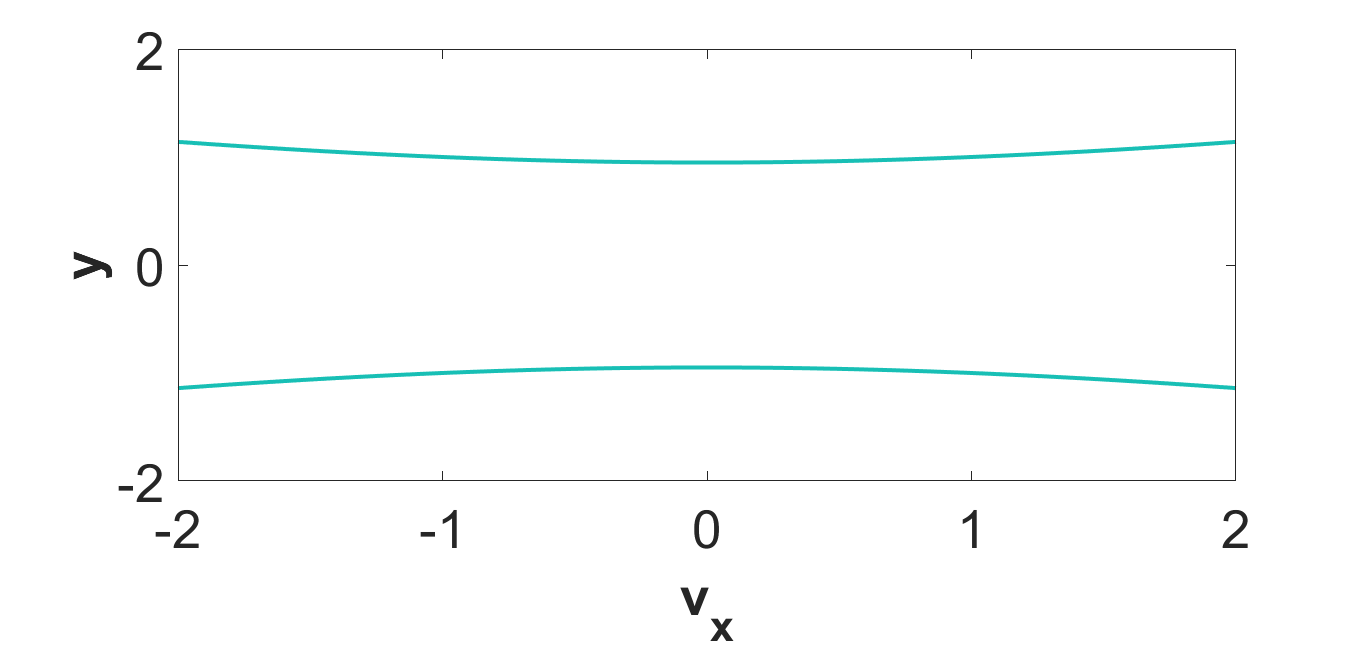}
(d) $a=1 \,A=0,\,B=1 \,and \,F=1$
\quad \qquad (e) $a=1 \,A=-1,\,B=1 \,and \,F=1$
\quad \qquad (f) $a=1 \,A=-10,\,B=1 \,and \,F=1$
 
\includegraphics[width=5.6cm]{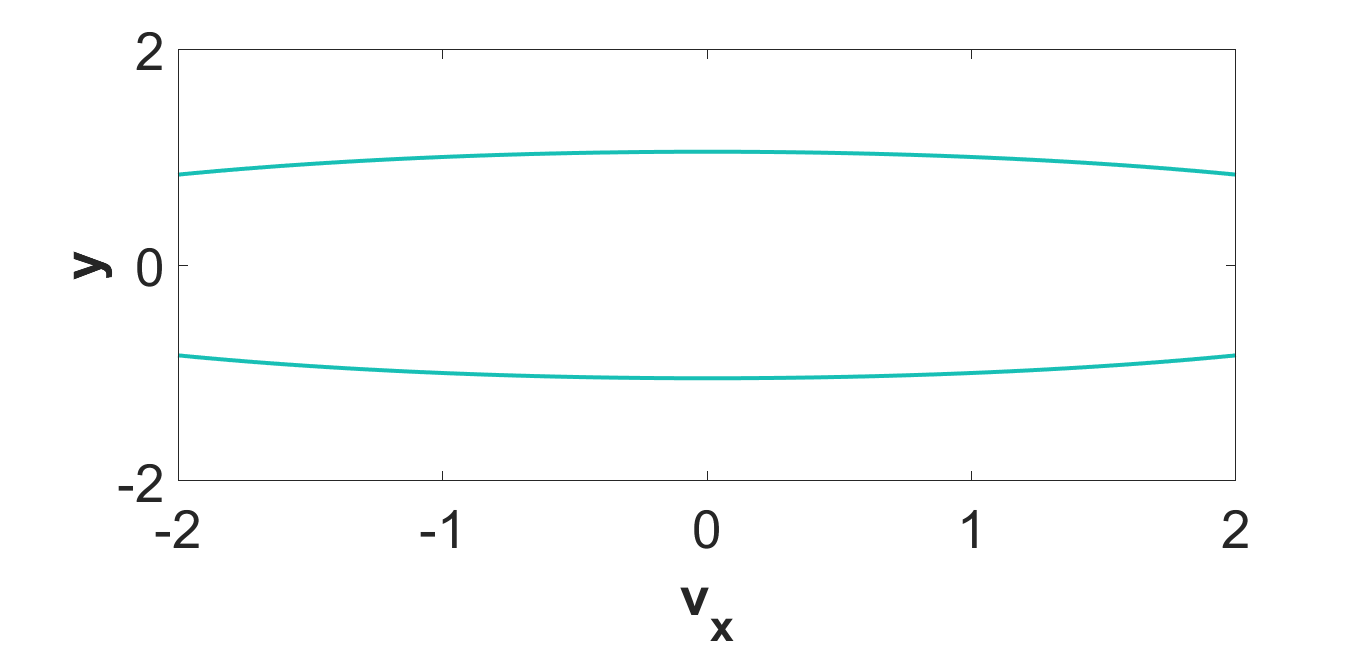}
 \ \ \ \
\includegraphics[width=5.6cm]{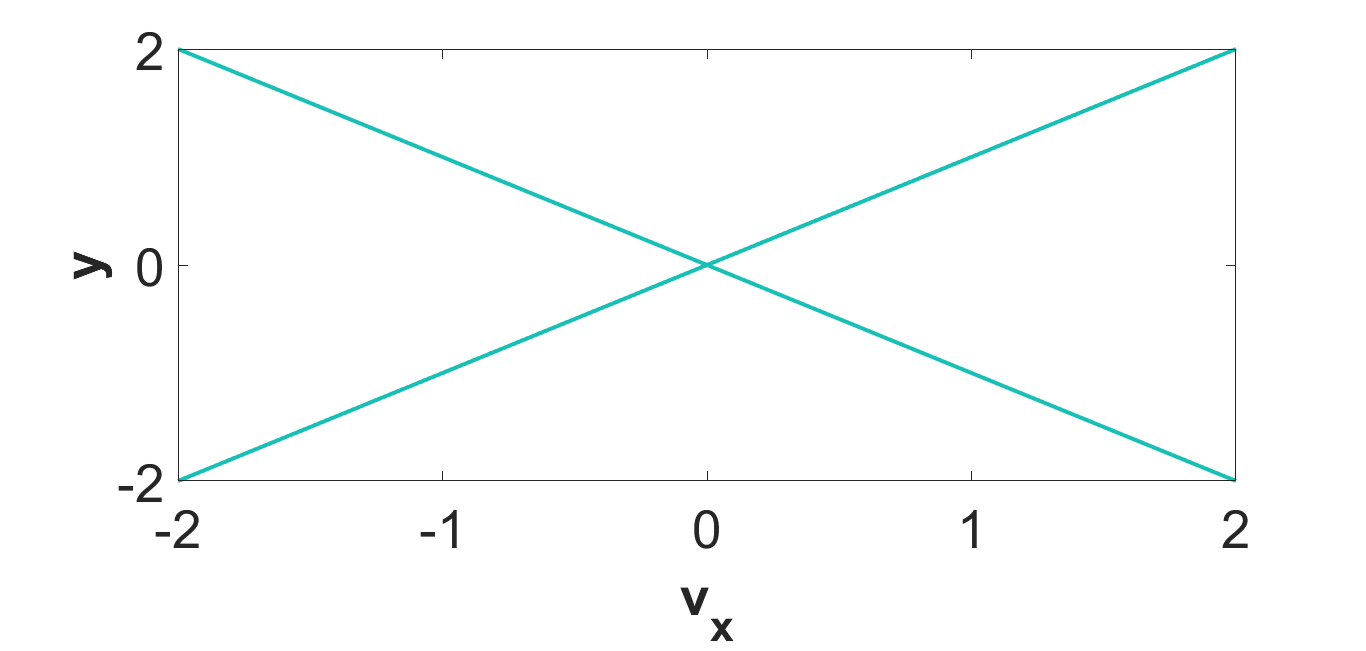}
 \ \ \ \
\includegraphics[width=5.6cm]{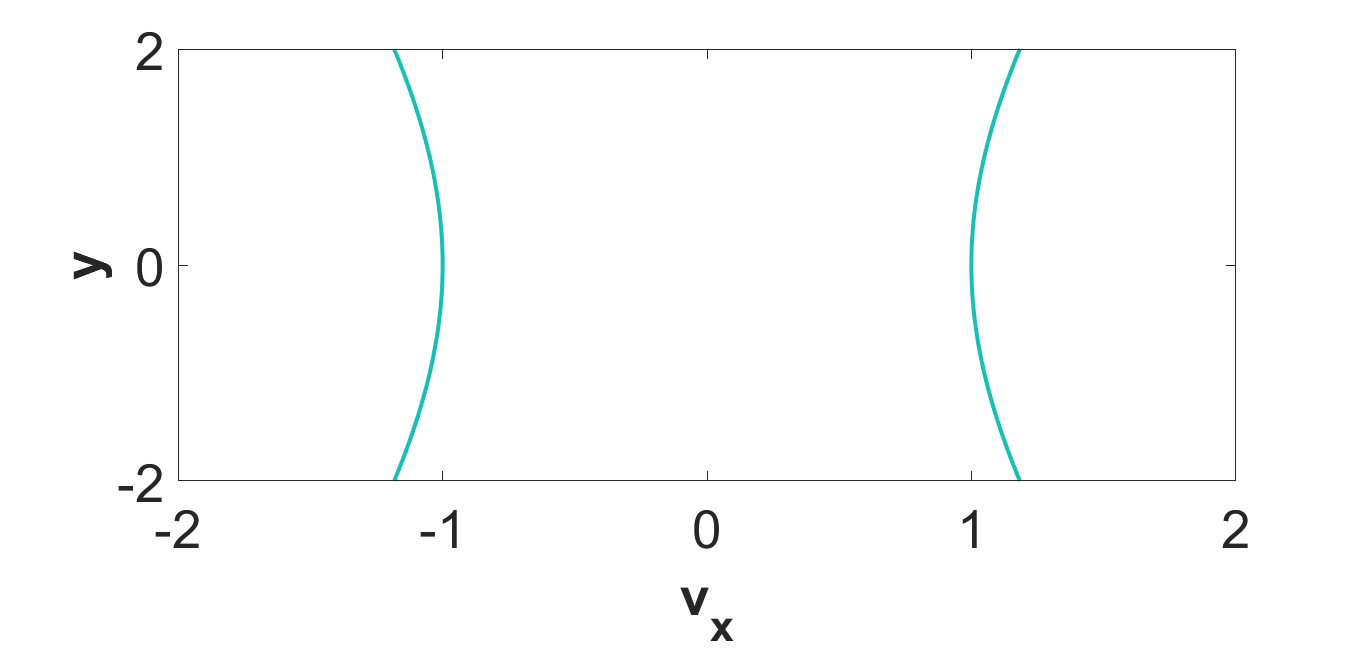}
(g) $a=1 \,A=10,\,B=1 \,and \,F=1$
\quad \qquad (h) $a=1 \,A=1,\,B=-1 \,and \,F=1$
\quad \qquad (i) $a=1 \,A=1,\,B=-10 \,and \,F=1$
\includegraphics[width=5.6cm]{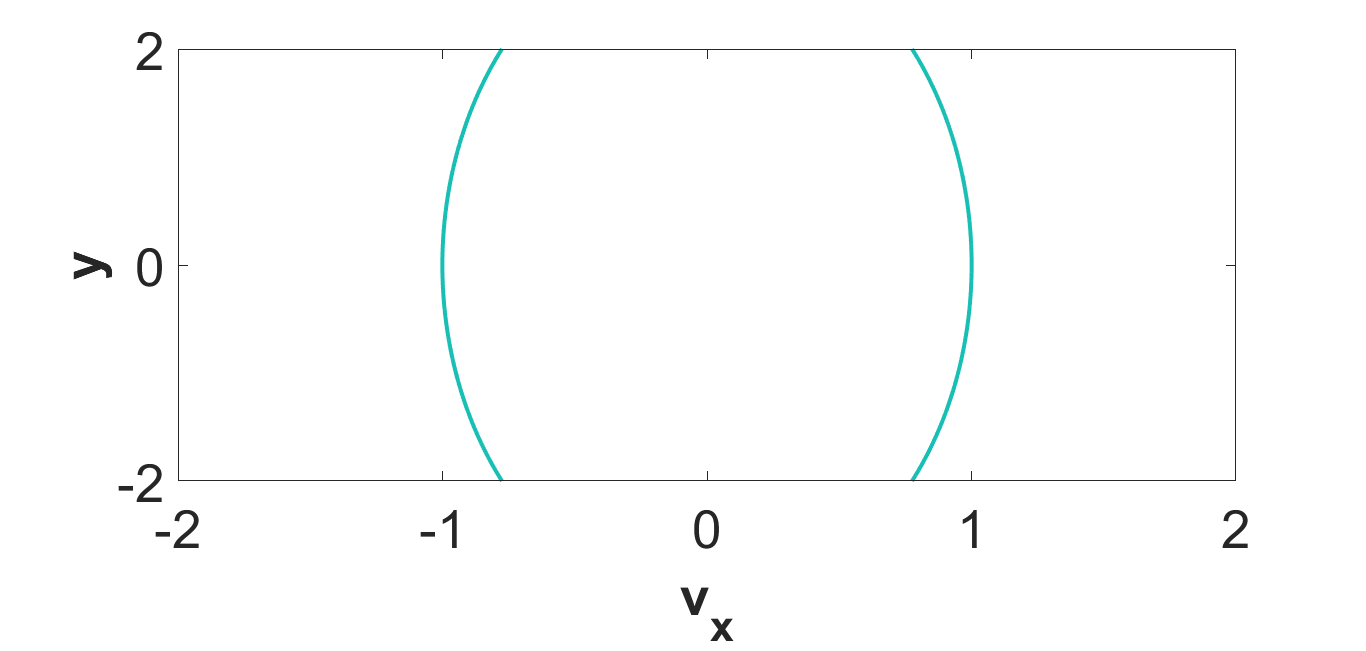}
 \ \ \ \
\includegraphics[width=5.6cm]{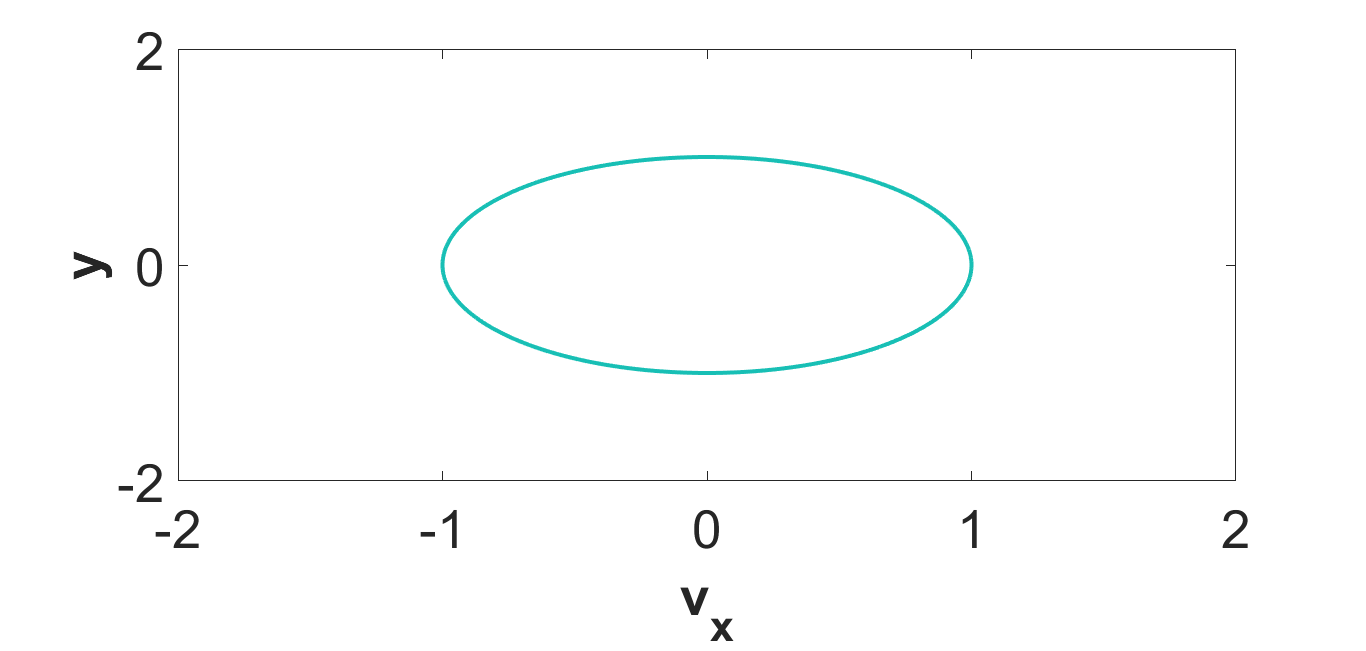}
 \ \ \ \
\includegraphics[width=5.6cm]{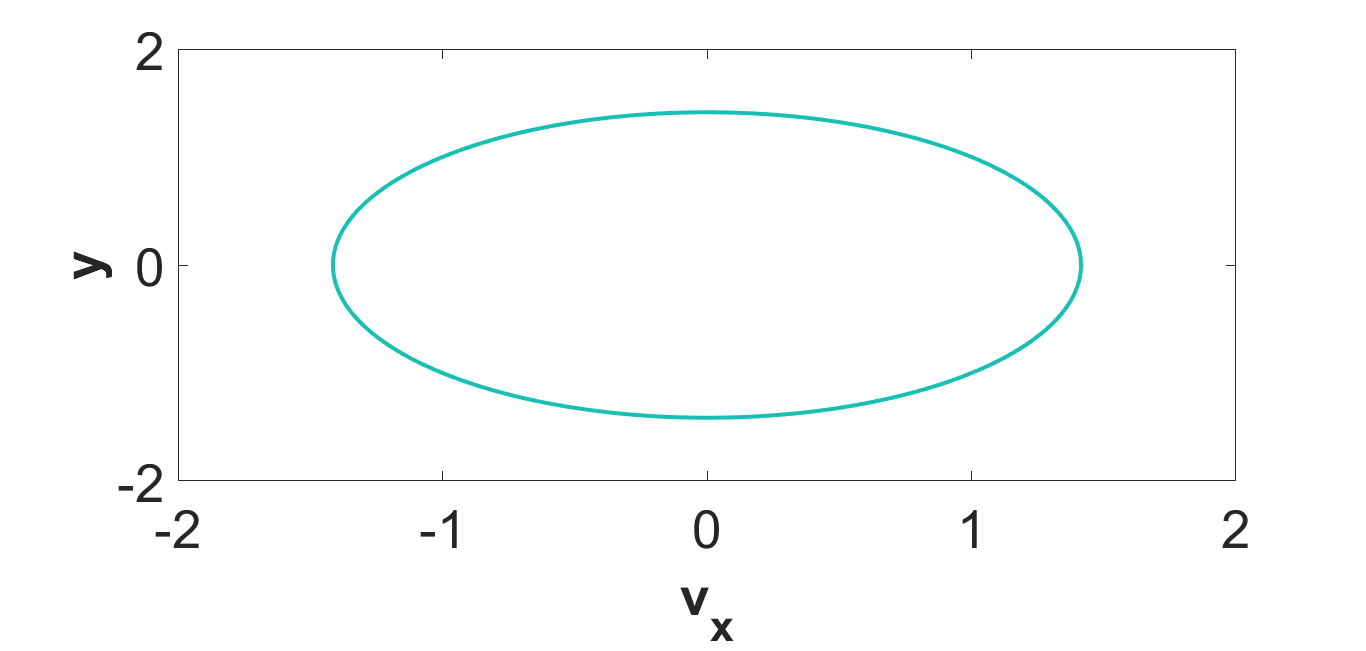}
(j) $a=1 \,A=1,\,B=10 \,and \,F=1$
\quad \qquad (k) $a=1 \,A=1,\,B=1 \,and \,F=0$
\quad \qquad (l) $a=1 \,A=1,\,B=1 \,and \,F=-1$
\caption{Level curves as represented by Eq. (\ref{lc3}) for some arbitrarily chosen parameter values corresponding to the FP $(v_x,y)=(a,-\frac{C}{A})$} \label{lc3fig}
\end{figure}

In this case for the FP3, the first integral Eq. (\ref{Ham}) becomes
\begin{equation}
H(v_x,y)=\frac{v^2_x}{2}+\frac{A}{2B}y^2, \label{Ham3}
\end{equation} 
where we have used Eq. (\ref{FP3cond}) in Eq. (\ref{Ham}) in writing the above Eq. (\ref{Ham3}). Now the value of the first integral Eq. (\ref{Ham3}) at the FP3 $(v_x,y)=(a,-\frac{F}{B})$ is given by
\begin{equation}
H(a,-\frac{F}{B})=\frac{a^2}{2}+\frac{AF^2}{2B^3}=h_3 \label{h3}
\end{equation}
where $h_3$ is a constant corresponding to the first integral Eq. (\ref{Ham}) for the FP3. Using Eqs. (\ref{Ham3}) and (\ref{h3}), the locus of points on $(v_x,y)$ plane or phase plane satisfying the constant value $h_3$ is given by
\begin{equation}
\frac{v^2_x}{2}+\frac{A}{2B}y^2=\frac{a^2}{2}+\frac{AF^2}{2B^3}. \label{Locus3}
\end{equation}
This implies
\begin{equation}
y^2=\frac{2B}{A}(\frac{a^2}{2}+\frac{AF^2}{2B^3}-\frac{v^2_x}{2}). \label{y3}
\end{equation}
From the first equation in the planar dynamical system Eq. (\ref{Phaseplane1}), we have
\begin{equation}
\frac{dv_x}{d\eta}=Ay^2+Cy. \label{Firstreg}
\end{equation}
Substituting the value of $y$ in terms of $v_x$ from Eq. (\ref{y3}) in the above Eq. (\ref{Firstreg}), we get
\begin{equation}
\int \frac{dv_x}{B(a^2+\frac{AF^2}{B^3}-v^2_x)+C\sqrt{\frac{B}{A}}\sqrt{a^2+\frac{AF^2}{B^3}-v^2_x}}=\int d\eta. \label{aFP3}
\end{equation}
After evaluating the above integrations in Eq. (\ref{aFP3}), we obtain
\begin{equation}
v_x=2\sqrt{a^2+\frac{AF^2}{B^3}}\frac{\lambda_1 \lambda_2 tan[\frac{B}{2}\sqrt{\frac{C^2}{AB}-a^2-\frac{AF^2}{B^3}}(\eta +k_3)]}{\lambda^2_1+\lambda^2_2 tan^2[\frac{B}{2}\sqrt{\frac{C^2}{AB}-a^2-\frac{AF^2}{B^3}}(\eta +k_3)]},\label{solfp3}
\end{equation}
where $\lambda_1 =\sqrt{C-\sqrt{AB(a^2+\frac{AF^2}{B^3})}}$, $\lambda_2=\sqrt{C+\sqrt{AB(a^2+\frac{AF^2}{B^3})}}$ and  $k_3$ is the constant of integration. The solution Eq. (\ref{solfp3}) typically looks like Fig. \ref{fp3so} for some arbitrarily chosen values of the parameters. Therefore, in this case for the FP3, the solutions can be found at the level curves given by
\begin{equation}
\frac{A}{B} y^2+v^2_x-a^2-\frac{AF^2}{B^3}=0, \label{lc3}
\end{equation}
which is evident from Eqs. (\ref{y3}) and (\ref{solfp3}). These level curves are plotted in Fig. \ref{lc3fig}.

.
\section{Exact solution of the reduced Eq. (\ref{fFinalNLEE2})} \label{ExSec}
\begin{figure}[hbt!]
\centering
\includegraphics[width=20cm]{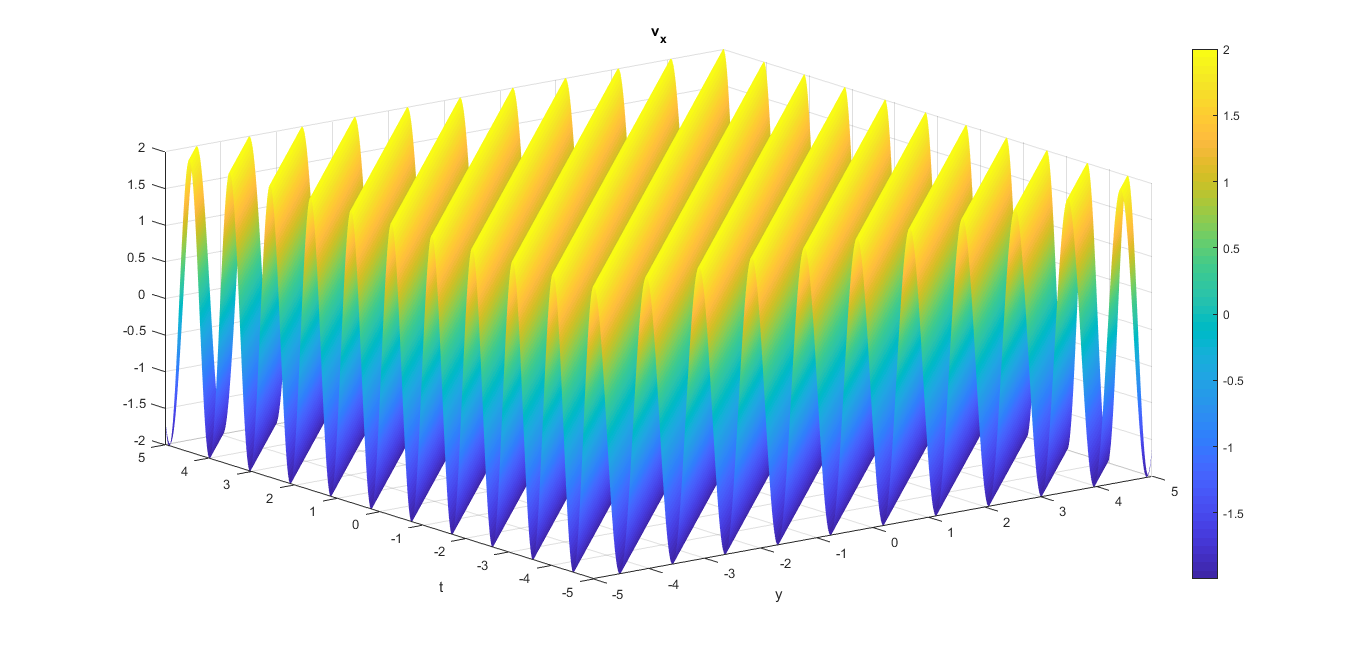}
\caption{Typical behaviour of the solution Eq. (\ref{OscSol1}) for the values of the parameters $\omega_{ci}=9.4kHz,\,c_s=5\times10^5\,cm/sec,\,v=2.3 \times 10^5 \,cm/sec$ and $L_n=13\,cm$ which are the same as observed in \cite{Ghosh2015} with the arbitrary parameters $P=1$ and $Q=1$.} \label{secone}
\end{figure}

\begin{figure}[hbt!]
\centering
\includegraphics[width=20cm]{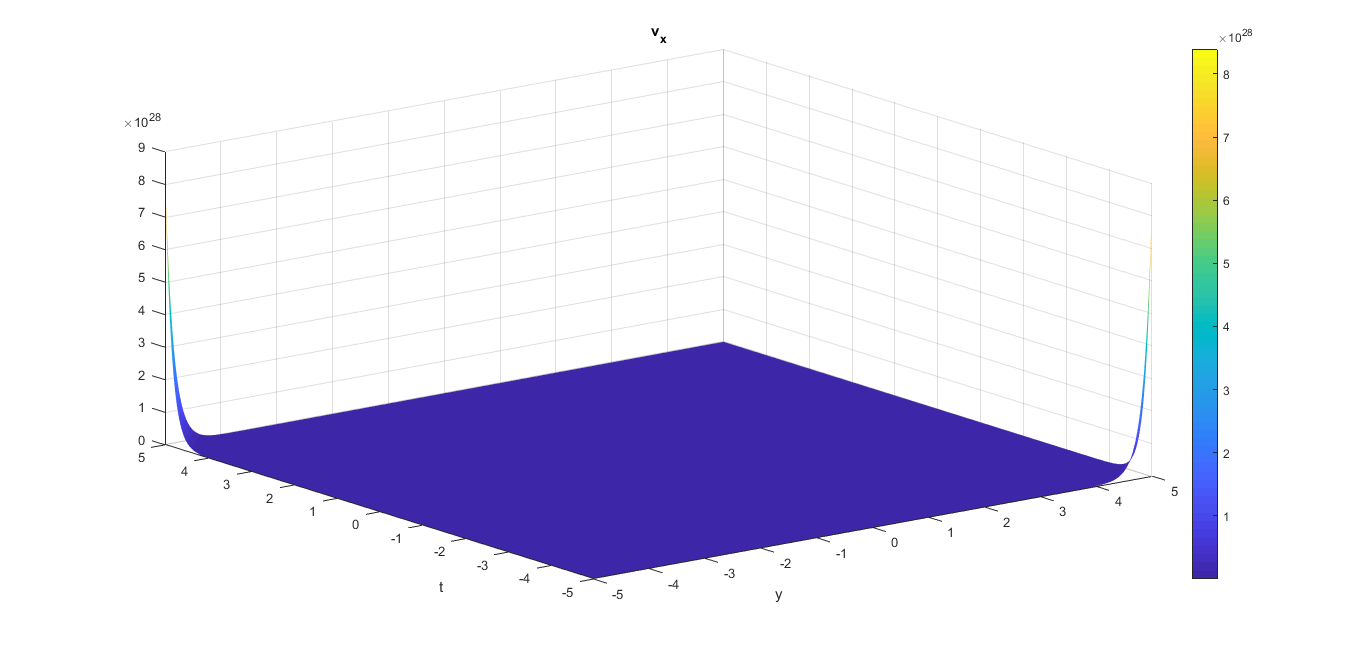}
\caption{Typical behaviour of the solution Eq. (\ref{ExpSol1}) for the values of the parameters $\omega_{ci}=9.4kHz,\,c_s=5\times10^5\,cm/sec,\,v=2.3 \times 10^5 \,cm/sec$ and $L_n=13\,cm$ which are the same as observed in \cite{Ghosh2015} with the arbitrary parameters $R=1$ and $S=1$.} \label{sectwo}
\end{figure}

In this section, we are going to explore the exact solutions of the reduced evolution Eq. (\ref{fFinalNLEE2}) which is similar to a simple harmonic oscillator (SHO). This Eq. (\ref{fFinalNLEE2}) can be rewritten as
\begin{equation}
\frac{d^2v_x}{d \xi^2} \pm \vert \frac{B-E}{A-D} \vert v_x=0. \label{fFinalNLEE2a} 
\end{equation}
It is very well-known that the above Eq. (\ref{fFinalNLEE2a}) admits oscillatory and exponential solutions for the considerations of positive and negative signs respectively as given below:
\begin{equation}
v_x=P\,exp [i\sqrt{\vert \frac{B-E}{A-D}\vert}\xi]+Q\,exp [-i\sqrt{\vert \frac{B-E}{A-D}\vert}\xi], \label{OS}
\end{equation}
 and
\begin{equation}
v_x=R\,exp [\sqrt{\vert \frac{B-E}{A-D}\vert}\xi]+S\,exp [-\sqrt{\vert \frac{B-E}{A-D}\vert}\xi], \label{ES}
\end{equation}
where Eq. (\ref{OS}) represents the oscillatory solution and Eq. (\ref{ES}) stands for the exponential solution. Since $\xi=y-vt$, we have

\begin{equation}
v_x=P\,exp [i\sqrt{\vert \frac{B-E}{A-D}\vert}(y-vt)]+Q\,exp [-i\sqrt{\vert \frac{B-E}{A-D}\vert}(y-vt)], \label{OscSol1}
\end{equation}
 and
\begin{equation}
v_x=R\,exp [\sqrt{\vert \frac{B-E}{A-D}\vert}(y-vt)]+S\,exp [-\sqrt{\vert \frac{B-E}{A-D}\vert}(y-vt)], \label{ExpSol1}
\end{equation}
These two above solutions (\ref{OscSol1}) and (\ref{ExpSol1}) typically look like as shown in Fig. \ref{secone} and Fig. \ref{sectwo} respectively for the experimental values of the parameters observed in \cite{Ghosh2015}. From these Fig. \ref{secone} and Fig. \ref{sectwo}, it is clear that the two solutions Eqs. (\ref{OscSol1}) and (\ref{ExpSol1}) behave in a completely different manner. The solution Eq. (\ref{OscSol1}) is somewhat similar in behaviour to the typical exact and approximate solutions of the reduced planar dynamical system Eq. (\ref{fFinalNLEE1}) which can be seen by comparing Fig. \ref{secone} with Fig. \ref{fp1so}, Fig. \ref{fp2so} and Fig. \ref{fp3so} whereas the solution Eq. (\ref{ExpSol1}) is completely different from these solutions of Eq. (\ref{fFinalNLEE1}).

\section{Solutions of the novel nonlinear evolution Eq. (\ref{FinalNLEE})} \label{Intersection}


We have explored the exact as well as approximate solutions of the two reduced second order Eqs. (\ref{fFinalNLEE1}) and (\ref{fFinalNLEE2}) in Sec. \ref{Solution} and Sec. \ref{ExSec} respectively. In particular, Eqs. (\ref{solfp1}), (\ref{solfp2}) and (\ref{solfp3}) represent the exact as well as approximate solutions of the reduced second order nonlinear Eq. (\ref{fFinalNLEE2}); Eqs. (\ref{OscSol1}) and (\ref{ExpSol1}) represent the exact oscillatory and exponential solutions of the reduced second order linear Eq. (\ref{fFinalNLEE1}). But we are interested in exploring the solutions of the original third order nonlinear evolution Eq. (\ref{FinalNLEE}) which is to be satisfied by the nonlinear drift wave modes in our model.

In order to find the solutions of this Eq. (\ref{FinalNLEE}), it is clear to be understood that the intersections of the solutions of Eqs. (\ref{fFinalNLEE1}) and (\ref{fFinalNLEE2}) can represent the solutions of Eq. (\ref{FinalNLEE}). This is because these intersections of the solutions of Eqs. (\ref{fFinalNLEE1}) and (\ref{fFinalNLEE2}) are supposed to satisfy both the Eqs. (\ref{fFinalNLEE1}) and (\ref{fFinalNLEE2}). This, in turn, implies that these intersections can satisfy any linear combinations Eqs. (\ref{fFinalNLEE1}) and (\ref{fFinalNLEE2}). This is a very standard mathematical point. From Eqs. (\ref{FinalNLEE}) and (\ref{FinalNLEEdec}), it is clear that Eq. (\ref{FinalNLEE}) can be regarded as a particular linear combinantion of Eqs. (\ref{fFinalNLEE1}) and (\ref{fFinalNLEE2}). Therefore, the intersections of the solutions of Eqs. (\ref{fFinalNLEE1}) and (\ref{fFinalNLEE2}) can also satisfy Eq. (\ref{FinalNLEE}). These intersections of the solutions of Eqs. (\ref{fFinalNLEE1}) and (\ref{fFinalNLEE2}) can be regarded as some specific solutions of Eq. (\ref{FinalNLEE}) which are being explored in this section explicitly as given below.

It is easier to infer that the solutions of Eq. (\ref{fFinalNLEE1}) given in Sec. \ref{Solution} are subsets of the oscillatory solution Eq. (\ref{OscSol1}) of Eq. (\ref{fFinalNLEE2}) and the intersections of the solutions of Eqs. (\ref{fFinalNLEE1}) and (\ref{fFinalNLEE2}) are different if we consider the exponential solution Eq. (\ref{ExpSol1}) of Eq. (\ref{fFinalNLEE2}). This fact can be verified by comparing Fig. \ref{secone} with Fig. \ref{fp1so}, Fig. \ref{fp2so} and Fig. \ref{fp3so} in the case of the oscillatory solution Eq. (\ref{OscSol1}); similarly, Fig. \ref{sectwo} can be compared with Fig. \ref{fp1so}, Fig. \ref{fp2so} and Fig. \ref{fp3so} in the case of the exponential solution Eq. (\ref{ExpSol1}) in order to visualize the intersection of the solutions of Eqs. (\ref{fFinalNLEE1}) and (\ref{fFinalNLEE2}). It is observed that the intersections of the solutions of Eqs. (\ref{fFinalNLEE1}) and (\ref{fFinalNLEE2}) can be visualized as solely the solutions of Eq. (\ref{fFinalNLEE1}) when we consider the oscillatory solution of Eq. (\ref{fFinalNLEE2}) or equivalently $(B-E)$ and $(A-D)$ are of the same sign with the coefficients given in Eq. (\ref{abcdef}). If these coefficient relations $(B-E)$ and $(A-D)$ are of the opposite signs, then Eq. (\ref{fFinalNLEE2}) admits the exponential solution Eq. (\ref{ExpSol1}) and the intersection of the solutions of the Eqs. (\ref{fFinalNLEE1}) and (\ref{fFinalNLEE2}) can be directly visualized by comparing Fig. \ref{sectwo} with Fig. \ref{fp1so}, Fig. \ref{fp2so} and Fig. \ref{fp3so}. These intersections of the solutions of the reduced second order Eqs. (\ref{fFinalNLEE1}) and (\ref{fFinalNLEE2}) provide the final solutions of the original third order nonlinear Eq. (\ref{FinalNLEE}).

These types of solutions represent some specific solutions of Eq. (\ref{FinalNLEE}); these are not the general solutions of Eq. (\ref{FinalNLEE}). These specific solutions of Eq. (\ref{FinalNLEE}) have been obtained under the condition that Eq. (\ref{FinalNLEE}) can be decomposable into two reduced second order Eqs. (\ref{fFinalNLEE1}) and (\ref{fFinalNLEE2}). This means that the two parts in the left hand side of Eq. (\ref{FinalNLEEdec}) specified by square brackets vanish separately. This separate vanishing condition is just one possibility which can be implied by Eq. (\ref{FinalNLEE}). In general, there can be many such possibilities apart from the above-mentioned possibility. We have chosen this peculiar possibilty as the order of Eq. (\ref{FinalNLEE}) is reduced from three to two and one reduced equation becomes nonlinear whereas the other reduced equation becomes linear in this possibility. On the other hand, the reduced linear equation is also equivalent to the one-dimensional stationary HM equation as explored in Appendix A. Similarly the other reduced nonlinear equation can be regarded as a nonlinear generalization of the one-dimensional stationary HM equation. Therefore, under this possibiollity, the novel nonlinear evolution Eq. (\ref{FinalNLEE}) can be reduced in confrontation with the HM equation. This again implies that the excitations of both the conventional low frequency and the recently observed high frequency electrostatic drift waves can be correlated by such a particular reduction of Eq. (\ref{FinalNLEE}) into Eqs. (\ref{fFinalNLEE1}) and (\ref{fFinalNLEE2}). Taking into account these facts, it can be seen that the general solution of Eq. (\ref{FinalNLEE}) will be a superset of these solutions based on the intersections of the solutions of Eqs. (\ref{fFinalNLEE1}) and (\ref{fFinalNLEE2}); this is planned to be investigated in future.


One important aspect of the solutions Eqs. (\ref{solfp1}), (\ref{solfp2}) and (\ref{solfp3}) of the novel nonlinear evolution Eq. (\ref{FinalNLEE}) can be visualized by carefully observing Fig. \ref{fp1so}, Fig. \ref{fp2so} and Fig. \ref{fp3so}. These figures imply the presence of lattices of vortex-like structures; one can also interpret these as a series of localized solitary wave structures on $yt$ plane. Inspired by these ideas, we plan to explore the possibility of obtaining multiple solitary wave solutions of Eq. (\ref{FinalNLEE}) in some special limiting cases in future. In this context, it should be emphasized that the investigations on multiple solitary wave solutions of different types of nonlinear differential equations have been reported extensively by many researchers. In the work of Wazwaz \cite{Wazwaz2009}, multiple soliton solutions for coupled Korteweg-de Vries (KdV) and coupled Kadomtsev-Petviashvili (KP) systems have been studied by applying the Hirota bilinear method.  Multiple soliton solutions for a new generalization of the associated Camassa-Holm equation are reported in \cite{Long} using exp-function method. Two Boussinesq equations where the fourth order terms come with minus and plus signs have been investigated in \cite{Wazwaz2016nld} to derive multiple soliton solutions in case of the minus sign and multiple complex soliton solutions in case of the plus sign. Similarly multiple soliton solutions of NLS and coupled Burgers equations are explored in \cite{Akbar2019}.

In addition, for integrable equations, Riemann-Hilbert problems \cite{MaW2022, MaWW2022, MaWWW2022} are very important in the sense that they are efficient in establishing inverse scattering transforms (ISTs) and presenting soliton solutions. The recent work of Ma \cite{MaW2022} deals with the formulation of Riemann-Hilbert problems and soliton solutions of nonlocal reverse-time NLS hierarchies associated with the higher order matrix spectral problems. In particular, he \cite{MaW2022} proposed a novel formulation of solutions for dealing with the special Riemann-Hilbert problems with the identity jump matrix corresponding to the reflectionless ISTs; which are subsequently applied to each system in the considered nonlocal reverse-time NLS hierarchies to construct $N$-soliton solutions. Similarly the Riemann-Hilbert problems associated with the nonlocal real reverse-spacetime integrable hierarchies of Ablowitz–Kaup–Newell–Segur (AKNS) equations have been formulated in \cite{MaWW2022} to determine the generalized Jost solutions of arbitrary-order matrix spectral problems. N-soliton solutions are also derived for the system of AKNS hierarchies \cite{MaWW2022} after solving the Riemann-Hilbert problems corresponding to the reflectionless case explicitly. In another work of Ma \cite{MaWWW2022}, N-soliton solutions of reduced nonlocal integrable modified KdV (mKdV) hierarchies have been derived by solving the associated Riemann-Hilbert problems. Apart from these soliton solutions, it is also interesting to explore different kinds of exact solutions such as lump wave solutions, Rossby wave solutions, solitonless solutions, algebro-geometric solutions, dromions etc. in detail using the Riemann-Hilbert technique \cite{MaW2022} for different integrable differential equations.

\section{Discussions and applications} \label{DisApp}
 In this section, we are going to recapitulate some crucial findings of our theoretical study on the phenomenon of the excitations of nonlinear high frequency electrostatic drift waves using the plasma fluid model along with elaborate discussions on these results and possible applications. This is explained below in detail.

We have derived a novel nonlinear evolution Eq. (\ref{FinalNLEE}) of third order in Sec. \ref{Der} which is typically of the form of a Jerk equation \cite{Wharton}; this kind of nonlinear equation has not been reported till now for dealing with the drift waves as far as our knowledge goes. Thereafter, we have decomposed this novel nonlinear evolution Eq. (\ref{FinalNLEE}) into two parts as provided in Eq. (\ref{FinalNLEEdec}). This is due to the fact that these two parts are separately integrable once explicitly; the final form of these reduced equations are given in Eqs. (\ref{fFinalNLEE1}) and (\ref{fFinalNLEE2}). These two reduced equations are further consolidated for executing the analysis. One of the most crucial findings of our work is that we have proved the existence of the high frequency electrostatic drift waves using the dispersion relation Eq. (\ref{cubic_dr}) in Sec. \ref{DR}, which is derived using the derived novel nonlinear evolution Eq. (\ref{FinalNLEE}). This can be ascertained by carefully observing Fig. \ref{ExR2} where the variations of $\frac{\omega}{\omega_{ci}}$ have been shown against the variations of the density gradient scale length $L_n$ using the three exact roots of the dispersion relation Eq. (\ref{cubic_dr}). In this figure, the red and blue lines correspond to the ion cyclotron wave whereas the green line represents the drift wave. In this context, it has to be emphasized that the green line in this figure corresponds to both the conventional low frequency as well as the recently observed high frequency electrostatic drift waves; the low frequency electrostatic drift waves correspond to the region specified by thre limit $\frac{\omega}{\omega_{ci}}\rightarrow 0$ on this green line whereas the high frequency electrostatic drift waves correspond to the region specified by the limit $\frac{\omega}{\omega_{ci}}\rightarrow 1$. Now it is clear from Fig. \ref{ExR2} that both the regions corresponding to the conventional low frequency as well as the recently observed high frequency electrostatic drift waves vary significantly with respect to $L_n$. This justifies their drift wave nature appreciably.

Apart from this, we have also analyzed the dispersiopn relation Eq. (\ref{cubic_dr}) in Appendix B in detail applying various approximations. The well-known drift wave frequency $\omega_{lfd}$ representing the conventional low frequency electrostatic drift waves has been provided in Eq. (\ref{Lowf}) and a novel drift wave frequency $\omega_{hfd}$ corresponding to the recently observed high frequency electrostatic drift waves has been derived for the first time as shown in Eq. (\ref{Highf1a}). This approximated dispersion relation represented by Eq. (\ref{Highf1a}) corresponds to this recently observed high frequency electrostatic drift waves \cite{Ghosh2015, Ghosh2017} which is also studied theoretically in the present work. It is also clear from the expressions for $\omega_{lfd}$ and $\omega_{hfd}$ given in Eqs. (\ref{Lowf}) and (\ref{Highf1a}) respectively that the denominator of $\omega_{hfd}$ is less than that of $\omega_{lfd}$ by unity; this implies that $\lvert \omega_{hfd} \rvert > \lvert \omega_{lfd} \rvert$ as desired. The dependences of the exact root corresponding to the drift wave of the dispersion relation Eq. (\ref{cubic_dr}) on the density gradient scale length $L_n$ in both the conventional low frequency electrostatic drift wave region and the  recently observed high frequency electrostatic drift wave region along with the dependences of both the frequencies $\omega_{lfd}$ and $\omega_{hfd}$, which are similar in their mathematical expressions, on $L_n$ support appreciably the drift wave nature of $\omega_{hfd}$ especially as the drift wave nature of $\omega_{lfd}$ is already very well-known. We have also obtained the dispersion relation for the electrostatic ion cyclotron wave which is given in Eq. (\ref{Highf3}). One important aspect in this regard is that the coupling between the ion cyclotron and drift waves is not applicable in our model due to the fact that this kind of coupling generally occurs when the wavelength in our system becomes comparable with the ion Larmor radius \cite{Vranjes2008AA} whereas we are far removed from this regime by considering a fluid model in this work.

Then we proceed to analyse the two reduced Eqs. (\ref{fFinalNLEE1}) and (\ref{fFinalNLEE2}) explicitly. The fixed point analysis leading to the bifurcations of the phase portraits of the reduced Eq. (\ref{fFinalNLEE1}) have been performed in Sec. \ref{PPBifur}. Here the theory of planar dynamical systems has been employed for characterizing the bifurcations of the fixed points using the first integral or Hamiltonian function $H(v_x,y)$ given in Eq. (\ref{Ham}). It has to be mentioned here that the two fixed points $(0,0)$ and $(0,-\frac{C}{A})$ show characteristic bifurcation behaviours whereas the other fixed point $(a,-\frac{F}{B})$ does not show any bifurcation. In particular, the bifurcations of the phase portraits have been analysed explicitly for the two fixed points which are also accompanied with the bifurcation curves in the parameter space. The bifurcation curves are obtained using the linear stability analysis of the fixed points as shown in Fig. \ref{BCOne} and Fig. \ref{BCTwo} whereas the bifurcations of the phase portraits have been done using the first integral in case of all the fixed points as shown in Fig. \ref{BifPP1} and Fig. \ref{BifPP2}. The typical regular strutures of the solutions of the reduced second order nonlinear Eq. (\ref{fFinalNLEE1}) can also be visualized from the phase portraits shown in Fig. \ref{BifPP1} and Fig. \ref{BifPP2}. Then some exact as well as approximate solutions of this reduced nonlinear Eq. (\ref{fFinalNLEE1}) have been derived using the Hamiltonian function and considering all the fixed points explicitly. Subsequently the exact oscillatory and exponential solutions of the reduced second order linear Eq. (\ref{fFinalNLEE2}) have been derived in Sec. \ref{ExSec}.

Now it can be seen that the solutions of Eq. (\ref{fFinalNLEE1}) given in Sec. \ref{Solution} are subsets of the oscillatory solution Eq. (\ref{OscSol1}) whereas the the intersection of the solutions of Eqs. (\ref{fFinalNLEE1}) and (\ref{fFinalNLEE2}) are different if we consider the exponential solution Eq. (\ref{ExpSol1}) of Eq. (\ref{fFinalNLEE1}). This peculiar behaviour of the solutions can be explored in detail by comparing Fig.  \ref{secone} with Fig. \ref{fp1so}, Fig. \ref{fp2so} and Fig. \ref{fp3so} in the case of the oscillatory solution  Eq. (\ref{OscSol1}). Similarly, Fig. \ref{sectwo} can be compared with Fig. \ref{fp1so}, Fig. \ref{fp2so} and Fig. \ref{fp3so} in the case of the exponential solution Eq. (\ref{ExpSol1}) in order to visualize the intersection of the solutions of Eqs. (\ref{fFinalNLEE1}) and (\ref{fFinalNLEE2}). Therefore, the intersections of the solutions of Eqs. (\ref{fFinalNLEE1}) and (\ref{fFinalNLEE2}) can be visualized as solely the solutions of Eq. (\ref{fFinalNLEE1}), which is also explored in Sec. \ref{Intersection}, if we consider the oscillatory solution of Eq. (\ref{fFinalNLEE2}) or equivalently the signs of the coefficient relations $(B-E)$ and $(A-D)$ are the same; the coefficients are given in Eq. (\ref{abcdef}). If these coefficient relations $(B-E)$ and $(A-D)$ are of the opposite signs, then Eq. (\ref{fFinalNLEE2}) admits the exponential solution Eq. (\ref{ExpSol1}) and the intersection of Eqs. (\ref{fFinalNLEE1}) and (\ref{fFinalNLEE2}) can directly be visualized by comparing Fig. \ref{sectwo} with Fig. \ref{fp1so}, Fig. \ref{fp2so} and Fig. \ref{fp3so}. These tnersections of the solutions of the reduced second order Eqs. (\ref{fFinalNLEE1}) and (\ref{fFinalNLEE2}) provide the final solutions of the original third order nonlinear Eq. (\ref{FinalNLEE}). These solutions of Eq. (\ref{FinalNLEE}) are discussed in Sec. \ref{Intersection} in detail. In addition, it is to be noted here that thess kinds of solutions of Eq. (\ref{FinalNLEE}) represent specific solutions which are obtained under the condition that Eq. (\ref{FinalNLEE}) can be decomposable into two reduced second order Eqs. (\ref{fFinalNLEE1}) and (\ref{fFinalNLEE2}). The general solution of Eq. (\ref{FinalNLEE}) will be a superset of these specific solutions of Eq. (\ref{FinalNLEE}) based on the intersections of the solutions of Eqs. (\ref{fFinalNLEE1}) and (\ref{fFinalNLEE2}).

One of the most important findings of our work is that certain vortex-like nonlinear structures can be excited as the high frequency drift wave modes. These vortex-like structures can be ascetrtained by observing the solutions of Eq. (\ref{FinalNLEE}) at constant times. In particular, these structures can be visualized more precisely by carefully observing Fig. \ref{fp1so}, Fig. \ref{fp2so} and Fig. \ref{fp3so} along with  Fig. \ref{secone} and Fig. \ref{sectwo} at constant times. It is obvious that certain regular structures in one-dimension are to be resulted at constant times. If the coefficient relations $(B-E)$ and $(A-D)$ are of the same sign with the coefficients given in Eq. (\ref{abcdef}), then these time projections in Fig. \ref{fp1so}, Fig. \ref{fp2so} and Fig. \ref{fp3so} can directly represent the vortex-like structures of Eq. (\ref{FinalNLEE}) whereas the intersections of the time projections in Fig. \ref{sectwo} with those in Fig. \ref{fp1so}, Fig. \ref{fp2so} and Fig. \ref{fp3so} represent the required vortex-like structures of Eq. (\ref{FinalNLEE}). The theoretical analysis of such vortex-like nonlinear structures presented in this article in $(1+1)$ dimensions representing the recently observed high frequency electrostatic drift waves using the plasma fluid model is desirably novel and warrant further theoretical investigations of the same in $(2+1)$ dimensions. This is because vortices are basically two-dimensional structures that can be explored more suitably in $(2+1)$ dimensions. This is in our future plan.


We know that the drift waves are extremely useful for the understanding of transport phenomena in different plasma systems ranging from laboratory to space and astrophysical situations. In particular, for anomalous transport, the conventional low frequency electrostatic drift waves play a very crucial role where both electrons and ions can participate. Although electrons predominantly participate in the excitations of the high frequency drift waves, there are also chances that ions may involve in it according to Detyna and Wooding \cite{Detyna1972}. According to Detyna and Wooding \cite{Detyna1972}, ions in the regions of steep density gradients such as edge regions in a tokamak can excite ultra high frequency drift waves which may become unstable. Therefore, in our system consisting of the high frequency electrostatic drift waves, both electrons and ions may be supposed to participate in the regions of steep density gradients in a plasma, thus, providing even nore significant contributions to the anomalous transport phenomenon than that of the conventional low frequency electrostatic drift waves. But this fact is subject to experimental verifications. High frequency electrostatic drift waves are also extremely useful in fusion plasmas. Besides anomalous transport, they can have significant contributions in plasma heating and current drive leading to nuclear fusion. Due to their high frequency nature, these electrostatic drift waves may have effectively more significance than the conventional low frequency drift waves in these plasma phenomena, viz. anomalous transport, heating, current drive etc. These characteristics make the high frequency electrostatic drift waves special as they are inevitably significant even without a magnetic field perturbation in the plasma system considered.

The nonlinear dynamical modelling of the recently observed \cite{Ghosh2015,Ghosh2017} high frequency electrostatic drift waves using a novel nonlinear evolution Eq. (\ref{FinalNLEE}) presented in this article can enact as a milestone in the research on the nonlinear drift waves. This is because this nonlinear evolution Eq. (\ref{FinalNLEE}) generalizes the Hasegawa-Mima Eq. \cite{Hasegawa,Horton} nonlinearly in one-dimension to the high frequency regime in the case of the electrostatic drift waves. The results on the nonlinear dynamics of the high frequency electrostatic drift wave modes presented in this article may be verified by equivalent experiments on drift waves. These kinds of high frequency electrostatic drift waves can be found, in particular, in experimental situations where the magnetic field is comparatively low as discussed in \cite{Ghosh2015,Ghosh2017}. Apart from this, in many space and astrophysical plasmas, the high frequency electrostatic drift waves can occur abundantly. Our theoretical formulation of the high frequency electrostatis drift waves can have potential applications in such situations. This further initiates an emerging active field for research on the high frequency electrostatic drift waves in analytical, computational and experimental perspectives in different laboratory, space and astrophysical plasma conditions.


\section{Concluding remarks} \label{Con}
To conclude, we have studied theoretically the excitations of the recently observed high frequency electrostatic drift waves in an inhomogeneous plasma with a constant magnetic field in the framework of a plasma fluid model yielding a novel nonlinear evolution equation of third order. Subsequently, for simplicity of calculations, it is seen that this derived third order nonlinear evolution equation can be reduced to two second order equations for further analysis. The most interesting fact in this context is that one reduced equation is linear admitting exact solutions whereas the other reduced equation is nonlinear admitting both exact and approximate solutions which are explored in detail. The cubic linear dispersion relation derived after linearizing the fluid equations is then analysed in detail explaining the excitations of the conventional low frequency as well as the recently observed high frequency electrostatic drift waves along with the ion cyclotron waves. A detailed investigation of the fixed points along with the bifurcations of the phase portraits from the reduced second order nonlinear equation has been carried out through the first integral or equivalent Hamiltonian function of the system using the well-known theory of planar dynamical systems. The bifurcation curves of the fixed points are also obtained in the parameter space. Then, some exact as well as approximate vortex-like solutions of the reduced second order nonlinear equation are derived by considering the fixed points explicitly; precisely, the vortex-like solutions can be visualized from the plots of these exact as well as approximate solutions. The exact oscillatory and exponential solutions of the other reduced second order linear equation are discussed subsequently. The solutions of the original derived nonlinear evolution equation of third order which can be provided by the intersections of the solutions of the two reduced second order equations have also been explored. Then the the novel results of our theoretical work presented in this study are recapitulated along with detailed discussions and possible applications accompanied with future research plans. The nonlinear dynamical modelling of this recently observed high frequency electrostatic drift waves using the fluid theoretical approach presented in this article can have the potential to initiate a new regime for extensive research on the high frequency electrostatic drift waves.
 \begin{center}
 {\bf APPENDIX}
 \end{center}
\begin{center}
 {\bf A) Comparison of novel nonlinear evolution equation with HM equation}
\end{center}

We are now going to discuss some similarities between the derived novel nonlinear evolution Eq. (\ref{FinalNLEE}) associated with the high frequency electrostatic drift waves and the well-known Hasegawa-Mima (HM) Eq. \cite{Hasegawa, Horton} associated with the conventional low frequency electrostatic drift waves. The HM equation is two-dimensional in nature \cite{Horton, Hasegawa} which is given by
\begin{equation}
\frac{\partial}{\partial t}({\nabla}^2\phi-\phi)-[(\vec{\nabla}\phi\times \hat{z}).\vec{\nabla}][{\nabla}^2\phi-ln(\frac{n_0}{{\omega}_{ci}})]=0, \label{HM_eqn}
\end{equation}
In terms of components, it can be written as
\begin{equation}
\frac{\partial^3\phi}{\partial t \partial x^2}+\frac{\partial^3\phi}{\partial t \partial y^2}-\frac{\partial \phi}{\partial t}+\frac{\partial \phi}{\partial x}\frac{\partial^3\phi}{\partial y \partial x^2}+\frac{\partial \phi}{\partial x}\frac{\partial^3\phi}{ \partial y^3}-\frac{\partial \phi}{\partial x}\frac{\partial}{\partial y}[ln\,(\frac{n_0}{\omega_{ci}})]-\frac{\partial \phi}{\partial y}\frac{\partial^3\phi}{ \partial x^3}-\frac{\partial \phi}{\partial y}\frac{\partial^3\phi}{\partial x \partial y^2}-\frac{1}{L_n}\frac{\partial \phi}{\partial y}=0. \label{HM_eqn1}
\end{equation}

It can be easily seen that the above Eq. (\ref{HM_eqn1}), in one-dimension, becomes
\begin{equation}
\frac{\partial}{\partial t}(\frac{\partial^2\phi}{\partial y^2}-\phi) -\frac{1}{L_n}\frac{\partial \phi}{\partial y}=0; if \frac{\partial}{\partial x}=0. \label{HM1D}
\end{equation}
If we apply the frame transformation given by Eq. (\ref{FramTr}) to Eq. (\ref{HM1D}), we get
\begin{equation}
\frac{d^3\phi}{d\xi^3}-(1+\frac{1}{L_n})\frac{d\phi}{d\xi}=0 \,or\,\frac{d^2\phi}{d\xi^2}-(1+\frac{1}{L_n})\phi=K_1, \label{1HM1D}
\end{equation}
where $K_1$ is the constant of integration. If we take $K_1=0$ for simplicity, then we have
\begin{equation}
\frac{d^2\phi}{d\xi^2}-(1+\frac{1}{L_n})\phi=0. \label{2HM1D}
\end{equation}
Now it is clear that the reduced Eq. (\ref{fFinalNLEE2}) is similar to Eq. (\ref{2HM1D}) for some specific values of the coefficients $A,\,B,\,D$ and $E$ whereas the other reduced Eq. (\ref{fFinalNLEE1}) can be regarded as a specific nonlinear generalization of Eq. (\ref{2HM1D}) due to the presence of the function $f(\frac{dv_x}{d\xi})$. Therefore, Eq. (\ref{FinalNLEE}) can be regarded as a nonlinear high frequency generalization of the stationary HM equation Eq. (\ref{2HM1D}) in $(1+1)$ dimensions. Therefore, this novel nonlinear evolution Eq. (\ref{FinalNLEE}) can satisfactorily explain the excitations of the high frequency electrostatic drift waves in correspondence with the HM equation in stationary $(1+1)$ dimensions; the excitations of these high frequency electrostatic drift waves are also experimentally observed by Ghosh et al. \cite{Ghosh2015, Ghosh2017} in the last decade. In particular, it should be emphasized that, in $(1+1)$ dimensions followed by a stationary frame transformation like Eq. (\ref{FramTr}), only some specific drift wave modes are to be excited which need to satisfy  Eq. (\ref{FinalNLEE}) in the high frequency regime whereas the low frequency conventional drift wave modes need to satisfy only the stationary HM Eq. (\ref{2HM1D}). If we intend to correlate these excitations of both the low and high frequency electrostatic drift waves, it should be signified that the excited high frequency electrostatic drift wave modes are required to satisfy both Eqs. (\ref{fFinalNLEE1}) and (\ref{fFinalNLEE2}), which represent a pair of equations as a specific nonlinear generalization of Eq. (\ref{2HM1D}); this stationary HM Eq. (\ref{2HM1D}) is the equation required to be satisfied by the excited low frequency electrostatic drift wave modes. In addition, the excitations of certain other high frequency electrostatic drift wave modes can be possible excluding the above-mentioned drift modes satisfying Eqs. (\ref{fFinalNLEE1}) and (\ref{fFinalNLEE2}); these modes can be analyzed using the direct solutions of Eq. (\ref{FinalNLEE}) which is in our future plan.

\begin{center}
{\bf B) Limiting cases of linear dispersion relation}
\end{center} 

We now explore certain limiting cases of the linear dispersion relation Eq. (\ref{cubic_dr}) in order to explain the excitations of the conventional low frequency electrostatic drift waves as well as the recently observed high frequency electrostatic drift waves along with the electrostatic ion cyclotron waves. These are explained in the following two limiting cases.
\begin{center}
{\bf Case-1: $\omega<<\omega_{ci}$}
\end{center}

In this case, the first term in the dispersion relation Eq. (\ref{cubic_dr}) is negligible yielding
\begin{equation}
\omega =\omega_{lfd}= \frac{k_yv_d}{1+k^2_y \rho^2_i};\,v_d=\frac{c^2_s}{L_n \omega_{ci}};\,\rho_i=\frac{c_s}{\omega_{ci}}, \label{Lowf}
\end{equation}
where $\omega_{lfd}$ is the frequency corresponding to the low frequency conventional drift waves, $v_d$ denotes the diamagnetic drift velocity and $\rho_i$ represents the ion Larmor radius. The above dispersion relation Eq. (\ref{Lowf}) represents the well-known drift wave frequency for the conventional low frequency electrostatic drift waves.
\begin{center}
{\bf Case-2: $\omega \approx \omega_{ci}$}
\end{center}

In this case, there can be two predominant possibilities according to different terms in the left hand side of the dispersion relation Eq. (\ref{cubic_dr}). 
\begin{center}
{\bf 1. High frequency electrostatic drift wave}
\end{center}

We substitute $\omega$ in place of $\omega_{ci}$ only in the second term in the left hand side of the dispersion relation Eq. (\ref{cubic_dr}) as $\omega \approx \omega_{ci}$. This implies
\begin{equation}
\omega=\omega_{hfd}=\frac{\omega_{ci}}{k_yL_n}, \label{Highf1}
\end{equation}
where $\omega_{hfd}$ represents the high frequency electrostatic drift waves. In order to compare the dispersion relations corresponding to the low frequency conventional electrostatic drift waves and the recently observed high frequency electrostatic drift waves given in Eqs. (\ref{Lowf}) and (\ref{Highf1}) respectively, we now proceed to write the above dispersion relation Eq. (\ref{Highf1}) in terms of the diamagnetic drift velocity $v_d$ and ion Larmor radius $\rho_i$ as:
\begin{equation}
\omega_{hfd} =\frac{v_d}{k_y\rho^2_i}=\frac{k_yv_d}{k^2_y\rho^2_i}.\label{Highf1a}
\end{equation}

Therefore, from Eqs. (\ref{Lowf}) and (\ref{Highf1a}), it is clear that the only difference between the dispersion relations corresponding to the conventional low frequency electrostatic drift waves and the recently observed high frequency electrostatic drift waves is that the denominator in the dispersion relation of the high frequency drift waves is reduced by unity from that in the case of the conventional low frequency drift waves. Mathematically, this relation can be represented as:
\begin{equation}
\frac{1}{\omega_{lfd}}-\frac{1}{\omega_{hfd}}=\frac{1}{k_yv_d}. \label{RelLFHF}
\end{equation}

The above Eq. (\ref{RelLFHF}) effectively implies that the magnitude of the difference between the reciprocals of the low and high frequency electrostatic drift waves is equal to the reciprocal of the product of the perpendicular (with respect to the magnetic field direction) wave number $k_y$ and drift velocity $v_d$. The most important point is that both the expressions for $\omega_{lfd}$ and $\omega_{hfd}$ as given in Eqs. (\ref{Lowf}) and (\ref{Highf1a}) respectively are similar in nature mathematically, and vary significantly with respect to the variations of the density gradient scale length $L_n$; the equivalent behaviours of $\omega_{lfd}$ and $\omega_{hfd}$ against $L_n$ can be verified from Fig. \ref{ExR2} in the limits $\omega<<\omega_{ci}$ and $\omega \approx \omega_{ci}$. This justifies the drift wave nature of both $\omega_{lfd}$ and $\omega_{hfd}$. We know that the drift wave nature of $\omega_{lfd}$ is already very well established whereas that of $\omega_{hfd}$ is sufficiently novel and has not been reported till now as far as our knowledge goes.

\begin{center}
{\bf 2. Electrostatic ion cyclotron wave}
\end{center}

In the limit $\omega \approx \omega_{ci}$, there can be one more possibility implying that the third term in the dispersion relation Eq. (\ref{cubic_dr}) is negligible in comparison to the other terms. This yields
\begin{equation}
\omega^2=\omega_{ci}^2+k^2_yc^2_s. \label{Highf3}
\end{equation}
The above dispersion relation Eq. (\ref{Highf3}) represents the electrostatic ion cyclotron waves. This is because this mode exists even in the absence of the density gradient, i.e. when $L_n=0$. In addition, it should be noted that if we keep the third term in the dispersion relation, it may imply the coupling between the ion cyclotron and drift waves; but, this is not applicable in our model as the coupling may occur only in a narrow range of wavelengths when the wavelength becomes comparable to the ion Larmor radius as explored by Vranjes and Poedts \cite{Vranjes2008AA} whereas, in our fluid theoretical model, the wavelength is always greater than the ion Larmor radius.

 
\section{Acknowledgements}
Siba Prasad Acharya gratefully acknowledges the financial help received from Department of Atomic Energy (DAE) of Government of India through institute fellowship scheme. The authors are thankful to Prof. Nikhil Chakrabarti for some illuminating discussions on the physics of drift waves during the progress of this work. We also thank Dr. Pankaj Kumar Shaw for his useful comments on the nature of the novel nonlinear evolution equation derived in this work. Finally the authors are grateful to the two reviewers for their insightful comments to improvise this manuscript.
  
\end{document}